\documentclass[12pt,a4paper,english]{paper}
\usepackage{ae,aecompl}

\usepackage[T1]{fontenc}
\usepackage[latin9]{inputenc}
\usepackage{array}
\usepackage{float}
\usepackage{amstext}
\usepackage{amssymb}

\makeatletter

\providecommand{\tabularnewline}{\\}

\date{}
\usepackage{textcomp}
\usepackage[T1]{fontenc}
\usepackage{lmodern}
\@addtoreset{equation}{section}
\makeatother

\usepackage{cite}
\usepackage[left=1.5cm,top=1.cm,right=1cm]{geometry}
\usepackage{multicol}
\usepackage{appendix}
\usepackage{amssymb, amsmath}

\makeatother

\usepackage{babel}

\begin{document}
\setlength\abovedisplayskip{0pt} 
\setlength\belowdisplayskip{0pt}

\title{Irreducible tensor form of three-particle operator for open-shell
atoms}

\author{R. Jur\v{s}\.{e}nas and G. Merkelis}

\institution{Institute of Theoretical Physics and Astronomy of Vilnius University,
A. Go\v{s}tauto 12, LT-01108, Vilnius, Lithuania}
\maketitle
\begin{abstract}
The three-particle operator in a second quantized form is studied.
The operator is transformed into irreducible tensor form. Possible
coupling schemes, distinguished by the classes of symmetric group
$\mathrm{S_{6}}$, are presented. Recoupling coefficients, which allow
one to transform given scheme into another, are produced by using
the angular momentum theory, combined with quasispin formalism. The
classification of three-particle operator, which acts on $n=1,2,\ldots,6$
open shells of equivalent electrons of atom, is considered. The procedure
to construct three-particle matrix elements are examined.\end{abstract}

\section{Introduction}

The mathematical technique of effective operators plays an important
role in the studies of either atomic open-shell many-body perturbation
theories \cite{Brandow,Kvasnicka,Lindgren,Lindgren2} or atomic inner-shell
physics, when calculating transition probabilities or widths of levels
\cite{Karazija,Merkelis}. This approach provides an opportunity to
include electron-electron correlation effects in a flexible form.
The irreducible tensor forms of one-, two-particle effective operators
have been widely studied in many works \cite{Judd,Judd2,Rudzikas,Rudzikas2,Merkelis2}.
Starting from the second-order MBPT, in a particle-hole representation
one has to deal with expansion terms, which include three-particle
operators. In this paper we namely concentrate on classification of
effective three-particle operator, written in irreducible tensor form.
Special attention is paid for many open-shell aspects of the problem.
The group-theoretical classification of irreducible three-particle
operator for equivalent electrons (particularly, $f$ electrons) has
been considered by Judd \cite{Judd}. In our considerations we do
not classify the states of equivalent electrons. Our aim is the operator
by itself, providing a general classification of all possible coupling
schemes and determining connection among them.

The three-particle operator may inscribe on six open-shell electrons
at once. It is assumed that irreducible tensor operator matrix element
is constructed on the basis of multi-shell wave functions, coupled
in a consequent order. For this reason, tensor operator is also expressed
in a consequent order with respect to creation and annihilation operators,
which act on the first, second, etc. open-shell. All other mixed arrangements
are obtained by the corresponding permutation operations of $\mathrm{S_{6}}$
group. Recoupling coefficients, which appear due to such transformations,
are presented too. 

Besides the tensor structure, effective three-particle operator includes
weight coefficient. In MBPT this coefficient denotes miscellaneous
products of one- or two-particle matrix elements (with energy denominator
included). The systematic study of two-particle matrix elements can
be found in \cite{Gaigalas,Gaigalas2,Jursenas}. The application of
this methodology for the second-order effective Hamiltonian one can
find in \cite{Jursenas2}.

\section{Preliminaries\label{definitions}}

Let a generalized three-particle operator be

\begin{equation}
\widehat{L}_{3}={\displaystyle \sum_{ijk}}{\displaystyle \sum_{lpq}}a_{i}a_{j}a_{k}a_{l}^{\dagger}a_{p}^{\dagger}a_{q}^{\dagger}\;\ell_{ijkqpl}.\label{eq:2.1}\end{equation}

\noindent{}where $\ell_{ijkqpl}$ denotes weight coefficient, which
depends on concrete operator. The subscripts $x=i,j,\ldots$ indicate
the sets of one-electron quantum numbers $\left\{ n_{x}\lambda_{x}\mu_{x}\right\} $.
Let $\Lambda$ be either reducible ($LS$-coupling) or irreducible
($jj$-coupling) tensor space. Then each creation operator $a_{x}$
is labeled by irreducible representations $\lambda_{x}=l_{x}\frac{1}{2}$
(in $LS$-coupling) or $\lambda_{x}=j_{x}$ (in $jj$-coupling), denoting
$a_{\mu_{x}}^{\lambda_{x}}$. The reduction of Eq. (\ref{eq:2.1})
is performed transposing annihilation operator by $\tilde{a}_{\mu_{x}}^{\lambda_{x}}=(-1)^{\lambda_{x}-\mu_{x}}a_{-\mu_{x}}^{\lambda_{x}\dagger}$,
and exploiting known reduction rules for the Kronecker product $\lambda_{x}\times\lambda_{\bar{x}}\rightarrow\lambda_{x\bar{x}}$
of $\mathrm{SU(2)}$-irreducible representations. For instance,

\begin{equation}
\widehat{L}_{3}={\displaystyle \sum_{n_{x}\lambda_{x}\mu_{x}}}{\displaystyle \sum_{\lambda_{ij}\lambda_{ijk}}}{\displaystyle \sum_{\lambda_{lp}\lambda_{lpq}}}{\displaystyle \sum_{\lambda\mu}}(-1)^{\lambda_{l}+\lambda_{p}+\lambda_{q}+\mu_{l}+\mu_{p}+\mu_{q}}\ell_{ijkqpl}I\:_{\Lambda}\widehat{T}_{\mu}^{\lambda}\left(\lambda_{i}\lambda_{j}\lambda_{k}\lambda_{l}\lambda_{p}\lambda_{q}\right),\label{eq:2.3}\end{equation}

\[
I\equiv I\left[\lambda_{i}\lambda_{j}\left(\lambda_{ij}\right)\lambda_{k}\left(\lambda_{ijk}\right),\lambda_{l}\lambda_{p}\left(\lambda_{lp}\right)\lambda_{q}\left(\lambda_{lpq}\right)\lambda\mu\left\{ \mu_{x}\right\} \right]\]

\[
={\displaystyle \sum_{\mu_{ij}\mu_{ijk}}}{\displaystyle \sum_{\mu_{lp}\mu_{lpq}}}\left\langle \lambda{}_{i}\mu_{i}\lambda_{j}\mu_{j}|\lambda_{ij}\mu_{ij}\right\rangle \left\langle \lambda_{ij}\mu_{ij}\lambda_{k}\mu_{k}|\lambda_{ijk}\mu_{ijk}\right\rangle \left\langle \lambda_{l}\mu_{l}\lambda_{p}\mu_{p}|\lambda_{lp}\mu_{lp}\right\rangle \]

\begin{equation}
\times\left\langle \lambda_{lp}\mu_{lp}\lambda_{q}\mu_{q}|\lambda_{lpq}\mu_{lpq}\right\rangle \left\langle \lambda_{ijk}\mu_{ijk}\lambda_{lpq}\mu_{lpq}|\lambda\mu\right\rangle ,\label{eq:2.4}\end{equation}

\begin{equation}
\:_{\Lambda}\widehat{T}_{\mu}^{\lambda}\left(\lambda_{i}\lambda_{j}\lambda_{k}\lambda_{l}\lambda_{p}\lambda_{q}\right)=\left[\left[\left[a^{\lambda_{i}}\times a^{\lambda_{j}}\right]^{\lambda_{ij}}\times a^{\lambda_{k}}\right]^{\lambda_{ijk}}\times\left[\left[\tilde{a}^{\lambda_{l}}\times\tilde{a}^{\lambda_{p}}\right]^{\lambda_{lp}}\times\tilde{a}^{\lambda_{q}}\right]^{\lambda_{lpq}}\right]_{\mu}^{\lambda}.\label{eq:2.5}\end{equation}

\noindent{}In Eq. (\ref{eq:2.4}) the quantities $\left\langle \lambda_{x}\mu_{x}\lambda_{y}\mu_{y}|\lambda_{xy}\mu_{xy}\right\rangle $
denote Clebsch-Gordan coefficients of $\mathrm{SU\left(2\right)}$
\cite{Jucys}. Set $Q$ a quasispin space. Then the presentation of
$\:_{\Lambda}\widehat{T}_{\mu}^{\lambda}$ in $q$-space, where $q\equiv Q\times\Lambda$,
is

\begin{equation}
\widehat{T}_{\iota}^{\varsigma}\left(\lambda_{i}\lambda_{j}\lambda_{k}\lambda_{l}\lambda_{p}\lambda_{q}\right)=\left\langle ijklpq\right\rangle =\left[\left[\left[a^{\varsigma_{i}}\times a^{\varsigma_{j}}\right]^{\varsigma_{ij}}\times a^{\varsigma_{k}}\right]^{\varsigma_{ijk}}\times\left[\left[a^{\varsigma_{l}}\times a^{\varsigma_{p}}\right]^{\varsigma_{lp}}\times a^{\varsigma_{q}}\right]^{\varsigma_{lpq}}\right]_{\iota}^{\varsigma}.\label{eq:2.6}\end{equation}

\noindent{}The notation $\left\langle ijklpq\right\rangle $, as
a shorter description of $\widehat{T}_{\iota}^{\varsigma}$, will
be used later. The ranks $\varsigma_{x}=\kappa_{x}\lambda_{x}$. In
the next section it will be showed, that there are $42$ ways at all
to form $\widehat{T}_{\iota}^{\varsigma}$.

\section{Coupling schemes\label{schemes}}

Suppose there is constructed tensor

\begin{equation}
\widehat{O}_{n}=T_{\beta_{1}\beta_{2}\ldots\beta_{n}}^{\alpha_{1}\alpha_{2}\ldots\alpha_{n}}=a_{\beta_{1}}^{\alpha_{1}}a_{\beta_{2}}^{\alpha_{2}}\ldots a_{\beta_{n}}^{\alpha_{n}}\label{eq:3.1}\end{equation}

\noindent{}with $n\geq2$, and $\alpha_{i}$, $i\in\mathcal{I}=\{1,2,\ldots,6\}$
labeling irreducible representations. In general, for any $n\geq2$
there are different ways to reduce Kronecker products of $\alpha_{i}$.
We suggest a procedure, suitable for any $n$, to obtain possible
coupling schemes. Suppose there is a $n$-length string, which is
partitioned into the sum of $\lambda_{\varkappa}$-length strings,
where $n=\lambda_{1}+\lambda_{2}+\ldots+\lambda_{n}$. Consequently,
the partitions can be characterized by the irreducible representations
$\lambda\equiv\left[\lambda_{1},\lambda_{2},\ldots\lambda_{\varkappa},\ldots,\lambda_{n}\right]$
of $\mathrm{S}_{n}$. We exclude two partitions of length $1$ and
$2$. The $1$-partition, including only $1$ operator $a_{\beta_{i}}^{\alpha_{i}}$,
appears $h_{1}$ times in the coupling scheme, the $2$-partition
appears $h_{2}$ times in the same coupling scheme and so on. Furthermore,
in our considerations the ordering of partitions is important, while
the ordering of partitions $\lambda_{\varkappa}$ in the sum $\lambda_{1}+\lambda_{2}+\ldots+\lambda_{n}$
does not play any role%
\footnote{By definition, $\lambda_{1}\geq\lambda_{2}\geq\ldots\geq\lambda_{n}$.
However, in our case $\lambda_{i}+\lambda_{j}$ has different meaning
than $\lambda_{j}+\lambda_{i}$.%
}. In order to classify irreducible tensor operators by coupling schemes,
we use $n$-tuples which, by default, represent the notation of an
ordered list of elements in the scheme. Such elements will be one-
and two-operator partitions, denoted by the numbers $1$ and $2$.
For example, the notation of the scheme $\left[\bullet\times\bullet\right]$
is $\left[\left[2\right]\right]$; the schemes $\left[\left[\bullet\times\bullet\right]\times\bullet\right]$
and $\left[\bullet\times\left[\bullet\times\bullet\right]\right]$
are labeled by $\left[\left[21\right]\right]$ and $\left[\left[12\right]\right]$,
respectively. Here it is assumed that in the bullets $\bullet$ there
are placed the operators $a_{\beta_{i}}^{\alpha_{i}}$ and the subscript
$i$ marks the position of given operator (or bullet) starting from
the left. One may notice, the $3$-tuples $\left[\left[21\right]\right]$
and $\left[\left[12\right]\right]$ are characterized by the same
irreducible representation $\left[21\right]$, thus they belong to
the same class $\left(\alpha\right)=\left(1^{1}2^{1}\right)$ of $\mathrm{S}_{3}$.

{\footnotesize\begin{multicols}{2}

\begin{center}
\begin{tabular}{|c|c|c|}
\hline 
$\left(\alpha\right)$ & Tuples & Symbol\tabularnewline
\hline
\hline 
$\left(2^{1}\right)$ & $\left[\left[2\right]\right]$ & $\widehat{\mathrm{T}}^{[1^{2}]}$\tabularnewline
\hline
\hline 
\multicolumn{1}{|c|}{$\left(1^{1}2^{1}\right)$} & \multicolumn{1}{c|}{$\left[\left[12\right]\right]$} & \multicolumn{1}{c|}{$\widehat{\mathrm{T}}_{1}^{[21]}$}\tabularnewline
\cline{2-3} 
 & $\left[\left[21\right]\right]$ & $\widehat{\mathrm{T}}_{2}^{[21]}$\tabularnewline
\hline
\hline 
$\left(2^{2}\right)$ & $\left[\left[22\right]\right]$ & $\widehat{\mathrm{T}}^{[2^{2}]}$\tabularnewline
\hline 
 & $\left[\left[211\right]\right]\ltimes\widehat{\mathrm{T}}_{2}^{[21]}$ & $\widehat{\mathrm{T}}_{1}^{[21^{2}]}$\tabularnewline
\cline{2-3} 
$\left(1^{1}3^{1}\right)$ & $\left[\left[121\right]\right]\ltimes\widehat{\mathrm{T}}_{2,1}^{[21]}$ & $\widehat{\mathrm{T}}_{2,3}^{[21^{2}]}$\tabularnewline
\cline{2-3} 
 & $\left[\left[112\right]\right]\ltimes\widehat{\mathrm{T}}_{1}^{[21]}$ & $\widehat{\mathrm{T}}_{4}^{[21^{2}]}$\tabularnewline
\hline
\hline 
 & $\left[\left[221\right]\right]\ltimes\widehat{\mathrm{T}}_{2,1}^{[21]}$ & $\widehat{\mathrm{T}}_{1,2}^{[2^{2}1]}$\tabularnewline
\cline{2-3} 
$\left(2^{1}3^{1}\right)$ & $\left[\left[122\right]\right]\ltimes\widehat{\mathrm{T}}_{1,2}^{[21]}$ & $\widehat{\mathrm{T}}_{3,4}^{[2^{2}1]}$\tabularnewline
\cline{2-3} 
 & $\left[\left[212\right]\right]\ltimes\widehat{\mathrm{T}}_{2,1}^{[21]}$ & $\widehat{\mathrm{T}}_{5,6}^{[2^{2}1]}$\tabularnewline
\hline
 & $\left[\left[2111\right]\right]\ltimes\widehat{\mathrm{T}}_{1}^{[21^{2}]}$ & $\widehat{\mathrm{T}}_{1}^{[21^{3}]}$\tabularnewline
\cline{2-3} 
$\left(1^{1}4^{1}\right)$ & $\left[\left[1211\right]\right]\ltimes\widehat{\mathrm{T}}_{1,3,2}^{[21^{2}]}$ & $\widehat{\mathrm{T}}_{2,3,4}^{[21^{3}]}$\tabularnewline
\cline{2-3} 
 & $\left[\left[1121\right]\right]\ltimes\widehat{\mathrm{T}}_{4,2,3}^{[21^{2}]}$ & $\widehat{\mathrm{T}}_{5,6,7}^{[21^{3}]}$\tabularnewline
\cline{2-3} 
 & $\left[\left[1112\right]\right]\ltimes\widehat{\mathrm{T}}_{4}^{[21^{2}]}$ & $\widehat{\mathrm{T}}_{8}^{[21^{3}]}$\tabularnewline
\hline
\end{tabular}%
\begin{table}[H]
\caption{\label{Tab1}The coupling schemes for $\widehat{O}_{1-5}$}

\end{table}

\par\end{center}

\begin{center}
\begin{tabular}{|c|c|c|}
\hline 
$\left(\alpha\right)$ & Tuples & Symbol\tabularnewline
\hline
\hline 
$\left(3^{2}\right)$ & $\left[\left[222\right]\right]\ltimes\widehat{\mathrm{T}}_{2,1}^{[21]}$  & $\widehat{\mathrm{T}}_{1,2}^{[2^{3}]}$\tabularnewline
\hline 
 & $\left[\left[2211\right]\right]\ltimes\widehat{\mathrm{T}}_{1,2,3}^{[21^{2}]}$ & $\widehat{\mathrm{T}}_{1,2,3}^{[2^{2}1^{2}]}$\tabularnewline
\cline{2-3} 
 & $\left[\left[1221\right]\right]\ltimes\widehat{\mathrm{T}}^{[2^{2}]}$ & $\widehat{\mathrm{T}}_{4}^{[2^{2}1^{2}]}$\tabularnewline
\cline{2-3} 
 & $\left[\left[1221\right]\right]\ltimes\widehat{\mathrm{T}}_{1-4}^{[21^{2}]}$ & $\widehat{\mathrm{T}}_{5-8}^{[2^{2}1^{2}]}$\tabularnewline
\cline{2-3} 
 & $\left[\left[1122\right]\right]\ltimes\widehat{\mathrm{T}}_{2,3,4}^{[21^{2}]}$ & $\widehat{\mathrm{T}}_{9,10,11}^{[2^{2}1^{2}]}$\tabularnewline
\cline{2-3} 
$\left(2^{1}4^{1}\right)$ & $\left[\left[2121\right]\right]\ltimes\widehat{\mathrm{T}}^{[2^{2}]}$ & $\widehat{\mathrm{T}}_{12}^{[2^{2}1^{2}]}$\tabularnewline
\cline{2-3} 
 & $\left[\left[2121\right]\right]\ltimes\widehat{\mathrm{T}}_{1-4}^{[21^{2}]}$ & $\widehat{\mathrm{T}}_{13-16}^{[2^{2}1^{2}]}$\tabularnewline
\cline{2-3} 
 & $\left[\left[2112\right]\right]\ltimes\widehat{\mathrm{T}}^{[2^{2}]}$ & $\widehat{\mathrm{T}}_{17}^{[2^{2}1^{2}]}$\tabularnewline
\cline{2-3} 
 & $\left[\left[2112\right]\right]\ltimes\widehat{\mathrm{T}}_{1,4}^{[21^{2}]}$ & $\widehat{\mathrm{T}}_{18,19}^{[2^{2}1^{2}]}$\tabularnewline
\cline{2-3} 
 & $\left[\left[1212\right]\right]\ltimes\widehat{\mathrm{T}}^{[2^{2}]}$ & $\widehat{\mathrm{T}}_{20}^{[2^{2}1^{2}]}$\tabularnewline
\cline{2-3} 
 & $\left[\left[1212\right]\right]\ltimes\widehat{\mathrm{T}}_{1-4}^{[21^{2}]}$ & $\widehat{\mathrm{T}}_{21-24}^{[2^{2}1^{2}]}$\tabularnewline
\hline 
 & $\left[\left[21111\right]\right]\ltimes\widehat{\mathrm{T}}_{1}^{[21^{3}]}$ & $\widehat{\mathrm{T}}_{1}^{[21^{4}]}$\tabularnewline
\cline{2-3} 
 & $\left[\left[12111\right]\right]\ltimes\widehat{\mathrm{T}}_{1-4}^{[21^{3}]}$ & $\widehat{\mathrm{T}}_{2-5}^{[21^{4}]}$\tabularnewline
\cline{2-3} 
$\left(1^{1}5^{1}\right)$ & $\left[\left[11211\right]\right]\ltimes\widehat{\mathrm{T}}_{2-7}^{[21^{3}]}$ & $\widehat{\mathrm{T}}_{6-11}^{[21^{4}]}$\tabularnewline
\cline{2-3} 
 & $\left[\left[11121\right]\right]\ltimes\widehat{\mathrm{T}}_{5-8}^{[21^{3}]}$ & $\widehat{\mathrm{T}}_{12-15}^{[21^{4}]}$\tabularnewline
\cline{2-3} 
 & $\left[\left[11112\right]\right]\ltimes\widehat{\mathrm{T}}_{8}^{[21^{3}]}$ & $\widehat{\mathrm{T}}_{16}^{[21^{4}]}$\tabularnewline
\hline
\end{tabular}%
\begin{table}[H]
\caption{\label{Tab1a}The coupling schemes for $\widehat{O}_{6}$}

\end{table}

\par\end{center}

\end{multicols}}

Suppose $n=2$. Then $\lambda_{1}=\left[2\right]$ and $\lambda_{2}=[1^{2}]$.
We distinguish the antisymmetric $\mathrm{S}_{2}$-irreducible representation.
The tuple $\left[\left[2\right]\right]$ is denoted by $\widehat{\mathrm{T}}^{[1^{2}]}$
(see Tab. \ref{Tab1}). If $n=3$, $\mathrm{S}_{3}$-irreducible representation,
conforming to $1$- and $2$-partitions, is $\lambda=[21]$ (the full
set of partitions equals to $\left[3\right]$, $\left[21\right]$,
$\left[1^{3}\right]$). It corresponds to Kronecker products $(\alpha_{1}\times\alpha_{2})\times\alpha_{3}$
or $\alpha_{1}\times(\alpha_{2}\times\alpha_{3})$ (previously studied
example of $3$-tuples $[[21]]$, $[[12]]$). Further, when $n=4$,
the suitable representations are $\left[2^{2}\right]$ and $\left[21^{2}\right]$.
Let us study the schemes of class $\left(1^{1}3^{1}\right)$. Suppose
there is a tuple $\widehat{\mathrm{T}}_{1}^{[21^{2}]}=\left[\left[211\right]\right]\ltimes\widehat{\mathrm{T}}_{2}^{[21]}$,
where $\widehat{\mathrm{T}}_{2}^{[21]}=\left[\left[21\right]\right]$
(Tab. \ref{Tab1}). The tuple $\widehat{\mathrm{T}}_{2}^{[21]}$ in
an explicit form reads $\left[\left[\bullet\times\bullet\right]\times\bullet\right]$.
The (semijoin) notation $\ltimes$ denotes that the scheme $\widehat{\mathrm{T}}_{1}^{[21^{2}]}$
consists of only those partitions, which are in the scheme $\widehat{\mathrm{T}}_{2}^{[21]}$.
The tuple $\left[\left[211\right]\right]$ reads $\left[\bullet\times\bullet\right]\times\bullet\times\bullet$.
If we rename $\left[\bullet\times\bullet\right]=\bullet^{\prime}$,
then it becomes of the form $\bullet^{\prime}\times\bullet\times\bullet$.
We obtained the case of $n=3$. Then, according to $\widehat{\mathrm{T}}_{2}^{[21]}$,
the form $\bullet^{\prime}\times\bullet\times\bullet$ is partitioned
as $\left[\left[\bullet^{\prime}\times\bullet\right]\times\bullet\right]$.
Thus, the final scheme $\widehat{\mathrm{T}}_{1}^{[21^{2}]}$ reads
$\left[\left[\left[\bullet\times\bullet\right]\times\bullet\right]\times\bullet\right]$.

Analogously possible schemes are obtained for higher $n$. When $n=5$,
the suitable representations of $\mathrm{S}_{5}$ are $\left[2^{2}1\right]$
and $\left[21^{3}\right]$. Finally, when $n=6$, the representations
are $\left[2^{3}\right]$, $\left[2^{2}1^{2}\right]$ and $[21^{4}]$
(see Tab. \ref{Tab1a}). Particularly, the tuple $\widehat{\mathrm{T}}_{12}^{[2^{2}1^{2}]}$
corresponds to coupling scheme of ranks $\varsigma_{x}$ in $\widehat{T}^{\varsigma}$
in Eq. (\ref{eq:2.6}). In Tabs. \ref{Tab1}-\ref{Tab1a} notations
$\widehat{\mathrm{T}}_{a,b,c,\ldots}^{\lambda}=\left[\left[x\right]\right]\ltimes\widehat{\mathrm{T}}_{a^{\prime},b^{\prime},c^{\prime},\ldots}^{\lambda^{\prime}}$
signify that $\widehat{\mathrm{T}}_{a}^{\lambda}=\left[\left[x\right]\right]\ltimes\widehat{\mathrm{T}}_{a^{\prime}}^{\lambda^{\prime}}$,
$\widehat{\mathrm{T}}_{b}^{\lambda}=\left[\left[x\right]\right]\ltimes\widehat{\mathrm{T}}_{b^{\prime}}^{\lambda^{\prime}}$
and so on. The procedure considered is easily extended for any $n$,
if there are defined all $m$-tuples with $m<n$. To perform this
task, the following steps should be accomplished: (i) each $n$-length
string is distinguished by representations of $\mathrm{S}_{n}$, excluding
only those, which include $1$- and $2$-partitions; (ii) obtained
partitions $\lambda_{\varkappa}$ are written in all possible ways
in the sum $\lambda_{1}+\lambda_{2}+\ldots+\lambda_{n}$ (e.g., $2+1+1$,
$1+2+1$, $1+1+2$); (iii) renaming $2$ by $1^{\prime}$, one obtains
$(n-h_{2})$-length string ($h_{2}$ is a multiplicity of $2$), which
has already been classified by possible partitions; those partitions
are included in studied string (by using $\ltimes$), and in the final
step $1^{\prime}$ is reversed to $2$ back again. Note, for obtained
$(n-h_{2})$-length string there are suitable only those partitions,
which do not couple $1$-partitions; otherwise one would get a tuple,
characterized by another representation of $\mathrm{S}_{n}$ (e.g.,
$[[1^{\prime}\times1]\times1]$, $[[1\times1^{\prime}]\times1]$ or
$[1\times[1^{\prime}\times1]]$, $[1\times[1\times1^{\prime}]]$,
but not $[1^{\prime}\times[1\times1]]$, $[[1\times1]\times1^{\prime}]$,
which correspond to $\lambda=[2^{2}]$, rather than $\lambda=[21^{2}]$).
Finally, substituting in the bullets operators $a^{\alpha_{i}}$,
and each $2$-partition renaming by tensor operator $W^{\alpha_{ij}}(\lambda_{i}\lambda_{j})$,
where $W^{\alpha_{ij}}(\lambda_{i}\lambda_{j})=[a^{\alpha_{i}}\times a^{\alpha_{j}}]^{\alpha_{ij}}$,
we would gain irreducible tensor operators for given reducible product
$\widehat{O}_{n}$.

In general, there are $42\cdot42=1764$ recoupling coefficients which
can be generated among each of the scheme (written in a consequent
order) when $n=6$ (Tab. \ref{Tab1a}), including coefficients for
the same scheme and for the conjugate schemes. This number can be
reduced to only $42$. We will call them the basis coefficients. All
other recoupling coefficients are generated from basis coefficients.
Let it be the coefficients generated when transforming $\widehat{T}_{\iota}^{\varsigma}\equiv\widehat{\mathrm{T}}_{12}^{[2^{2}1^{2}]}$
into all other $41$ scheme. Here 

\begin{equation}
\widehat{\mathrm{T}}_{12}^{[2^{2}1^{2}]}\equiv\left[\left[\left[a^{\alpha_{1}}\times a^{\alpha_{2}}\right]^{\alpha_{12}}\times a^{\alpha_{3}}\right]^{\alpha_{123}}\times\left[\left[a^{\alpha_{4}}\times a^{\alpha_{5}}\right]^{\alpha_{45}}\times a^{\alpha_{6}}\right]^{\alpha_{456}}\right]_{\beta}^{\alpha}.\label{eq:3.2a}\end{equation}

\noindent{}Then the tensor $\widehat{O}_{6}$ (see Eq. (\ref{eq:3.1}))
is reduced as follows

\begin{equation}
\widehat{O}_{6}=T_{\beta_{1}\beta_{2}\ldots\beta_{6}}^{\alpha_{1}\alpha_{2}\ldots\alpha_{6}}={\displaystyle \sum_{\alpha_{12}\alpha_{123}}}{\displaystyle \sum_{\alpha_{45}\alpha_{456}}}\widehat{\mathrm{T}}_{12}^{[2^{2}1^{2}]}\: I.\label{eq:3.2}\end{equation}

\noindent{}The multiplier $I$ is defined in Eq. (\ref{eq:2.4})
making the replacements $\lambda_{x}\mu_{x}\rightarrow\varsigma_{x}\iota_{x}$,
where $\varsigma_{i}=\alpha_{1}$, $\varsigma_{j}=\alpha_{2}$, $\varsigma_{k}=\alpha_{3}$,
$\varsigma_{l}=\alpha_{4}$, $\varsigma_{p}=\alpha_{5}$, $\varsigma_{q}=\alpha_{6}$,
$\varsigma=\alpha_{123456}\equiv\alpha$, $\iota=\beta$. It has to
be taken into account that the numbering of the ranks and of the operators
differs. For example, the rank $\alpha_{12}$ is the total rank of
the ranks $\alpha_{1}$ and $\alpha_{2}$ of the operators which are
placed in the $1$st and $2$nd positions of the given scheme. On
the other hand, the ranks $\varsigma_{i}=\alpha_{1}$ and $\varsigma_{j}=\alpha_{2}$
can acquire the same value $\varsigma_{1}$. To avoid further misunderstanding,
the following notations will be used. The rank $\alpha_{i}=\kappa_{i}\lambda_{i}$
will denote the rank of the $i$th operator $a^{\alpha_{i}}$ in $q$-space;
the rank $\varsigma_{x}=\kappa_{x}\lambda_{x}$ will denote some value
of any of the operators in the sequence. Generally, if the notations
$\alpha_{i}$ and $\alpha_{j}$ mark the $i$th and the $j$th positions
of the corresponding operators, respectively, the rank $\varsigma_{x}$
can be the same for both $\alpha_{i}$ and $\alpha_{j}$.

The basis coefficients will be denoted by $\epsilon_{\xi}$, where

\begin{equation}
\epsilon_{\xi}={\displaystyle \sum_{\beta_{i\in\mathcal{I}}}}I\cdot I_{\xi}.\label{eq:3.3}\end{equation}

\noindent{}The integers $\xi\in\left\{ 1,2,\ldots,42\right\} $;
$I_{\xi}=I_{\varkappa}=I_{\varkappa}^{[2^{3}]}$ for $\varkappa=1,2$
and $\lambda=[2^{3}]$; $I_{\xi}=I_{\varkappa+2}=I_{\varkappa}^{[2^{2}1^{2}]}$
for $\varkappa\in\left[1,24\right]$ and $\lambda=\left[2^{2}1^{2}\right]$;
$I_{\xi}=I_{\varkappa+24}=I_{\varkappa}^{[21^{4}]}$ for $\varkappa\in\left[1,16\right]$
and $\lambda=[21^{4}]$. Then follows $I_{14}=I$. The coefficient
$I_{\varkappa}^{\lambda}$, in general, depends on the partition $\widehat{\mathrm{T}}_{\varkappa}^{\lambda}$
in the following way

\begin{equation}
\widehat{\mathrm{T}}_{\varkappa}^{\lambda}={\displaystyle \sum_{\beta_{i\in\mathcal{I}}}}\widehat{O}_{6}\: I_{\varkappa}^{\lambda}.\label{eq:3.4}\end{equation}

\noindent{}According to Eqs. (\ref{eq:3.2}), (\ref{eq:3.4}), the
schemes $\widehat{\mathrm{T}}_{\varkappa}^{\lambda}$ and $\widehat{\mathrm{T}}_{12}^{[2^{2}1^{2}]}$
are transforming among each other by the formulas

\begin{equation}
\widehat{\mathrm{T}}_{\varkappa}^{\lambda}={\displaystyle \sum_{\alpha_{\eta\in\Upsilon}}}\widehat{\mathrm{T}}_{12}^{[2^{2}1^{2}]}\:\epsilon_{\xi},\quad\widehat{\mathrm{T}}_{12}^{[2^{2}1^{2}]}={\displaystyle \sum_{\alpha_{\zeta\in\Gamma_{\xi}}}}\widehat{\mathrm{T}}_{\varkappa}^{\lambda}\:\epsilon_{\xi},\label{eq:3.5}\end{equation}

\noindent{}where the summations are performed over the internal ranks
$\alpha_{\eta}$, $\alpha_{\zeta}$. The indices $\eta\in\Upsilon=\left\{ 12,123,45,456\right\} $;
the indices $\zeta\in\Gamma_{\xi}=\left\{ \zeta_{1},\zeta_{2},\zeta_{3},\zeta_{4}\right\} $
depend on the concrete scheme $\widehat{\mathrm{T}}_{\varkappa}^{\lambda}$.
It directly follows from Eq. (\ref{eq:3.5}) that

\begin{equation}
\widehat{\mathrm{T}}_{\varkappa}^{\lambda}={\displaystyle \sum_{\alpha_{\eta\in\Upsilon}}}\:{\displaystyle \sum_{\alpha_{\zeta^{\prime}\in\Gamma_{\xi^{\prime}}}}}\widehat{\mathrm{T}}_{\varkappa^{\prime}}^{\lambda^{\prime}}\:\epsilon_{\xi}\epsilon_{\xi^{\prime}},\quad{\displaystyle \sum_{\alpha_{\eta\in\Upsilon}}}\epsilon_{\xi}^{2}=1.\label{eq:3.6}\end{equation}

\noindent{}In Eq. (\ref{eq:3.6}) the basis coefficients $\epsilon_{\xi}$
are defined in Eq. (\ref{eq:3.3}), while the basis coefficients $\epsilon_{\xi^{\prime}}$
correspond to the recoupling from $\widehat{\mathrm{T}}_{12}^{[2^{2}1^{2}]}$
to $\widehat{\mathrm{T}}_{\varkappa^{\prime}}^{\lambda^{\prime}}$.
The coefficients $\epsilon_{\xi}$ are presented in an explicit form
in Appendix \ref{B}.

\section{Permutations\label{permutations}}

In previous section the possible schemes to reduce operator $\widehat{O}_{6}$
have been studied. In present section we develop the procedure of
transformation for arbitrary operator $\widehat{T}^{\varsigma}(\lambda_{i}\lambda_{j}\lambda_{k}\lambda_{l}\lambda_{p}\lambda_{q})$
with $i,j,k,l,p,q$ labeling open shells, into $\widehat{T}^{\varsigma}(\lambda_{i^{\prime}}\lambda_{j^{\prime}}\lambda_{k^{\prime}}\lambda_{l^{\prime}}\lambda_{p^{\prime}}\lambda_{q^{\prime}})$,
where $i^{\prime}\leq j^{\prime}\leq k^{\prime}\leq l^{\prime}\leq p^{\prime}\leq q^{\prime}$.
It is assumed, both operators are coupled by the scheme $\widehat{\mathrm{T}}_{12}^{[2^{2}1^{2}]}$
(see Eqs. (\ref{eq:2.6}), (\ref{eq:3.2a})). Recoupling coefficients,
which appear due to the permutations of ranks of $T_{\beta_{1}\ldots\beta_{6}}^{\alpha_{1}\ldots\alpha_{6}}$,
written in irreducible tensor form $\widehat{T}_{\beta}^{\alpha}$,
will be called the permutation coefficients, denoted $\varepsilon$.
It is clear, there are $6!$ possible permutations for given distributions
$i,j,k,l,p,q$, if $i\neq j\neq k\neq l\neq p\neq q$. Consequently,
there are $6!$ permutation coefficients $\varepsilon$, in general.
This number will be reduced up to $15$ permutations which are described
by the $2$-cycles $\left(ij\right)\in\mathrm{S}_{6}$ ($i,j\in\mathcal{I}$).
This appears from the general theory which states that all $n$-cycles
($n\geq2$) can be expanded into the composition of $n-1$ number
of $2$-cycles. Hence, the rest of permutation coefficients are obtained
from the permutation coefficients $\varepsilon_{ij}$. Moreover, the
coefficients $\varepsilon_{ij}$ will be constructed of the basis
coefficients $\epsilon_{\xi}$ in such a way, that the desirable ordering
of ranks of the tensor operator $\widehat{T}_{\beta}^{\alpha}$ would
be obtained making minimal number of recouplings between the schemes
$\widehat{\mathrm{T}}_{\varkappa}^{\lambda}$. The coefficients $\varepsilon_{ij}$
are determined from the equation%
\footnote{Hereafter we formally distinguish permutation operators $\widehat{\pi}$
on $n$-particle Hilbert space and elements $\pi$ of $\mathrm{S}_{n}$.
However, a cycle notation $(ij\ldots k)$ suits for both cases.%
}

\begin{equation}
\left(ij\right)\widehat{\mathrm{T}}_{12}^{[2^{2}1^{2}]}={\displaystyle \sum_{\alpha_{\varrho\in\mathcal{F}}}}\varepsilon_{ij}\widehat{\mathrm{T}}_{12}^{[2^{2}1^{2}]}.\label{eq:4.5}\end{equation}

\noindent{}The summation is performed over $\alpha_{\varrho}$, where
the indices $\varrho\in\mathcal{F}\subset\mathcal{H}=\left\{ 12,123,45,46,456\right\} $
depend on the concrete permutation $\left(ij\right)$. These coefficients
$\varepsilon_{ij}$ are invariant under the operation $\left(ij\right)$

\begin{equation}
\left(ij\right)\varepsilon_{ij}=\varepsilon_{ij}.\label{eq:4.6}\end{equation}

\noindent{}However, in general, $\left(ij\right)\varepsilon_{ik}\neq\varepsilon_{jk}$.
Some of the coefficients $\left(ij\right)\varepsilon_{ik}$ and $\varepsilon_{jk}$
differ only by the phase factor

\begin{equation}
\left\{ \begin{array}{l}
\left(12\right)\varepsilon_{1k}=f_{1k}f_{2k}\varepsilon_{2k},\quad k\in\left\{ 3,4,5,6\right\} ,\\
\left(45\right)\varepsilon_{34}=f_{34}f_{35}\varepsilon_{35},\\
\left(45\right)\varepsilon_{46}=f_{46}f_{56}\varepsilon_{56}.\end{array}\right.\label{eq:4.7}\end{equation}

\noindent{}Here

\begin{equation}
f_{ij}={\displaystyle \prod_{k=1}^{j-i}}\:\:{\displaystyle \prod_{l=1}^{j-i-1}}\varpi_{i\: i+k}\:\varpi_{i+l\: j}\qquad\forall j>i;\quad\varpi_{ii}=1,\quad\varpi_{ij}=(-1)^{\alpha_{i}+\alpha_{j}-\alpha_{ij}+1}.\label{eq:4.8}\end{equation}

\noindent{}Since $\varpi_{ij}=\varpi_{ji}=\varpi_{ij}^{-1}$, the
condition $j>i$ can always be satisfied. We conclude that any operation
of the $n$-length ($2\leq n\leq6$) permutation $\left(i_{1}i_{2}\ldots i_{n}\right)$
on the scheme $\widehat{\mathrm{T}}_{12}^{[2^{2}1^{2}]}$ can be decomposed
into the $n-1$ power sums of the operations of $2$-cycles $\left(ij\right)$
on given scheme by applying Eq. (\ref{eq:4.5}) in the following way

\[
\left(ij\right)\ldots\left(kl\right)\left(pq\right)\widehat{\mathrm{T}}_{12}^{[2^{2}1^{2}]}=\left(ij\right)\ldots\left(kl\right){\displaystyle \sum_{\alpha_{\varrho\in\mathcal{F}}}}\varepsilon_{pq}\widehat{\mathrm{T}}_{12}^{[2^{2}1^{2}]}=\left(ij\right)\ldots{\displaystyle \sum_{\alpha_{\varrho\in\left(kl\right)\mathcal{F}}}}\:{\displaystyle \sum_{\alpha_{\varrho^{\prime}\in\mathcal{F}^{\prime}}}}\varepsilon_{kl}\left\{ \left(kl\right)\varepsilon_{pq}\right\} \widehat{\mathrm{T}}_{12}^{[2^{2}1^{2}]}\]

\[
=\left(ij\right){\displaystyle \sum_{\alpha_{\varrho\in\ldots\left(kl\right)\mathcal{F}}}}\:\:{\displaystyle \sum_{\alpha_{\varrho^{\prime}\in\ldots\mathcal{F}^{\prime}}}}\ldots{\displaystyle \sum_{\alpha_{\varrho^{\prime\prime}\in\mathcal{F}^{\prime\prime}}}}\ldots\left\{ \ldots\varepsilon_{kl}\right\} \left\{ \ldots\left(kl\right)\varepsilon_{pq}\right\} \widehat{\mathrm{T}}_{12}^{[2^{2}1^{2}]}\]

\begin{equation}
={\displaystyle \sum_{\alpha_{\varrho\in\left(ij\right)\ldots\left(kl\right)\mathcal{F}}}}\:\:{\displaystyle \sum_{\alpha_{\varrho^{\prime}\in\left(ij\right)\ldots\mathcal{F}^{\prime}}}}\ldots{\displaystyle \sum_{\alpha_{\varrho^{\prime\prime}\in\ldots\mathcal{F}^{\prime\prime}}}}\:\:{\displaystyle \sum_{\alpha_{\varrho^{\prime\prime\prime}\in\mathcal{F}^{\prime\prime\prime}}}}\varepsilon_{ij}\ldots\left\{ \left(ij\right)\ldots\varepsilon_{kl}\right\} \left\{ \left(ij\right)\ldots\left(kl\right)\varepsilon_{pq}\right\} \widehat{\mathrm{T}}_{12}^{[2^{2}1^{2}]}.\label{eq:4.9}\end{equation}

\noindent{}In Tab. \ref{TabEPS} the explicit forms of $\varepsilon_{ij}$
are presented for all $15$ cases. In the last column the set $\mathcal{F}$
of $\alpha_{\varrho}$ indices is defined in Eq. (\ref{eq:4.5}).
The number $\xi>1$ signifies, that there are more than one equivalent
opportunities to recouple given operator $\widehat{T}^{\varsigma}(\lambda_{i}\lambda_{j}\lambda_{k}\lambda_{l}\lambda_{p}\lambda_{q})$
into $\widehat{T}^{\varsigma}(\lambda_{i^{\prime}}\lambda_{j^{\prime}}\lambda_{k^{\prime}}\lambda_{l^{\prime}}\lambda_{p^{\prime}}\lambda_{q^{\prime}})$.
By default, the coefficients $\varepsilon_{ij}$ appear when permuting
the operators $a_{\iota_{x}}^{\varsigma_{x}}$ ($x=i,j,k,l,p,q\in\mathcal{I}$)
of $\widehat{\mathrm{T}}_{12}^{[2^{2}1^{2}]}$ with different ranks
$\varsigma_{x}$. It is noticeable, some of the indices $x$ can acquire
the same values, i.e., some ranks can be equal. In this case one assumes
that the same rank operators $a_{\iota_{x}}^{\varsigma_{x}}$ are
not permuted because in a contrary case the Kronecker delta function
appears. However, we will demonstrate, that there is a simple way
to avoid arising deltas. Suppose there is a tensor operator $\left\langle jiilpq\right\rangle $
(see Eq. (\ref{eq:2.6})) which has to be transformed to $\left\langle iijlpq\right\rangle $.
The set of possible operators $\widehat{\pi}$ satisfying such a transformation
equals to $\left\{ \left(13\right),\left(132\right)\right\} $. If
$\widehat{\pi}=(13)=\bigl(\begin{smallmatrix}1 & 2 & 3\\ 3 & 2 & 1\end{smallmatrix}\bigr)$,
then $\left(13\right)\left\langle iijlpq\right\rangle =\left\langle jiilpq\right\rangle =\sum_{\varsigma_{ii}=\mathrm{odd}}\varepsilon_{13}\left\langle iijlpq\right\rangle $,
where $\varepsilon_{13}=f_{13}\left\{ \left(12\right)\epsilon_{6}\right\} \left\{ \left(123\right)\epsilon_{6}\right\} $,
$f_{13}=\varpi_{12}\varpi_{13}\varpi_{23}=1$, since $\varpi_{12}=1$,
$\varpi_{13}=\varpi_{23}=(-1)^{\varsigma_{i}+\varsigma_{j}+\varsigma_{ij}+1}$,
and $\epsilon_{6}$ is presented in Eq. (\ref{eq:A6}).

{\footnotesize

\begin{center}
\begin{tabular}{|c|c|c|c|c|}
\hline 
$(ij)$ & $\varepsilon_{ij}$ & $\xi$ & $\Gamma_{\xi}$ & $\mathcal{F}$\tabularnewline
\hline
\hline 
$(12)$ & $f_{12}$ & $1$ & $\emptyset$ & $\emptyset$\tabularnewline
\hline 
$(23)$ & $\sum_{\alpha_{23}}f_{23}\epsilon_{6}\left\{ \left(23\right)\epsilon_{6}\right\} $ & $1$ & $\emptyset$ & $\{12\}$\tabularnewline
\hline 
$(34)$ & $\sum_{\alpha_{\zeta\in\Gamma_{\xi}}}f_{34}\epsilon_{\xi}\left\{ \left(34\right)\epsilon_{\xi}\right\} $ & $\begin{array}{c}
1\\
2\\
3\\
4\\
5\end{array}$ & $\begin{array}{c}
\left\{ 34,1234,56\right\} \\
\left\{ 34,56,3456\right\} \\
\left\{ 34,1234,12345\right\} \\
\left\{ 34,345,12345\right\} \\
\left\{ 34,345,3456\right\} \end{array}$ & $\left\{ 123,45\right\} $\tabularnewline
\hline 
$\left(45\right)$ & $f_{45}$ & $1$ & $\emptyset$ & $\emptyset$\tabularnewline
\hline 
$\left(56\right)$ & $\sum_{\alpha_{56}}f_{56}\epsilon_{19}\left\{ \left(56\right)\epsilon_{19}\right\} $ & $1$ & $\emptyset$ & $\left\{ 45\right\} $\tabularnewline
\hline 
$(13)$ & $\sum_{\alpha_{13}}f_{13}\left\{ \left(12\right)\epsilon_{6}\right\} \left\{ \left(123\right)\epsilon_{6}\right\} $ & $1$ & $\emptyset$ & $\{12\}$\tabularnewline
\hline 
$\left(24\right)$ & $\begin{array}{l}
\sum_{\alpha_{13}\alpha_{23}\alpha_{34}}\:\sum_{\alpha_{\zeta\in\Gamma_{\xi}}}f_{24}\epsilon_{6}\left\{ \left(24\right)\epsilon_{6}\right\} \\
\times\left\{ \left(23\right)\left[\epsilon_{6}\epsilon_{\xi}\right]\right\} \left\{ \left(234\right)\left[\epsilon_{6}\epsilon_{\xi}\right]\right\} \end{array}$ & $\begin{array}{c}
1\\
2\\
3\\
4\\
5\end{array}$ & $\begin{array}{c}
\left\{ 24,1234,56\right\} \\
\left\{ 24,56,2456\right\} \\
\left\{ 24,1234,12345\right\} \\
\left\{ 24,245,12345\right\} \\
\left\{ 24,245,2456\right\} \end{array}$ & $\left\{ 12,123,45,456\right\} $\tabularnewline
\hline 
$\left(35\right)$ & $\sum_{\alpha_{\zeta\in\Gamma_{\xi}}}f_{35}\left\{ \left(45\right)\epsilon_{\xi}\right\} \left\{ \left(354\right)\epsilon_{\xi}\right\} $ & $\begin{array}{c}
1\\
2\\
3\\
4\\
5\end{array}$ & $\begin{array}{c}
\left\{ 35,46,1235\right\} \\
\left\{ 35,46,3456\right\} \\
\left\{ 35,1235,12345\right\} \\
\left\{ 35,345,12345\right\} \\
\left\{ 35,345,3456\right\} \end{array}$ & $\left\{ 123,45,456\right\} $\tabularnewline
\hline 
$\left(46\right)$ & $\sum_{\alpha_{46}}f_{46}\left\{ \left(45\right)\epsilon_{19}\right\} \left\{ \left(456\right)\epsilon_{19}\right\} $ & $1$ & $\emptyset$ & $\left\{ 45\right\} $\tabularnewline
\hline 
$\left(14\right)$ & $\begin{array}{l}
\sum_{\alpha_{13}\alpha_{23}\alpha_{34}}\:\sum_{\alpha_{\zeta\in\Gamma_{\xi}}}f_{14}\left\{ \left(12\right)\epsilon_{6}\right\} \left\{ \left(124\right)\epsilon_{6}\right\} \\
\times\left\{ \left(123\right)\left[\epsilon_{6}\epsilon_{\xi}\right]\right\} \left\{ \left(1234\right)\left[\epsilon_{6}\epsilon_{\xi}\right]\right\} \end{array}$ & $\begin{array}{c}
1\\
2\\
3\\
4\\
5\end{array}$ & $\begin{array}{c}
\left\{ 14,1234,56\right\} \\
\left\{ 14,56,1456\right\} \\
\left\{ 14,1234,12345\right\} \\
\left\{ 14,145,12345\right\} \\
\left\{ 14,145,1456\right\} \end{array}$ & $\left\{ 12,123,45,456\right\} $\tabularnewline
\hline 
$\left(25\right)$ & $\begin{array}{l}
\sum_{\alpha_{13}\alpha_{23}\alpha_{35}}\:\sum_{\alpha_{\zeta\in\Gamma_{\xi}}}f_{25}\epsilon_{6}\left\{ \left(254\right)\epsilon_{6}\right\} \\
\times\left\{ \left(2354\right)\left[\epsilon_{6}\epsilon_{\xi}\right]\right\} \left\{ \left(23\right)\left(45\right)\left[\epsilon_{6}\epsilon_{\xi}\right]\right\} \end{array}$ & $\begin{array}{c}
1\\
2\\
3\\
4\\
5\end{array}$ & $\begin{array}{c}
\left\{ 25,46,1235\right\} \\
\left\{ 25,46,2456\right\} \\
\left\{ 25,1235,12345\right\} \\
\left\{ 25,245,12345\right\} \\
\left\{ 25,245,2456\right\} \end{array}$ & $\left\{ 12,123,45,456\right\} $\tabularnewline
\hline 
$\left(36\right)$ & $\begin{array}{l}
\sum_{\alpha_{34}\alpha_{35}\alpha_{56}}\:\sum_{\alpha_{\zeta\in\Gamma_{\xi}}}f_{36}\epsilon_{19}\left\{ \left(36\right)\epsilon_{19}\right\} \left\{ \left(56\right)\epsilon_{19}\right\} \\
\times\left\{ \left(365\right)\epsilon_{19}\right\} \left\{ \left(465\right)\epsilon_{1}\right\} \left\{ \left(3654\right)\epsilon_{1}\right\} \end{array}$ & $\begin{array}{c}
1\\
2\end{array}$ & $\begin{array}{c}
\left\{ 36,1236\right\} \\
\left\{ 36,3456\right\} \end{array}$ & $\left\{ 123,46,456\right\} $\tabularnewline
\hline 
$\left(15\right)$ & $\begin{array}{l}
\sum_{\alpha_{13}\alpha_{23}\alpha_{35}}\:\sum_{\alpha_{\zeta\in\Gamma_{\xi}}}f_{15}\left\{ \left(12\right)\epsilon_{6}\right\} \left\{ \left(1254\right)\epsilon_{6}\right\} \\
\times\left\{ \left(12354\right)\left[\epsilon_{6}\epsilon_{\xi}\right]\right\} \left\{ \left(123\right)\left(45\right)\left[\epsilon_{6}\epsilon_{\xi}\right]\right\} \end{array}$ & $\begin{array}{c}
1\\
2\\
3\\
4\\
5\end{array}$ & $\begin{array}{c}
\left\{ 15,46,1235\right\} \\
\left\{ 15,46,1456\right\} \\
\left\{ 15,1235,12345\right\} \\
\left\{ 15,145,12345\right\} \\
\left\{ 15,145,1456\right\} \end{array}$ & $\left\{ 12,123,45,456\right\} $\tabularnewline
\hline 
$\left(26\right)$ & $\begin{array}{l}
\sum_{\alpha_{13}\alpha_{23}\alpha_{36}}\:\sum_{\alpha_{24}\alpha_{25}\alpha_{56}}\:\sum_{\alpha_{\zeta\in\Gamma_{\xi}}}f_{26}\epsilon_{6}\\
\times\left\{ \left(23\right)\left[\epsilon_{6}\epsilon_{19}\right]\right\} \left\{ \left(26\right)\epsilon_{6}\right\} \left\{ \left(236\right)\left[\epsilon_{6}\epsilon_{19}\right]\right\} \\
\times\left\{ \left(23\right)\left(56\right)\epsilon_{19}\right\} \left\{ \left(2365\right)\epsilon_{19}\right\} \\
\times\left\{ \left(23\right)\left(465\right)\epsilon_{\xi}\right\} \left\{ \left(23654\right)\epsilon_{\xi}\right\} \end{array}$ & $\begin{array}{c}
1\\
2\end{array}$ & $\begin{array}{c}
\left\{ 26,1236\right\} \\
\left\{ 26,2456\right\} \end{array}$ & $\left\{ 12,123,46,456\right\} $\tabularnewline
\hline 
$\left(16\right)$ & $\begin{array}{l}
\sum_{\alpha_{13}\alpha_{23}\alpha_{36}}\:\sum_{\alpha_{14}\alpha_{15}\alpha_{56}}\:\sum_{\alpha_{\zeta\in\Gamma_{\xi}}}f_{16}\left\{ \left(12\right)\epsilon_{6}\right\} \\
\times\left\{ \left(126\right)\epsilon_{6}\right\} \left\{ \left(123\right)\left[\epsilon_{6}\epsilon_{19}\right]\right\} \left\{ \left(1236\right)\left[\epsilon_{6}\epsilon_{19}\right]\right\} \\
\times\left\{ \left(123\right)\left(56\right)\epsilon_{19}\right\} \left\{ \left(12365\right)\epsilon_{19}\right\} \\
\times\left\{ \left(123\right)\left(465\right)\epsilon_{\xi}\right\} \left\{ \left(123654\right)\epsilon_{\xi}\right\} \end{array}$ & $\begin{array}{c}
1\\
2\end{array}$ & $\begin{array}{c}
\left\{ 16,1236\right\} \\
\left\{ 16,1456\right\} \end{array}$ & $\left\{ 12,123,46,456\right\} $\tabularnewline
\hline
\end{tabular}%
\begin{table}[H]
\caption{\label{TabEPS}Recoupling coefficients $\varepsilon_{ij}$}

\end{table}

\par\end{center}

}

\noindent{}If $\widehat{\pi}=\left(132\right)$, a $3$-cycle has
to be decomposed into the composition of two $2$-cycles. Such a decomposition
is the set of operators $\widehat{\mathcal{E}}_{\pi}\in\left\{ \left(13\right)\left(23\right),\left(12\right)\left(13\right),\left(23\right)\left(12\right)\right\} $.
The quantity $\pi$ in the subscript in the notation of the operator
$\widehat{\mathcal{E}}_{\pi}$ shows that the present set has been
obtained from the operator $\widehat{\pi}$. The operators $\widehat{\mathcal{E}}_{\pi}$
will be called the equivalent operators (or equivalent permutations)
for given $\widehat{\pi}$. Particularly, $\widehat{\mathcal{E}}_{ij}=\left(ij\right)$,
i.e., for a $2$-cycle $\left(ij\right)$ there is only one equivalent
operator $\widehat{\mathcal{E}}_{ij}$ which coincides with $\left(ij\right)$.
Hence, $\widehat{\mathcal{E}}_{132}\left\langle iijlpq\right\rangle =\left\langle jiilpq\right\rangle =\sum_{\varsigma_{ii}=\mathrm{odd}}\varepsilon_{132}\left\langle iijlpq\right\rangle $,
where $\widehat{\mathcal{E}}_{132}=\bigl(\begin{smallmatrix}1 & 2 & 3\\ 3 & 1 & 2\end{smallmatrix}\bigr)$.
The task is to find out $\varepsilon_{132}$. If $\widehat{\mathcal{E}}_{132}=\left(13\right)\left(23\right)$,
the third operator in $\left\langle iijlpq\right\rangle $ is permuted
with the second one and after that the permuted third operator is
replaced with the first one (none of the permutations of the same
rank operators are performed). According to Eq. (\ref{eq:4.5}), $\left(13\right)\left(23\right)\left\langle iijlpq\right\rangle $
equals to $\sum_{\varsigma_{ii}=\mathrm{odd}}\varepsilon_{13}\left\{ \left(13\right)\varepsilon_{23}\right\} \left\langle iijlpq\right\rangle $,
where $\varepsilon_{23}=\varpi_{23}\epsilon_{6}\left\{ \left(23\right)\epsilon_{6}\right\} $.
Besides, in this case identities $\varpi_{13}=\varpi_{23}=1$. Then
$\left(13\right)\varepsilon_{23}=\left\{ \left(13\right)\epsilon_{6}\right\} \left\{ \left(132\right)\epsilon_{6}\right\} $.
But $\alpha_{1}=\alpha_{2}=\varsigma_{i}$, $\alpha_{3}=\varsigma_{j}$.
Then, according to Eq. (\ref{eq:A6}) we obtain $\left(13\right)\varepsilon_{23}=-1$,
and the coefficient $\varepsilon_{132}=-\varepsilon_{13}$. If $\widehat{\mathcal{E}}_{132}=\left(12\right)\left(13\right)$,
then, in accordance with Eqs. (\ref{eq:4.5}), (\ref{eq:4.7}), $\left(12\right)\left(13\right)\left\langle iijlpq\right\rangle =\left\langle jiilpq\right\rangle =\sum_{\varsigma_{ii}=\mathrm{odd}}\left\{ \left(12\right)\varepsilon_{13}\right\} \left\langle iijlpq\right\rangle =\sum_{\varsigma_{ii}=\mathrm{odd}}\varepsilon_{23}\left\langle iijlpq\right\rangle $.
The last equality has been obtained exploiting Eq. (\ref{eq:4.7}),
which says that $\left(12\right)\varepsilon_{13}=f_{13}f_{23}\varepsilon_{23}$.
But $f_{13}=1$, $f_{23}=\varpi_{23}=\varpi_{13}=1$, and $\left(12\right)\varepsilon_{13}=\varepsilon_{23}$.
On the other hand, $\varepsilon_{23}=\varepsilon_{13}\left\{ \left(13\right)\varepsilon_{23}\right\} =-\varepsilon_{13}$.
Thus, the coefficient $\varepsilon_{132}=-\varepsilon_{13}$. Finally,
if $\widehat{\mathcal{E}}_{132}=\left(23\right)\left(12\right)$,
then $\left(23\right)\left(12\right)\left\langle iijlpq\right\rangle =\sum_{\varsigma_{ii}=\mathrm{odd}}\varepsilon_{23}\left\langle iijlpq\right\rangle $.
Hence, according to the results of the previous composition, we get
that $\varepsilon_{132}=-\varepsilon_{13}$. This indicates identity
$\left(132\right)\left\langle iijlpq\right\rangle =-\left(13\right)\left\langle iijlpq\right\rangle $.
Note, $\left(132\right)$ does not permute same rank operators, while
$\left(13\right)$ does. This means every $n$-length permutation
$\widehat{\pi}$ with $n>2$ can be replaced by the shorter permutation
$\widehat{\pi}_{\min}$ (if such exists), i.e., the operation with
the less number of the permutations, in the following way

\begin{equation}
\widehat{\pi}\widehat{\mathrm{T}}_{\varkappa}^{\lambda}=\theta\widehat{\pi}_{\min}\widehat{\mathrm{T}}_{\varkappa}^{\lambda},\label{eq:n5.1.12}\end{equation}

\noindent{}where $\theta=\pm1$, and the sign depends whether the
number of permuted same rank operators is even or odd. For example,
$\left(1432\right)\left\langle iiijpq\right\rangle =\left(14\right)\left\langle iiijpq\right\rangle $,
where the operation $\left(1432\right)$ does not permute the same
rank $\varsigma_{i}$ operator. The operation $\left(14\right)$ permutes
the first operator with the second and the third one. In this case
$\theta=1$. Analogous expression can be obtained if we write $\left(243\right)\left\langle iiijpq\right\rangle =-\left(24\right)\left\langle iiijpq\right\rangle $.
Since $\left(14\right)\left(243\right)=\left(1432\right)$ and $\left(14\right)\left(24\right)=\left(142\right)$,
it follows that $\left(1432\right)\left\langle iiijpq\right\rangle =-\left(142\right)\left\langle iiijpq\right\rangle $
is valid. On the other hand $\left(142\right)\left\langle iiijpq\right\rangle =-\left(14\right)\left\langle iiijpq\right\rangle $.
Thus, we obtain identical result. Note, the procedures studied in
this example are valid for all $\widehat{\mathcal{E}}_{1432}$. The
equivalent operators $\widehat{\mathcal{E}}_{\pi}$ for $\widehat{\pi}=\left(ijk\right)$
and $\widehat{\pi}=\left(ijkl\right)$ are these

\begin{equation}
\begin{array}{l}
\widehat{\mathcal{E}}_{ijk}\in\left\{ \left(ij\right)\left(kj\right),\left(kj\right)\left(ik\right),\left(ik\right)\left(ij\right)\right\} ,\\
\widehat{\mathcal{E}}_{ijkl}\in\left\{ \left(ij\right)\left(jk\right)\left(kl\right),\left(jk\right)\left(il\right)\left(ij\right),\left(jk\right)\left(kl\right)\left(il\right),\left(kl\right)\left(il\right)\left(ij\right)\right\} .\end{array}\label{eq:Equiv}\end{equation}

The presented results, generalized in Eq. (\ref{eq:n5.1.12}), demonstrate
how any $n$-length permutation can be replaced by simpler permutation
even if it permutes same rank operators. Equivalently, this fact allows
one to reduce the number of summation parameters, which appear due
to performed permutations.

\section{Inclusion of electron-electron interaction in $n\leq6$ open shells\label{shells}}

In this section we address to the study of the three-particle operator
$\widehat{T}_{\iota}^{\varsigma}\left(\lambda_{i}\lambda_{j}\lambda_{k}\lambda_{l}\lambda_{p}\lambda_{q}\right)=\left\langle ijklpq\right\rangle $
(see Eq. (\ref{eq:2.6})), depending on how many shells $n_{x}\lambda_{x}^{N_{x}}$
it acts on. We distinguish operators $\left\langle ijklpq\right\rangle $
by the classes $X_{n}\left(\Delta_{1},\Delta_{2},\ldots,\Delta_{n}\right)$,
where $\Delta_{x}=N_{x}-\bar{N}_{x}$ denotes the difference of electron
numbers in the shells $n_{x}\lambda_{x}^{N_{x}}$ and $n_{x}\lambda_{x}^{\bar{N}_{x}}$
of bra $\langle\Psi_{f}^{x}\vert$ and ket $\left|\Psi_{i}^{x}\right\rangle $
wave functions, respectively. It is assumed that the conservation
law for electron number is satisfied

\begin{equation}
{\displaystyle \sum_{x=1}^{n}}\Delta_{x}=0.\label{eq:5.3}\end{equation}

\noindent{}In Eq. (\ref{eq:5.3}) the summation is performed over
all shells of bra and ket wave functions. The operators $\left\langle ijklpq\right\rangle $
belong to one class if the sum of numbers $x\in\{i,j,k\}$ and $\bar{x}\in\{-l,-p,-q\}$,
where $x=\vert\bar{x}\vert$ (same shell), coequals to $\Delta_{x}$.
The dimension of class $d=\mathrm{dim}\: X_{n}\left(\Delta_{1},\Delta_{2},\ldots,\Delta_{n}\right)$
is characterized by the number of operators $\left\langle ijklpq\right\rangle $,
which belong to given class. We consider three types of classes: (i)
parent; (ii) dual; (iii) derived. Each parent class $X_{n}\left(\Delta_{1},\Delta_{2},\ldots,\Delta_{n}\right)$
has its dual class $X_{n}^{*}\left(\Delta_{1},\Delta_{2},\ldots,\Delta_{n}\right)$,
if 

\begin{equation}
X_{n}^{*}\left(\Delta_{1},\Delta_{2},\ldots,\Delta_{n}\right)=X_{n}\left(-\Delta_{1},-\Delta_{2},\ldots,-\Delta_{n}\right),\label{eq:5.1.9}\end{equation}

\begin{equation}
\mathrm{dim}\: X_{n}^{*}\left(\Delta_{1},\Delta_{2},\ldots,\Delta_{n}\right)=\mathrm{dim}\: X_{n}\left(\Delta_{1},\Delta_{2},\ldots,\Delta_{n}\right).\label{eq:5.1.10}\end{equation}

\noindent{}The selection which class is parent and which one is dual,
is optional. It is evident, matrix elements of operators, which belong
to given class and its dual one, coincide. Dual class is a particular
case of derived class. Let us study derived classes in a more detail.

Suppose our task is to transform given operator $\left\langle ijklpq\right\rangle \equiv\left\langle x_{\pi}\right\rangle $
of $X_{n}\left(\Delta_{1},\Delta_{2},\ldots,\Delta_{n}\right)$ into
a consequent order operator $\left\langle i^{\prime}j^{\prime}k^{\prime}l^{\prime}p^{\prime}q^{\prime}\right\rangle \equiv\langle x\rangle$,
where $i^{\prime}\leq j^{\prime}\leq k^{\prime}\leq l^{\prime}\leq p^{\prime}\leq q^{\prime}$.
This transformation is depicted by permutation operator $\widehat{\pi}$
of $\mathrm{S}_{n}$

\begin{equation}
\widehat{\pi}\left\langle x\right\rangle =\left\langle x_{\pi}\right\rangle .\label{eq:5.3.2}\end{equation}

\noindent{}Further, we demonstrate how one can obtain acquainted
permutations $\widehat{\pi}^{\prime}$ for $\widehat{\pi}^{\prime}\langle y\rangle=\langle y_{\pi^{\prime}}\rangle$,
where $\langle y_{\pi^{\prime}}\rangle$ belongs to $X_{n}\left(\Delta_{1}^{\prime},\Delta_{2}^{\prime},\ldots,\Delta_{n}^{\prime}\right)$,
if the permutations $\widehat{\pi}$ of $X_{n}\left(\Delta_{1},\Delta_{2},\ldots,\Delta_{n}\right)$
are known. Suppose there is a map

\begin{equation}
\phi:\quad X_{n}\left(\Delta_{1},\Delta_{2},\ldots,\Delta_{n}\right)\mapsto X_{n}\left(\Delta_{1}^{\prime},\Delta_{2}^{\prime},\ldots,\Delta_{n}^{\prime}\right),\label{eq:5.3.3}\end{equation}

\noindent{}which is associated to the numbers $x_{\pi}$ and $y_{\pi^{\prime}}$
as follows

\begin{equation}
y_{\pi^{\prime}}=\phi\left(x_{\pi}\right)\Rightarrow\left\langle y_{\pi^{\prime}}\right\rangle =\phi\left\langle x_{\pi}\right\rangle .\label{eq:5.3.5}\end{equation}

\noindent{}Substituting Eq. (\ref{eq:5.3.2}) in Eq. (\ref{eq:5.3.5})
one gets

\begin{equation}
\widehat{\pi}\left\langle x\right\rangle =\phi^{-1}\widehat{\pi}^{\prime}\left\langle y\right\rangle .\label{eq:5.3.7}\end{equation}

\noindent{}The last equation means that for every operation $\widehat{\pi}$
on the operator $\left\langle x\right\rangle $ from the class $X_{n}\left(\Delta_{1},\Delta_{2},\ldots,\Delta_{n}\right)$
there exists the operation $\phi^{-1}\widehat{\pi}^{\prime}$ on the
operator $\left\langle y\right\rangle $ from the class $X_{n}\left(\Delta_{1}^{\prime},\Delta_{2}^{\prime},\ldots,\Delta_{n}^{\prime}\right)$
such that the resultant operator $\left\langle x_{\pi}\right\rangle $
is invariant. Further, let $\widehat{\Phi}$ be defined by

\begin{equation}
\left\langle x\right\rangle =\widehat{\Phi}\left\langle y\right\rangle .\label{eq:5.3.8}\end{equation}

\noindent{}Then it follows from Eqs. (\ref{eq:5.3.7}), (\ref{eq:5.3.8})

\begin{equation}
\widehat{\pi}^{\prime}=\phi\widehat{\pi}\widehat{\Phi}.\label{eq:5.3.9}\end{equation}

\noindent{}Obtained expression connects the known permutation $\widehat{\pi}$
of the class $X_{n}\left(\Delta_{1},\Delta_{2},\ldots,\Delta_{n}\right)$
with the permutation $\widehat{\pi}^{\prime}$ of $X_{n}\left(\Delta_{1}^{\prime},\Delta_{2}^{\prime},\ldots,\Delta_{n}^{\prime}\right)$,
called derived class. It is important to mention, that the number
of operators $\widehat{\Phi}$ is much more less than the number $\mathrm{dim}\: X_{n}\left(\Delta_{1}^{\prime},\Delta_{2}^{\prime},\ldots,\Delta_{n}^{\prime}\right)$
of operators $\widehat{\pi}^{\prime}$. For example, $\mathrm{dim}\: X_{3}\left(+2,-1,-1\right)=24$,
while the number of operators $\widehat{\Phi}$ equals to $3$. This
number coincides with amount of the ordered operators $\left\langle x\right\rangle $
(it directly follows from Eq. (\ref{eq:5.3.8})).

At this step we pick out special case $\phi=\Pi_{\mu\nu}$, where

\begin{equation}
\Pi_{\mu\nu}X_{n}\left(\Delta_{1},\Delta_{2},\ldots,\Delta_{\mu},\ldots,\Delta_{\nu},\ldots,\Delta_{n}\right)=X_{n}\left(\Delta_{1},\Delta_{2},\ldots,\Delta_{\nu},\ldots,\Delta_{\mu},\ldots,\Delta_{n}\right).\label{eq:Pimn}\end{equation}

\noindent{}Then

\begin{equation}
\widehat{\Phi}=\Pi_{\mu\nu}\widehat{\tilde{\pi}}.\label{eq:5.3.10}\end{equation}

\noindent{}For instance, if $X_{n}\left(\Delta_{1}^{\prime},\Delta_{2}^{\prime},\Delta_{3},\ldots,\Delta_{n}^{\prime}\right)=X_{n}\left(\Delta_{2},\Delta_{1},\Delta_{3},\ldots,\Delta_{n}\right)$,
then $\Pi_{\mu\nu}=\Pi_{12}$. Further, if $X_{n}\left(\Delta_{1},\Delta_{2},\ldots,\Delta_{n}\right)=X_{3}\left(+2,-1,-1\right)$,
then $X_{n}\left(\Delta_{1}^{\prime},\Delta_{2}^{\prime},\ldots,\Delta_{n}^{\prime}\right)=X_{3}\left(-1,+2,-1\right)$.
Notice, the signs of $\Delta_{1}$ and $\Delta_{2}$ have changed.
For this reason the operator $\Pi_{\mu\nu}\widehat{\pi}\widehat{\Phi}$$ $
has to be replaced by $\Pi_{\mu\nu}\widehat{\Phi}\widehat{\pi}$.
Then from Eqs. (\ref{eq:5.3.9})-(\ref{eq:5.3.10}) follows

\begin{equation}
\widehat{\pi}^{\prime}=\widehat{\tilde{\pi}}\widehat{\pi}\label{eq:5.3.11}\end{equation}

\noindent{}and the operator $\widehat{\tilde{\pi}}$ is generated
from the next equation

\begin{equation}
\widehat{\tilde{\pi}}\left\langle y\right\rangle =\Pi_{\mu\nu}\left\langle x\right\rangle ,\label{eq:5.3.12}\end{equation}

\noindent{}where $\Pi_{\mu\nu}$ permutes the numbers $\mu$ and
$\nu$ in $x$. In Eq. (\ref{eq:5.3.12}) $\Pi_{\mu\nu}$ can be substituted
by various products of $\Pi_{\mu_{1}\nu_{1}}\Pi_{\mu_{2}\nu_{2}}\ldots$
depending on the classes in which the operators $\left\langle x_{\pi}\right\rangle $
and $\left\langle y_{\pi^{\prime}}\right\rangle $ are prescribed.
Notice, Eq. (\ref{eq:5.3.11}) allows one to compute, particularly,
the permutations of dual class. If, e.g., $\left\langle x\right\rangle =\left\langle 111112\right\rangle $,
then $\left\langle y\right\rangle =\left\langle 122222\right\rangle $
and $\widehat{\tilde{\pi}}=\left(16\right)$. The representation of
this permutation in dual class is namely $\left(16\right)1_{6}=\left(16\right)$.
Suppose $\widehat{\pi}=-\left(13\right)$ and $\left\langle x\right\rangle =\left\langle 112223\right\rangle $.
Then $\left\langle x_{\pi}\right\rangle =\left\langle 211223\right\rangle $.
For $\Pi_{\mu\nu}=\Pi_{12}$, $\left\langle y_{\pi^{\prime}}\right\rangle =\Pi_{12}\left\langle x_{\pi}\right\rangle =\left\langle 122113\right\rangle $
and $\left\langle y\right\rangle =\left\langle 111223\right\rangle $,
where the operators $\left\langle x\right\rangle ,\left\langle x_{\pi}\right\rangle \in X_{3}\left(+2,-1,-1\right)$
and $\left\langle y\right\rangle ,\left\langle y_{\pi^{\prime}}\right\rangle \in X_{3}\left(-1,+2,-1\right)$.
It follows from Eq. (\ref{eq:5.3.12}), $\widehat{\tilde{\pi}}\left\langle 111223\right\rangle =\Pi_{12}\left\langle 112223\right\rangle =\left\langle 221113\right\rangle $
$\Rightarrow\widehat{\tilde{\pi}}=\left(14253\right)$. According
to Eq. (\ref{eq:n5.1.12}), the operator $\widehat{\tilde{\pi}}$
can be replaced by the less number permutation $\widehat{\tilde{\pi}}=\widehat{\tilde{\pi}}_{\min}=\left(14\right)\left(25\right)$.
Then $\widehat{\pi}^{\prime}=\widehat{\tilde{\pi}}\widehat{\pi}=-\left(14\right)\left(25\right)\left(13\right)=-\left(134\right)\left(25\right)=\left(25\right)\left(34\right)$,
where the last equality has been obtained exploiting Eq. (\ref{eq:n5.1.12}).
The result is checked by the equation $ $$\widehat{\pi}^{\prime}\left\langle y\right\rangle =\left\langle y_{\pi^{\prime}}\right\rangle $,
i.e., $\widehat{\pi}^{\prime}\left\langle 111223\right\rangle =\left\langle 122113\right\rangle \Rightarrow\widehat{\pi}^{\prime}=\left(24\right)\left(35\right)=\left(25\right)\left(34\right)$.
Hence, both methods produce identical results. Notice, $\widehat{\tilde{\pi}}$
is the same for all operators $\left\langle x_{\pi}\right\rangle $,
if the ordered operator is $\left\langle x\right\rangle $. It directly
follows from Eq. (\ref{eq:5.3.12}). That means it is enough to find
out operators $\widehat{\tilde{\pi}}$ - all other permutations $\widehat{\pi}^{\prime}$
of derived class $X_{n}\left(\Delta_{1}^{\prime},\Delta_{2}^{\prime},\ldots,\Delta_{n}^{\prime}\right)$
are found from Eq. (\ref{eq:5.3.11}), if the permutations $\widehat{\pi}$
of parent class $X_{n}\left(\Delta_{1},\Delta_{2},\ldots,\Delta_{n}\right)$
are known.

The classification of irreducible tensor operators $\left\langle ijklpq\right\rangle $
is presented in Appendix \ref{C} in Tabs. \ref{TabX1}-\ref{TabX21}.
In tables notation $\left\langle ijk\left\{ lpq\right\} \right\rangle $
denotes the set $\left\{ \left\langle ijklpq\right\rangle ,\left\langle ijklqp\right\rangle ,\left\langle ijkplq\right\rangle ,\left\langle ijkpql\right\rangle ,\left\langle ijkqlp\right\rangle ,\left\langle ijkqpl\right\rangle \right\} $.
Every permutation of the numbers in the set $\left\{ lpq\right\} $
is prescribed by the operation $\widehat{\vartheta}$, where

\begin{equation}
\widehat{\vartheta}\in\left\{ 1_{6},\left(56\right),\left(45\right),\left(456\right),\left(465\right),\left(46\right)\right\} .\label{eq:5.4.2}\end{equation}

\noindent{}Similarly, $\left\langle \left\{ ijk\right\} lpq\right\rangle $
denotes the set $\left\{ \left\langle ijklpq\right\rangle ,\left\langle ikjlpq\right\rangle ,\left\langle jiklpq\right\rangle ,\left\langle jkilpq\right\rangle ,\left\langle kijlpq\right\rangle ,\left\langle kjilpq\right\rangle \right\} $,
and every permutation of the numbers in the set $\left\{ ijk\right\} $
is prescribed by the operation $\widehat{\eta}$, where

\begin{equation}
\widehat{\eta}\in\left\{ 1_{6},\left(23\right),\left(12\right),\left(123\right),\left(132\right),\left(13\right)\right\} .\label{eq:5.4.3}\end{equation}

\noindent{}The operators $\widehat{\eta}$ and $\widehat{\vartheta}$
commute with each other, i.e., $[\widehat{\eta},\widehat{\vartheta}]=0$.

One fact needs to be mentioned. If some of the differences $\Delta_{x}$
and $\Delta_{y}$ coequal, the expressions of $\widehat{\tilde{\pi}}$
can acquire different forms, though they are equivalent. Suppose (see
Tab. \ref{TabX16}), that $\left\langle y\right\rangle =\left\langle 122344\right\rangle \in X_{4}\left(\Delta_{1}^{\prime},\Delta_{2}^{\prime},\Delta_{3}^{\prime},\Delta_{4}^{\prime}\right)=X_{4}\left(-1,0,-1,+2\right)$.
Then operator $\widehat{\tilde{\pi}}=-\left(1634\right)\left(25\right)$
and it satisfies Eq. (\ref{eq:5.3.12}) with $\Pi_{\mu\nu}=\Pi_{23}\Pi_{14}\Pi_{34}$.
Further, if $\left\langle x_{\pi}\right\rangle =\left\langle 114234\right\rangle $,
then $\widehat{\pi}=\left(354\right)1_{6}=\left(354\right)$. Consequently,
permutation $\widehat{\pi}^{\prime}=\widehat{\tilde{\pi}}\widehat{\pi}=-\left(1634\right)\left(25\right)\left(354\right)=-\left(15\right)\left(26\right)$
and thus irreducible tensor operator equals to $\left\langle y_{\pi^{\prime}}\right\rangle =\left\langle 442132\right\rangle \in X_{4}\left(-1,0,-1,+2\right)$.
In this case $\Delta_{2}=\Delta_{3}=-1$ and $\Delta_{1}^{\prime}=\Delta_{3}^{\prime}=-1$.
Hence there is another $\widehat{\tilde{\pi}}=-\left(163\right)\left(25\right)$
value, which also satisfies Eq. (\ref{eq:5.3.12}) with $\Pi_{\mu\nu}=\Pi_{14}\Pi_{12}\Pi_{24}$.
This implies $\widehat{\pi}^{\prime}=\widehat{\tilde{\pi}}\widehat{\pi}=-\left(163\right)\left(25\right)\left(354\right)=-\left(154\right)\left(26\right)$
and the tensor operator $\left\langle y_{\pi^{\prime}}\right\rangle =\left\langle 442312\right\rangle \in X_{4}\left(-1,0,-1,+2\right)$.
It is seen, both operators $\left\langle 442132\right\rangle $ and
$\left\langle 442312\right\rangle $ belong to the same class $X_{4}\left(-1,0,-1,+2\right)$,
where $ $$\left\langle 442132\right\rangle =\left(45\right)\left\langle 442312\right\rangle $.
Hence, both $\widehat{\tilde{\pi}}$ operators generate the elements
of same class.

\section{Matrix elements\label{matrix}}

In this section we study matrix elements $\langle\Psi_{f}\vert\:_{\Lambda}\widehat{T}_{\mu}^{\lambda}\vert\Psi_{i}\rangle$
on the basis%
\footnote{The subscripts $i$ and $f$ in $\Psi_{i}$ and $\Psi_{f}$ designate
initial and final states.%
}

\begin{equation}
\left|\Psi_{i}\right\rangle =\left|\lambda_{1}^{N_{1}}\lambda_{2}^{N_{2}}\ldots\lambda_{n}^{N_{n}}\Gamma_{1}\Lambda_{1}\Gamma_{2}\Lambda_{2}\Lambda_{12}\Gamma_{3}\Lambda_{3}\Lambda_{123}\ldots\Gamma_{u}\Lambda_{u}\Lambda M_{\Lambda}\right\rangle ,\label{eq:6.2}\end{equation}

\noindent{}composed of $u$ shells of equivalent electrons, where
single-shell function in $\Lambda$-space is defined by

\begin{equation}
\left|\Psi_{i}^{x}\right\rangle =\left|\lambda_{x}^{N_{x}}\Gamma_{x}\Lambda_{x}M_{\Lambda_{x}}\right\rangle ,\label{eq:6.1}\end{equation}

\noindent{}with $\Gamma_{x}$ indicating additional quantum numbers.
In $q$-space $\vert\Psi_{i}^{x}\rangle$ is denoted $\left|\lambda_{x}\Gamma_{x}q_{x}m_{q_{x}}\right\rangle $
($q\equiv Q\Lambda$). Then the total wave function $\left|\Psi_{i}\right\rangle $
in $q$-space (for $n_{x}=\textrm{const.}$ $\forall x=1,2,\ldots,u$)
is represented by superposition of functions $\left|\lambda_{x}\Gamma_{x}q_{x}m_{q_{x}}\right\rangle $,
coupled by the same scheme as in Eq. (\ref{eq:6.2}). Namely,

\begin{equation}
\left|\left(\lambda_{1}+\lambda_{2}+\ldots+\lambda_{u}\right)\Gamma_{1}q_{1}\Gamma_{2}q_{2}q_{12}\ldots\Gamma_{u}q_{u}qm_{q}\right\rangle .\label{eq:6.1.18}\end{equation}

\noindent{}The momenta coupling is performed by using known rules
of angular momentum theory. The coupling coefficients are usual Clebsch-Gordan
coefficients. In Eq. (\ref{eq:6.1.18}) the representations $\lambda_{x}$
are written in a sum differently from Eq. (\ref{eq:6.2}). This is
because the total basis index of $Q$ in Eq. (\ref{eq:6.1.18}) is
$M_{Q}=M_{Q_{1}}+M_{Q_{2}}+\ldots+M_{Q_{u}}=\left(2N-\left[\lambda\right]\right)/4$,
where $N=N_{1}+N_{2}+\ldots+N_{u}$ and $\lambda=\lambda_{1}+\lambda_{2}+\ldots+\lambda_{u}$,
i.e., the representation of constructed function in $\Lambda$-space
is the function of the shell $n_{x}\lambda^{N}$ \cite{Rudzikas}.

The expressions of reduced matrix elements in quasispin formalism
can be found in various literature. See, for instance, \cite{Judd,Rudzikas2}.
Reduced matrix element of operator $\:_{\Lambda}\widehat{T}^{\lambda}$
in $\Lambda$-space will be marked by $\bigl[\lambda_{x}^{N_{x}}\Gamma_{x}\Lambda_{x}\Vert\:_{\Lambda}\widehat{T}^{\lambda}\Vert\lambda_{x}^{\bar{N}_{x}}\bar{\Gamma}_{x}\bar{\Lambda}_{x}\bigr]$,
while reduced matrix element of operator $\widehat{T}^{\varsigma}$
in $q$-space will be denoted by $\bigl[\lambda_{x}\Gamma_{x}q_{x}|||\widehat{T}^{\varsigma}|||\lambda_{x}\bar{\Gamma}_{x}\bar{q}_{x}\bigr]$. 

The relationship between reduced matrix elements in $\Lambda$ and
$q$ spaces of the corresponding operators $\:_{\Lambda}\widehat{T}^{\lambda}$
and $\widehat{T}^{\varsigma}=\widehat{T}^{\kappa\lambda}$ (see Eqs.
(\ref{eq:2.5})-(\ref{eq:2.6})) is associated by the next equation

\[
\bigl[\lambda_{1}^{N_{1}}\lambda_{2}^{N_{2}}\ldots\lambda_{u}^{N_{u}}\Gamma_{1}\Lambda_{1}\Gamma_{2}\Lambda_{2}\Lambda_{12}\ldots\Gamma_{u}\Lambda_{u}\Lambda\Vert\:_{\Lambda}\widehat{T}^{\lambda}\Vert\lambda_{1}^{\bar{N}_{1}}\lambda_{2}^{\bar{N}_{2}}\ldots\lambda_{u}^{\bar{N}_{u}}\bar{\Gamma}_{1}\bar{\Lambda}_{1}\bar{\Gamma}_{2}\bar{\Lambda}_{2}\bar{\Lambda}_{12}\ldots\bar{\Gamma}_{u}\bar{\Lambda}_{u}\bar{\Lambda}\bigr]\]

\[
=6\delta_{N,\bar{N}}\delta_{\pi,0}\delta_{\kappa_{ij},1}\delta_{\kappa_{ij},\kappa_{lp}}\delta_{\kappa_{ijk},\frac{3}{2}}\delta_{\kappa_{ijk},\kappa_{lpq}}{\displaystyle \sum_{\kappa=0}^{3}}\frac{\left[\kappa\right]^{\frac{1}{2}}}{\sqrt{\left(3-\kappa\right)!\left(4+\kappa\right)!}}\left\langle \bar{Q}M_{Q}\kappa0|QM_{Q}\right\rangle \]

\begin{equation}
\times\bigl[\left(\lambda_{1}+\lambda_{2}+\ldots+\lambda_{u}\right)\Gamma_{1}q_{1}\Gamma_{2}q_{2}q_{12}\ldots\Gamma_{u}q_{u}q|||\widehat{T}^{\kappa\lambda}|||\left(\lambda_{1}+\lambda_{2}+\ldots+\lambda_{u}\right)\bar{\Gamma}_{1}\bar{q}_{1}\bar{\Gamma}_{2}\bar{q}_{2}\bar{q}_{12}\ldots\bar{\Gamma}_{u}\bar{q}_{u}\bar{q}\bigr].\label{eq:6.6}\end{equation}

\noindent{}Note, the condition $\pi=0$ has already been used in
Sec. \ref{shells}, when classifying tensor operators $\left\langle ijklpq\right\rangle $.
It directly follows from Eq. (\ref{eq:6.6}) that $N=\bar{N}$, where
$N=\sum_{x=1}^{u}N_{x}$ and $\bar{N}=\sum_{x=1}^{u}\bar{N}_{x}$.
This coincides with Eq. (\ref{eq:5.3}).

Reduced matrix element of $\widehat{T}^{\varsigma}\left(\lambda_{i}\lambda_{j}\lambda_{k}\lambda_{l}\lambda_{p}\lambda_{q}\right)$
with $i,j,k,l,p,q\in\left\{ 1,2,\ldots,n\right\} \subseteq\mathcal{I}$
when $n>1$ is constructed as follows

\[
\bigl[\left(\lambda_{1}+\ldots+\lambda_{n}\right)\Gamma_{1}q_{1}\Gamma_{2}q_{2}q_{12}\ldots\Gamma_{n}q_{n}q|||\widehat{T}^{\varsigma}\left(\lambda_{i}\lambda_{j}\lambda_{k}\lambda_{l}\lambda_{p}\lambda_{q}\right)|||\left(\lambda_{1}+\ldots+\lambda_{n}\right)\bar{\Gamma}_{1}\bar{q}_{1}\bar{\Gamma}_{2}\bar{q}_{2}\bar{q}_{12}\ldots\bar{\Gamma}_{n}\bar{q}_{n}\bar{q}\bigr]\]

\[
=(-1)^{\Phi_{n}}{\displaystyle \sum_{\varsigma_{w\in\mathcal{L}_{\xi}}}\epsilon_{\xi}{\displaystyle \prod_{x=1}^{n}}\left[q_{12\ldots x},q_{x+1},\bar{q}_{12\ldots x+1},\varsigma_{p_{x+1}}\right]^{\frac{1}{2}}\left\{ \begin{array}{ccc}
\bar{q}_{12\ldots x} & \bar{q}_{x+1} & \bar{q}_{12\ldots x+1}\\
\varsigma_{p_{x}} & \varsigma_{p_{x+1\: x+1}} & \varsigma_{p_{x+1}}\\
q_{12\ldots x} & q_{x+1} & q_{12\ldots x+1}\end{array}\right\} }\]

\begin{equation}
\times\left[\lambda_{x}\Gamma_{x}q_{x}|||\widehat{\mathcal{O}}^{\varsigma_{p_{xx}}}\left(\lambda_{x}\lambda_{x}\ldots\lambda_{x}\right)|||\lambda_{x}\bar{\Gamma}_{x}\bar{q}_{x}\right];\quad q_{12\ldots n}=q,\quad\bar{q}_{12\ldots n}=\bar{q},\quad\varsigma_{p_{n}}=\varsigma.\label{eq:6.4.new11}\end{equation}

\noindent{}The numbers $p_{x}$ and $p_{\ell\ell}$ are defined by

\begin{equation}
p_{x}=11\ldots122\ldots2xx\ldots x,\quad p_{\ell\ell}=\ell\ell\ldots\ell\quad\left(\mathcal{N}_{\ell}\:\textrm{times}\right).\label{eq:6.4.new12}\end{equation}

\noindent{}The phase multiplier $\Phi_{n}$ is 

\begin{equation}
\Phi_{n}={\displaystyle \sum_{x=1}^{n}}\left[\left(N_{x}+\mathcal{N}_{x}+\Delta_{x}\right){\displaystyle \sum_{y>x}^{n}}N_{y}+\mathcal{N}_{x}{\displaystyle \sum_{z=1}^{x-1}}\left(N_{z}+\Delta_{z}\right)\right].\label{eq:6.4.12n}\end{equation}

\noindent{}Here it is assumed that $\widehat{T}^{\varsigma}\left(\lambda_{i}\lambda_{j}\lambda_{k}\lambda_{l}\lambda_{p}\lambda_{q}\right)$
is presented in a consequent order (Sec. \ref{shells}), i.e., it
is of the form $\left\langle x\right\rangle \in X_{n}\left(\Delta_{1},\Delta_{2},\ldots,\Delta_{n}\right)$.
In Eq. (\ref{eq:6.4.12n}) $\mathcal{N}_{x}$ denotes the number of
operators $a^{\varsigma_{x}}$ in $\widehat{T}^{\varsigma}\left(\lambda_{i}\lambda_{j}\lambda_{k}\lambda_{l}\lambda_{p}\lambda_{q}\right)$.
In general, the operator $\widehat{\mathcal{O}}$ reads

{\footnotesize\begin{multicols}{2}

\begin{center}
\begin{tabular}{|c|c|c|c|c|}
\hline 
$X_{2}\left(\Delta_{1},\Delta_{2}\right)$ & $\left\langle x\right\rangle $ & $\xi$ & $w_{1}$ & $w_{2}$\tabularnewline
\hline
\hline 
$X_{2}\left(0,0\right)$ & $\left\langle 111122\right\rangle $ & $20$ & $1111$ & $22$\tabularnewline
\cline{2-5} 
 & $\left\langle 112222\right\rangle $ & $5$ & $2222$ & $-$\tabularnewline
\hline 
 & $\left\langle 111112\right\rangle $ & $27$ & $1111$ & $11111$\tabularnewline
\cline{2-5} 
$X_{2}\left(+1,-1\right)$ & $\left\langle 111222\right\rangle $ & $14$ & $-$ & $-$\tabularnewline
\cline{2-5} 
 & $\left\langle 122222\right\rangle $ & $30$ & $2222$ & $22222$\tabularnewline
\hline
\end{tabular}%
\begin{table}[H]
\caption{\label{TabM1}The parameters for three-particle matrix element. Two-shell
case}

\end{table}

\par\end{center}

\begin{center}
\begin{tabular}{|c|c|c|c|c|c|}
\hline 
$X_{3}\left(\Delta_{1},\Delta_{2},\Delta_{3}\right)$ & $\left\langle x\right\rangle $ & $\xi$ & $w_{1}$ & $w_{2}$ & $w_{3}$\tabularnewline
\hline
\hline 
$X_{3}\left(0,0,0\right)$ & $\left\langle 112233\right\rangle $ & $1$ & $22$ & $33$ & $1122$\tabularnewline
\hline 
 & $\left\langle 111123\right\rangle $ & $27$ & $1111$ & $11112$ & $-$\tabularnewline
\cline{2-6} 
$X_{3}\left(+2,-1,-1\right)$ & $\left\langle 112223\right\rangle $ & $4$ & $222$ & $11222$ & $-$\tabularnewline
\cline{2-6} 
 & $\left\langle 112333\right\rangle $ & $14$ & $-$ & $-$ & $-$\tabularnewline
\hline 
$X_{3}\left(+3,-2,-1\right)$ & $\left\langle 111223\right\rangle $ & $15$ & $11122$ & $-$ & $-$\tabularnewline
\hline 
 & $\left\langle 123333\right\rangle $ & $5$ & $3333$ & $-$ & $-$\tabularnewline
\cline{2-6} 
$X_{3}\left(+1,-1,0\right)$ & $\left\langle 111233\right\rangle $ & $20$ & $1112$ & $33$ & $-$\tabularnewline
\cline{2-6} 
 & $\left\langle 122233\right\rangle $ & $24$ & $22$ & $33$ & $222$\tabularnewline
\hline
\hline 
$X_{3}\left(\Delta_{1}^{\prime},\Delta_{2}^{\prime},\Delta_{3}^{\prime}\right)$ & $\left\langle y\right\rangle $ & $\xi$ & $w_{1}$ & $w_{2}$ & $w_{3}$\tabularnewline
\hline
\hline 
$X_{3}\left(-1,+2,-1\right)$ & $\left\langle 122223\right\rangle $ & $31$ & $222$ & $2222$ & $12222$\tabularnewline
\cline{2-6} 
 & $\left\langle 122333\right\rangle $ & $6$ & $22$ & $-$ & $-$\tabularnewline
\hline
\end{tabular}%
\begin{table}[H]
\caption{\label{TabM2}The parameters for three-particle matrix element. Three-shell
case}

\end{table}

\par\end{center}

\end{multicols}}

\begin{equation}
\widehat{\mathcal{O}}^{\varsigma_{xy\ldots z}}\left(\lambda_{x}\lambda_{y}\ldots\lambda_{z}\right)=\left[\left[\ldots\left[a^{\varsigma_{x}}\times a^{\varsigma_{y}}\right]^{\varsigma_{xy}}\times\ldots\right]^{\varsigma_{xy\ldots}}\times a^{\varsigma_{z}}\right]^{\varsigma_{xy\ldots z}}.\label{eq:6.4.11}\end{equation}

\noindent{}The representation of $\widehat{\mathcal{O}}^{\varsigma}$
in $\Lambda$-space is $\widehat{\mathcal{O}}^{\lambda}$. Reduced
matrix elements of $\widehat{\mathcal{O}}^{\lambda}$ are known. Expressions
can be found, for instance, in \cite{Merkelis2}. Thus, the reduced
matrix element on the right hand side of Eq. (\ref{eq:6.4.new11})
is assumed to be known. In Eq. (\ref{eq:6.4.new11}) operators $\widehat{\mathcal{O}}^{\varsigma_{p_{xx}}}\left(\lambda_{x}\lambda_{x}\ldots\lambda_{x}\right)$
consist of $\mathcal{N}_{x}$ operators $a^{\varsigma_{x}}$. Representations
$\varsigma_{p_{\ell\ell}}$ with $p_{\ell\ell}$ being $\mathcal{N}_{\ell}$-length
number, are prescribed in $q_{\ell}$-space. The summation is performed
over $\varsigma_{w}$, where the indices $w\in\mathcal{L}_{\xi}=\left\{ w_{1},w_{2},w_{3}\right\} $.
The values of $w_{i\in\left\{ 1,2,3\right\} }$ depend on the concrete
operator $\left\langle x\right\rangle $. These are depicted in Tabs.
\ref{TabM1}-\ref{TabM4}.

{\footnotesize

\begin{center}
\begin{tabular}{|c|c|c|c|c|c|}
\hline 
$X_{4}\left(\Delta_{1},\Delta_{2},\Delta_{3},\Delta_{4}\right)$ & $\left\langle x\right\rangle $ & $\xi$ & $w_{1}$ & $w_{2}$ & $w_{3}$\tabularnewline
\hline
\hline 
 & $\left\langle 111234\right\rangle $ & $27$ & $1112$ & $11123$ & $-$\tabularnewline
\cline{2-6} 
$X_{4}\left(+1,+1,-1,-1\right)$ & $\left\langle 122234\right\rangle $ & $29$ & $222$ & $1222$ & $12223$\tabularnewline
\cline{2-6} 
 & $\left\langle 123334\right\rangle $ & $4$ & $333$ & $12333$ & $-$\tabularnewline
\cline{2-6} 
 & $\left\langle 123444\right\rangle $ & $14$ & $-$ & $-$ & $-$\tabularnewline
\hline 
$X_{4}\left(+2,-2,+1,-1\right)$ & $\left\langle 112234\right\rangle $ & $3$ & $22$ & $1122$ & $11223$\tabularnewline
\hline
\hline 
$X_{4}\left(\Delta_{1}^{\prime},\Delta_{2}^{\prime},\Delta_{3}^{\prime},\Delta_{4}^{\prime}\right)$ & $\left\langle y\right\rangle $ & $\xi$ & $w_{1}$ & $w_{2}$ & $w_{3}$\tabularnewline
\hline
\hline 
$X_{4}\left(+2,+1,-2,-1\right)$ & $\left\langle 112334\right\rangle $ & $15$ & $11233$ & $-$ & $-$\tabularnewline
\hline
$X_{4}\left(+2,+1,-1,-2\right)$ & $\left\langle 112344\right\rangle $ & $20$ & $44$ & $1123$ & $-$\tabularnewline
\hline
$X_{4}\left(+1,+2,-2,-1\right)$ & $\left\langle 122334\right\rangle $ & $7$ & $12233$ & $-$ & $-$\tabularnewline
\hline
$X_{4}\left(+1,+2,-1,-2\right)$ & $\left\langle 122344\right\rangle $ & $23$ & $44$ & $1223$ & $-$\tabularnewline
\hline
$X_{4}\left(+1,-1,+2,-2\right)$ & $\left\langle 123344\right\rangle $ & $1$ & $33$ & $44$ & $1233$\tabularnewline
\hline
\end{tabular}%
\begin{table}[H]
\caption{\label{TabM3}The parameters for three-particle matrix element. Four-shell
case}

\end{table}

\par\end{center}

\begin{center}
\begin{tabular}{|c|c|c|c|c|c|}
\hline 
$X_{5}\left(\Delta_{1},\Delta_{2},\Delta_{3},\Delta_{4},\Delta_{5}\right)$ & $\left\langle x\right\rangle $ & $\xi$ & $w_{1}$ & $w_{2}$ & $w_{3}$\tabularnewline
\hline
\hline 
$X_{5}\left(+2,+1,-1,-1,-1\right)$ & $\left\langle 112345\right\rangle $ & $27$ & $1123$ & $11234$ & $-$\tabularnewline
\hline 
$X_{5}\left(+1,+1,-1,-1,0\right)$ & $\left\langle 123455\right\rangle $ & $20$ & $55$ & $1234$ & $-$\tabularnewline
\hline
\hline 
$X_{5}\left(\Delta_{1}^{\prime},\Delta_{2}^{\prime},\Delta_{3}^{\prime},\Delta_{4}^{\prime},\Delta_{5}^{\prime}\right)$ & $\left\langle y\right\rangle $ & $\xi$ & $w_{1}$ & $w_{2}$ & $w_{3}$\tabularnewline
\hline
\hline 
$X_{5}\left(+1,+2,-1,-1,-1\right)$ & $\left\langle 122345\right\rangle $ & $28$ & $22$ & $1223$ & $12234$\tabularnewline
\hline
$X_{5}\left(-1,-1,+2,+1,-1\right)$ & $\left\langle 123345\right\rangle $ & $3$ & $33$ & $1233$ & $12334$\tabularnewline
\hline
$X_{5}\left(-1,+1,-1,+2,-1\right)$ & $\left\langle 123445\right\rangle $ & $15$ & $12344$ & $-$ & $-$\tabularnewline
\hline
\end{tabular}%
\begin{table}[H]
\caption{\label{TabM4}The parameters for three-particle matrix element. Five-shell
case}

\end{table}

\par\end{center}

}

When $n=6$, the parameters, suitable for reduced matrix element of
$\left\langle x\right\rangle =\left\langle 123456\right\rangle $,
where $\left\langle 123456\right\rangle \in X_{6}\left(+1,+1,+1,-1,-1,-1\right)$,
are these: $\xi=27$, $w_{1}=1234$, $w_{2}=12345$.

Let us take some examples, which show how one can easily obtain the
structure of $\left\langle x\right\rangle $ matrix element. Suppose
$\left\langle x\right\rangle =\left\langle 111122\right\rangle $.
Then $\mathcal{N}_{1}=4$, $\mathcal{N}_{2}=2$ and (see Tab. \ref{TabM1})

\[
\bigl[\left(\lambda_{1}+\lambda_{2}\right)\Gamma_{1}q_{1}\Gamma_{2}q_{2}q|||\widehat{T}^{\varsigma}\left(\lambda_{1}\lambda_{1}\lambda_{1}\lambda_{1}\lambda_{2}\lambda_{2}\right)|||\left(\lambda_{1}+\lambda_{2}\right)\bar{\Gamma}_{1}\bar{q}_{1}\bar{\Gamma}_{2}\bar{q}_{2}\bar{q}\bigr]=(-1)^{N_{1}N_{2}}\left[q_{1},q_{2},\bar{q},\varsigma\right]^{\frac{1}{2}}{\displaystyle \sum_{\varsigma_{1111}\varsigma_{22}}}\epsilon_{20}\]

\begin{equation}
\times\left\{ \begin{array}{ccc}
\bar{q}_{1} & \bar{q}_{2} & \bar{q}\\
\varsigma_{1111} & \varsigma_{22} & \varsigma\\
q_{1} & q_{2} & q\end{array}\right\} \bigl[\lambda_{1}\Gamma_{1}q_{1}|||\widehat{\mathcal{O}}^{\varsigma_{1111}}\left(\lambda_{1}\lambda_{1}\lambda_{1}\lambda_{1}\right)|||\lambda_{1}\bar{\Gamma}_{1}\bar{q}_{1}\bigr]\bigl[\lambda_{2}\Gamma_{2}q_{2}|||\widehat{\mathcal{O}}^{\varsigma_{22}}\left(\lambda_{2}\lambda_{2}\right)|||\lambda_{2}\bar{\Gamma}_{2}\bar{q}_{2}\bigr].\label{eq:6.4.13}\end{equation}

\noindent{}Other matrix elements of $\left\langle x_{\pi}\right\rangle =\widehat{\pi}\left\langle x\right\rangle $
are obtained using Eq. (\ref{eq:4.9}). For example, the operator
$\left\langle 121121\right\rangle =-\left(26\right)\left\langle 111122\right\rangle $.
This points to (see Eq. (\ref{eq:4.5}) and Tab. \ref{TabX1})

\[
\bigl[\left(\lambda_{1}+\lambda_{2}\right)\Gamma_{1}q_{1}\Gamma_{2}q_{2}q|||\widehat{T}^{\varsigma}\left(\lambda_{1}\lambda_{2}\lambda_{1}\lambda_{1}\lambda_{2}\lambda_{1}\right)|||\left(\lambda_{1}+\lambda_{2}\right)\bar{\Gamma}_{1}\bar{q}_{1}\bar{\Gamma}_{2}\bar{q}_{2}\bar{q}\bigr]\]

\begin{equation}
=-{\displaystyle \sum_{\varsigma_{11}\varsigma_{111}\varsigma_{122}}}\varepsilon_{26}\bigl[\left(\lambda_{1}+\lambda_{2}\right)\Gamma_{1}q_{1}\Gamma_{2}q_{2}q|||\widehat{T}^{\varsigma}\left(\lambda_{1}\lambda_{1}\lambda_{1}\lambda_{1}\lambda_{2}\lambda_{2}\right)|||\left(\lambda_{1}+\lambda_{2}\right)\bar{\Gamma}_{1}\bar{q}_{1}\bar{\Gamma}_{2}\bar{q}_{2}\bar{q}\bigr].\label{eq:6.4.15}\end{equation}

\section{Conclusions}

The comprehensive treatment of an effective three-particle operator
$\widehat{L}_{3}$ for open-shell atoms was presented. However, for
open-shell atoms the overall treatment of this operator is very complex
and tedious. In present paper effective three-particle operator was
studied requesting quasispin formalism. Due to the complexity and
high abundance of miscellaneous forms to reduce given operator $\widehat{L}_{3}$,
the procedure, suitable for any $\widehat{L}_{n}$, to classify irreducible
tensor operators were performed by exploiting irreducible representations
of $\mathrm{S}_{n}$, combined with proposed adaptation of $n$-tuple
concept. Recoupling coefficients, called the basis coefficients, which
arise when transforming one irreducible tensor form into another,
are expressed by $3nj$-coefficients (Appendix \ref{B}). As long
as $\widehat{L}_{3}$ consists of six operators $a^{\varsigma_{x}}$,
there are $46,656$ ways to express chosen coupling scheme with respect
to the ordering of ranks $\varsigma_{x}$. These possible distributions
were also determined proposing the so-called permutation coefficients,
i.e., the coefficients which arise when transforming one partition
with its own ordering of ranks into another (or the same) partition
with different ordering of ranks. Permutation coefficients, expressed
by the basis coefficients, appear when a $2$-cycle operation $\left(ij\right)$
acts on chosen tensor operator in such a way, that the ordering of
given operator ranks is permuted into the consequent order, labeled
by the sequence $1,2,\ldots,6$. The permutation considered makes
sense when one-shell wave functions, on the basis of which the matrix
element is calculated, are coupled in analogous consequent order.
Permutation coefficients were constructed making the least number
of recouplings, since, in general, every permutation stipulates the
summation over intermediate ranks. Besides, we have proved that every
permutation can be replaced by the less number permutation (if such
exists), irrespective of whether the ranks of permuted operators $a^{\varsigma_{x}}$
are equal or not. Present circumstance allows to reduce the amount
of summation parameters, which are found when performing recoupling
of ranks.

The three-particle operators, which act on $n=1,2,\ldots,6$ shells
of equivalent electrons, were distinguished by classes (see Sec. \ref{shells}
and Appendix \ref{C}). These classes are characterized by the number
of electrons in bra and ket wave functions, and by the number of shells.
The proposed classification allows us to find out momenta recoupling
coefficients, which appear when permuting the ranks of any tensor
operator from total number $46,656$. Consequently, proposed systematic
classification permits to account for electron-electron correlations
in $n$ open shells ($n\leq6$) in a convenient way. The inclusion
of these effects is estimated by three-particle matrix elements (Sec.
\ref{matrix}).

\appendix\appendixpage\section{Basis coefficients}\label{B}

{\footnotesize

\[
\epsilon_{1}=(-1)^{\wp_{1}}\Delta\left(\alpha_{1},\alpha_{2},\alpha_{12}\right)\left[\alpha_{34},\alpha_{45},\alpha_{56},\alpha_{123},\alpha_{456},\alpha_{1234}\right]^{\frac{1}{2}}\]

\begin{equation}
\times\left\{ \begin{array}{ccc}
\alpha_{6} & \alpha_{5} & \alpha_{56}\\
\alpha_{4} & \alpha_{456} & \alpha_{45}\end{array}\right\} \left\{ \begin{array}{ccc}
\alpha_{12} & \alpha_{3} & \alpha_{123}\\
\alpha_{4} & \alpha_{1234} & \alpha_{34}\end{array}\right\} \left\{ \begin{array}{ccc}
\alpha_{123} & \alpha_{4} & \alpha_{1234}\\
\alpha_{56} & \alpha & \alpha_{456}\end{array}\right\} ,\label{eq:A1}\end{equation}

\begin{equation}
\wp_{1}=\alpha_{5}+\alpha_{6}-\alpha_{56}+\alpha-\alpha_{12}+\alpha_{123}-\alpha_{1234}-\alpha_{3}-\alpha_{4}-\alpha_{456}.\label{eq:A1a}\end{equation}

\[
\epsilon_{2}=(-1)^{\wp_{2}}\Delta\left(\alpha_{1},\alpha_{2},\alpha_{12}\right)\left[\alpha_{34},\alpha_{45},\alpha_{56},\alpha_{123},\alpha_{456},\alpha_{3456}\right]^{\frac{1}{2}}\]

\begin{equation}
\times\left\{ \begin{array}{ccc}
\alpha_{6} & \alpha_{5} & \alpha_{56}\\
\alpha_{4} & \alpha_{456} & \alpha_{45}\end{array}\right\} \left\{ \begin{array}{ccc}
\alpha_{12} & \alpha_{3} & \alpha_{123}\\
\alpha_{456} & \alpha & \alpha_{3456}\end{array}\right\} \left\{ \begin{array}{ccc}
\alpha_{3} & \alpha_{4} & \alpha_{34}\\
\alpha_{56} & \alpha_{3456} & \alpha_{456}\end{array}\right\} ,\label{eq:A2}\end{equation}

\begin{equation}
\wp_{2}=\alpha_{5}+\alpha_{6}-\alpha_{56}+\alpha+\alpha_{12}+2\alpha_{3}+\alpha_{3456}.\label{eq:A2a}\end{equation}

\[
\epsilon_{3}=(-1)^{\wp_{3}}\Delta\left(\alpha_{1},\alpha_{2},\alpha_{12}\right)\left[\alpha_{34},\alpha_{45},\alpha_{123},\alpha_{456},\alpha_{1234},\alpha_{12345}\right]^{\frac{1}{2}}\]

\begin{equation}
\times\left\{ \begin{array}{ccc}
\alpha_{12} & \alpha_{3} & \alpha_{123}\\
\alpha_{4} & \alpha_{1234} & \alpha_{34}\end{array}\right\} \left\{ \begin{array}{ccc}
\alpha_{4} & \alpha_{5} & \alpha_{45}\\
\alpha_{12345} & \alpha_{123} & \alpha_{1234}\end{array}\right\} \left\{ \begin{array}{ccc}
\alpha_{45} & \alpha_{6} & \alpha_{456}\\
\alpha & \alpha_{123} & \alpha_{12345}\end{array}\right\} ,\label{eq:A3}\end{equation}

\begin{equation}
\wp_{3}=-\alpha_{45}+\alpha_{12345}-\alpha-\alpha_{5}+\alpha_{6}-\alpha_{12}-\alpha_{1234}-\alpha_{3}.\label{eq:A3a}\end{equation}

\[
\epsilon_{4}=(-1)^{\wp_{4}}\Delta\left(\alpha_{1},\alpha_{2},\alpha_{12}\right)\left[\alpha_{34},\alpha_{45},\alpha_{123},\alpha_{345},\alpha_{456},\alpha_{12345}\right]^{\frac{1}{2}}\]

\begin{equation}
\times\left\{ \begin{array}{ccc}
\alpha_{12} & \alpha_{3} & \alpha_{123}\\
\alpha_{45} & \alpha_{12345} & \alpha_{345}\end{array}\right\} \left\{ \begin{array}{ccc}
\alpha_{4} & \alpha_{5} & \alpha_{45}\\
\alpha_{345} & \alpha_{3} & \alpha_{34}\end{array}\right\} \left\{ \begin{array}{ccc}
\alpha_{45} & \alpha_{6} & \alpha_{456}\\
\alpha & \alpha_{123} & \alpha_{12345}\end{array}\right\} ,\label{eq:A4}\end{equation}

\begin{equation}
\wp_{4}=-\alpha_{4}-\alpha_{5}+\alpha_{345}-\alpha-\alpha_{12}-\alpha_{123}-\alpha_{12345}-\alpha_{6}.\label{eq:A4a}\end{equation}

\[
\epsilon_{5}=(-1)^{\wp_{5}}\Delta\left(\alpha_{1},\alpha_{2},\alpha_{12}\right)\left[\alpha_{34},\alpha_{45},\alpha_{123},\alpha_{345},\alpha_{456},\alpha_{3456}\right]^{\frac{1}{2}}\]

\begin{equation}
\times\left\{ \begin{array}{ccc}
\alpha_{12} & \alpha_{3} & \alpha_{123}\\
\alpha_{456} & \alpha & \alpha_{3456}\end{array}\right\} \left\{ \begin{array}{ccc}
\alpha_{4} & \alpha_{5} & \alpha_{45}\\
\alpha_{345} & \alpha_{3} & \alpha_{34}\end{array}\right\} \left\{ \begin{array}{ccc}
\alpha_{45} & \alpha_{6} & \alpha_{456}\\
\alpha_{3456} & \alpha_{3} & \alpha_{345}\end{array}\right\} ,\label{eq:A5}\end{equation}

\begin{equation}
\wp_{5}=-\alpha_{3}-\alpha_{4}+\alpha_{5}+\alpha_{6}+\alpha_{45}-\alpha_{12}+\alpha_{345}-\alpha_{456}-\alpha_{3456}-\alpha.\label{eq:A5a}\end{equation}

\[
\epsilon_{6}=(-1)^{\wp_{6}}\Delta\left(\alpha_{4},\alpha_{5},\alpha_{45}\right)\Delta\left(\alpha_{45},\alpha_{6},\alpha_{456}\right)\Delta\left(\alpha_{123},\alpha_{456},\alpha\right)\]

\begin{equation}
\times\left[\alpha_{12},\alpha_{23}\right]^{\frac{1}{2}}\left\{ \begin{array}{ccc}
\alpha_{1} & \alpha_{2} & \alpha_{12}\\
\alpha_{3} & \alpha_{123} & \alpha_{23}\end{array}\right\} ,\label{eq:A6}\end{equation}

\begin{equation}
\wp_{6}=\alpha_{1}+\alpha_{2}+\alpha_{3}+\alpha_{123}.\label{eq:A6a}\end{equation}

\[
\epsilon_{7}=(-1)^{\wp_{7}}\Delta\left(\alpha_{4},\alpha_{5},\alpha_{45}\right)\left[\alpha_{12},\alpha_{23},\alpha_{456},\alpha_{12345}\right]^{\frac{1}{2}}\]

\begin{equation}
\times\left\{ \begin{array}{ccc}
\alpha_{1} & \alpha_{2} & \alpha_{12}\\
\alpha_{3} & \alpha_{123} & \alpha_{23}\end{array}\right\} \left\{ \begin{array}{ccc}
\alpha_{45} & \alpha_{6} & \alpha_{456}\\
\alpha & \alpha_{123} & \alpha_{12345}\end{array}\right\} ,\label{eq:A7}\end{equation}

\begin{equation}
\wp_{7}=\alpha_{1}+\alpha_{2}+\alpha_{3}-\alpha_{6}-\alpha_{45}-\alpha.\label{eq:A7a}\end{equation}

\[
\epsilon_{8}=(-1)^{\wp_{8}}\Delta\left(\alpha_{4},\alpha_{5},\alpha_{45}\right)\left[\alpha_{12},\alpha_{23},\alpha_{123},\alpha_{456},\alpha_{2345},\alpha_{12345}\right]^{\frac{1}{2}}\]

\begin{equation}
\times\left\{ \begin{array}{ccc}
\alpha_{1} & \alpha_{2} & \alpha_{12}\\
\alpha_{3} & \alpha_{123} & \alpha_{23}\end{array}\right\} \left\{ \begin{array}{ccc}
\alpha_{1} & \alpha_{23} & \alpha_{123}\\
\alpha_{45} & \alpha_{12345} & \alpha_{2345}\end{array}\right\} \left\{ \begin{array}{ccc}
\alpha_{45} & \alpha_{6} & \alpha_{456}\\
\alpha & \alpha_{123} & \alpha_{12345}\end{array}\right\} ,\label{eq:A8}\end{equation}

\begin{equation}
\wp_{8}=\alpha_{2}+\alpha_{3}-\alpha_{12345}+2\alpha_{1}+\alpha_{23}+\alpha+\alpha_{6}.\label{eq:A8a}\end{equation}

\[
\epsilon_{9}=(-1)^{\wp_{9}}\Delta\left(\alpha_{4},\alpha_{5},\alpha_{45}\right)\left[\alpha_{12},\alpha_{23},\alpha_{123},\alpha_{456},\alpha_{2345},\alpha_{23456}\right]^{\frac{1}{2}}\]

\begin{equation}
\times\left\{ \begin{array}{ccc}
\alpha_{1} & \alpha_{2} & \alpha_{12}\\
\alpha_{3} & \alpha_{123} & \alpha_{23}\end{array}\right\} \left\{ \begin{array}{ccc}
\alpha_{1} & \alpha_{23} & \alpha_{123}\\
\alpha_{456} & \alpha & \alpha_{23456}\end{array}\right\} \left\{ \begin{array}{ccc}
\alpha_{45} & \alpha_{6} & \alpha_{456}\\
\alpha_{23456} & \alpha_{23} & \alpha_{2345}\end{array}\right\} ,\label{eq:A9}\end{equation}

\begin{equation}
\wp_{9}=\alpha_{2}+\alpha_{3}+\alpha_{123}+\alpha_{23456}+\alpha_{6}+\alpha_{45}-\alpha_{456}-\alpha.\label{eq:A9a}\end{equation}

\[
\epsilon_{10}=(-1)^{\wp_{10}}\Delta\left(\alpha_{4},\alpha_{5},\alpha_{45}\right)\Delta\left(\alpha_{45},\alpha_{6},\alpha_{456}\right)\left[\alpha_{12},\alpha_{23},\alpha_{123},\alpha_{23456}\right]^{\frac{1}{2}}\]

\begin{equation}
\times\left\{ \begin{array}{ccc}
\alpha_{1} & \alpha_{2} & \alpha_{12}\\
\alpha_{3} & \alpha_{123} & \alpha_{23}\end{array}\right\} \left\{ \begin{array}{ccc}
\alpha_{1} & \alpha_{23} & \alpha_{123}\\
\alpha_{456} & \alpha & \alpha_{23456}\end{array}\right\} ,\label{eq:A10}\end{equation}

\begin{equation}
\wp_{10}=\alpha-\alpha_{123}+\alpha_{2}-\alpha_{23}+\alpha_{3}+\alpha_{456}.\label{eq:A10a}\end{equation}

\[
\epsilon_{11}=(-1)^{\wp_{11}}\left[\alpha_{12},\alpha_{34},\alpha_{45},\alpha_{56},\alpha_{123},\alpha_{456},\alpha_{234},\alpha_{1234}\right]^{\frac{1}{2}}\]

\begin{equation}
\times\left\{ \begin{array}{ccc}
\alpha_{1} & \alpha_{2} & \alpha_{12}\\
\alpha_{34} & \alpha_{1234} & \alpha_{234}\end{array}\right\} \left\{ \begin{array}{ccc}
\alpha_{12} & \alpha_{3} & \alpha_{123}\\
\alpha_{4} & \alpha_{1234} & \alpha_{34}\end{array}\right\} \left\{ \begin{array}{ccc}
\alpha_{4} & \alpha_{5} & \alpha_{45}\\
\alpha_{6} & \alpha_{456} & \alpha_{56}\end{array}\right\} \left\{ \begin{array}{ccc}
\alpha_{4} & \alpha_{56} & \alpha_{456}\\
\alpha & \alpha_{123} & \alpha_{1234}\end{array}\right\} ,\label{eq:A11}\end{equation}

\begin{equation}
\wp_{11}=-\alpha_{1}+\alpha_{2}+\alpha_{3}+\alpha_{12}+\alpha_{34}+\alpha_{5}+\alpha_{6}-\alpha_{56}+\alpha_{456}-\alpha_{4}-\alpha_{123}-\alpha.\label{eq:A11a}\end{equation}

\[
\epsilon_{12}=(-1)^{\wp_{12}}\left[\alpha_{12},\alpha_{34},\alpha_{45},\alpha_{56},\alpha_{123},\alpha_{234},\alpha_{456},\alpha_{23456}\right]^{\frac{1}{2}}\]

\begin{equation}
\times\left\{ \begin{array}{ccc}
\alpha_{4} & \alpha_{5} & \alpha_{45}\\
\alpha_{6} & \alpha_{456} & \alpha_{56}\end{array}\right\} \left[\begin{array}{cccc}
\alpha_{2} & \alpha_{23456} & \alpha_{3} & \alpha_{456}\\
\alpha_{12} & \alpha_{34} & \alpha & \alpha_{56}\\
\alpha_{1} & \alpha_{123} & \alpha_{234} & \alpha_{4}\end{array}\right],\label{eq:A12}\end{equation}

\begin{equation}
\wp_{12}=\alpha_{5}+\alpha_{6}-\alpha_{123}-\alpha_{4}-\alpha_{234}+\alpha_{23456}+\alpha_{12}-\alpha_{34}.\label{eq:A12a}\end{equation}

\[
\epsilon_{13}=(-1)^{\wp_{13}}\left[\alpha_{12},\alpha_{34},\alpha_{45},\alpha_{56},\alpha_{123},\alpha_{456},\alpha_{3456},\alpha_{23456}\right]^{\frac{1}{2}}\]

\begin{equation}
\times\left\{ \begin{array}{ccc}
\alpha_{1} & \alpha_{2} & \alpha_{12}\\
\alpha_{3456} & \alpha & \alpha_{23456}\end{array}\right\} \left\{ \begin{array}{ccc}
\alpha_{12} & \alpha_{3} & \alpha_{123}\\
\alpha_{456} & \alpha & \alpha_{3456}\end{array}\right\} \left\{ \begin{array}{ccc}
\alpha_{4} & \alpha_{5} & \alpha_{45}\\
\alpha_{6} & \alpha_{456} & \alpha_{56}\end{array}\right\} \left\{ \begin{array}{ccc}
\alpha_{4} & \alpha_{56} & \alpha_{456}\\
\alpha_{3456} & \alpha_{3} & \alpha_{34}\end{array}\right\} ,\label{eq:A13}\end{equation}

\begin{equation}
\wp_{13}=\alpha_{5}+\alpha_{6}-\alpha_{56}-\alpha_{1}-\alpha_{2}+\alpha_{12}+2\alpha_{3}.\label{eq:A13a}\end{equation}

\begin{equation}
\epsilon_{14}=\Delta\left(\alpha_{1},\alpha_{2},\alpha_{12}\right)\Delta\left(\alpha_{12},\alpha_{3},\alpha_{123}\right)\Delta\left(\alpha_{4},\alpha_{5},\alpha_{45}\right)\Delta\left(\alpha_{45},\alpha_{6},\alpha_{456}\right)\Delta\left(\alpha_{123},\alpha_{456},\alpha\right).\label{eq:A14}\end{equation}

\[
\epsilon_{15}=(-1)^{\wp_{15}}\Delta\left(\alpha_{1},\alpha_{2},\alpha_{12}\right)\Delta\left(\alpha_{12},\alpha_{3},\alpha_{123}\right)\Delta\left(\alpha_{4},\alpha_{5},\alpha_{45}\right)\]

\begin{equation}
\times\left[\alpha_{456},\alpha_{12345}\right]^{\frac{1}{2}}\left\{ \begin{array}{ccc}
\alpha_{45} & \alpha_{6} & \alpha_{456}\\
\alpha & \alpha_{123} & \alpha_{12345}\end{array}\right\} ,\label{eq:A15}\end{equation}

\begin{equation}
\wp_{15}=\alpha+\alpha_{123}+\alpha_{6}+\alpha_{45}.\label{eq:A15a}\end{equation}

\[
\epsilon_{16}=(-1)^{\wp_{16}}\Delta\left(\alpha_{1},\alpha_{2},\alpha_{12}\right)\Delta\left(\alpha_{4},\alpha_{5},\alpha_{45}\right)\]

\begin{equation}
\times\left[\alpha_{123},\alpha_{345},\alpha_{456},\alpha_{12345}\right]^{\frac{1}{2}}\left\{ \begin{array}{ccc}
\alpha_{12} & \alpha_{3} & \alpha_{123}\\
\alpha_{45} & \alpha_{12345} & \alpha_{345}\end{array}\right\} \left\{ \begin{array}{ccc}
\alpha_{45} & \alpha_{6} & \alpha_{456}\\
\alpha & \alpha_{123} & \alpha_{12345}\end{array}\right\} ,\label{eq:A16}\end{equation}

\begin{equation}
\wp_{16}=\alpha-\alpha_{12}+\alpha_{123}-\alpha_{12345}-\alpha_{3}+\alpha_{6}.\label{eq:A16a}\end{equation}

\[
\epsilon_{17}=(-1)^{\wp_{17}}\Delta\left(\alpha_{1},\alpha_{2},\alpha_{12}\right)\Delta\left(\alpha_{4},\alpha_{5},\alpha_{45}\right)\]

\begin{equation}
\times\left[\alpha_{123},\alpha_{345},\alpha_{456},\alpha_{3456}\right]^{\frac{1}{2}}\left\{ \begin{array}{ccc}
\alpha_{12} & \alpha_{3} & \alpha_{123}\\
\alpha_{456} & \alpha & \alpha_{3456}\end{array}\right\} \left\{ \begin{array}{ccc}
\alpha_{45} & \alpha_{6} & \alpha_{456}\\
\alpha_{3456} & \alpha_{3} & \alpha_{345}\end{array}\right\} ,\label{eq:A17}\end{equation}

\begin{equation}
\wp_{17}=\alpha+\alpha_{12}-\alpha_{3456}+\alpha_{45}-\alpha_{456}+\alpha_{6}.\label{eq:A17a}\end{equation}

\[
\epsilon_{18}=(-1)^{\wp_{18}}\Delta\left(\alpha_{1},\alpha_{2},\alpha_{12}\right)\Delta\left(\alpha_{4},\alpha_{5},\alpha_{45}\right)\Delta\left(\alpha_{45},\alpha_{6},\alpha_{456}\right)\]

\begin{equation}
\times\left[\alpha_{123},\alpha_{3456}\right]^{\frac{1}{2}}\left\{ \begin{array}{ccc}
\alpha_{12} & \alpha_{3} & \alpha_{123}\\
\alpha_{456} & \alpha & \alpha_{3456}\end{array}\right\} ,\label{eq:A18}\end{equation}

\begin{equation}
\wp_{18}=\alpha+\alpha_{12}+\alpha_{3}+\alpha_{456}.\label{eq:A18a}\end{equation}

\[
\epsilon_{19}=(-1)^{\wp_{19}}\Delta\left(\alpha_{1},\alpha_{2},\alpha_{12}\right)\Delta\left(\alpha_{12},\alpha_{3},\alpha_{123}\right)\Delta\left(\alpha_{123},\alpha_{456},\alpha\right)\]

\begin{equation}
\times\left[\alpha_{45},\alpha_{56}\right]^{\frac{1}{2}}\left\{ \begin{array}{ccc}
\alpha_{4} & \alpha_{5} & \alpha_{45}\\
\alpha_{6} & \alpha_{456} & \alpha_{56}\end{array}\right\} ,\label{eq:A19}\end{equation}

\begin{equation}
\wp_{19}=\alpha_{4}+\alpha_{5}+\alpha_{6}+\alpha_{456}.\label{eq:A19a}\end{equation}

\[
\epsilon_{20}=(-1)^{\wp_{20}}\Delta\left(\alpha_{1},\alpha_{2},\alpha_{12}\right)\Delta\left(\alpha_{12},\alpha_{3},\alpha_{123}\right)\]

\begin{equation}
\times\left[\alpha_{45},\alpha_{56},\alpha_{456},\alpha_{1234}\right]^{\frac{1}{2}}\left\{ \begin{array}{ccc}
\alpha_{4} & \alpha_{5} & \alpha_{45}\\
\alpha_{6} & \alpha_{456} & \alpha_{56}\end{array}\right\} \left\{ \begin{array}{ccc}
\alpha_{4} & \alpha_{56} & \alpha_{456}\\
\alpha & \alpha_{123} & \alpha_{1234}\end{array}\right\} ,\label{eq:A20}\end{equation}

\begin{equation}
\wp_{20}=\alpha+\alpha_{123}-\alpha_{456}+\alpha_{5}+\alpha_{6}-\alpha_{56}.\label{eq:A20a}\end{equation}

\begin{equation}
\epsilon_{21}=(-1)^{\wp_{21}}\Delta\left(\alpha_{1},\alpha_{2},\alpha_{12}\right)\left[\alpha_{45},\alpha_{56},\alpha_{123},\alpha_{3456}\right]^{\frac{1}{2}}\left\{ \begin{array}{ccc}
\alpha_{4} & \alpha_{5} & \alpha_{45}\\
\alpha_{6} & \alpha_{456} & \alpha_{56}\end{array}\right\} \left\{ \begin{array}{ccc}
\alpha_{12} & \alpha_{3} & \alpha_{123}\\
\alpha_{456} & \alpha & \alpha_{3456}\end{array}\right\} ,\label{eq:A21}\end{equation}

\begin{equation}
\wp_{21}=-\alpha_{3}+\alpha_{4}+\alpha_{5}+\alpha_{6}-\alpha_{12}-\alpha.\label{eq:A21a}\end{equation}

\begin{equation}
\epsilon_{22}=(-1)^{\wp_{22}}\Delta\left(\alpha_{123},\alpha_{456},\alpha\right)\left[\alpha_{12},\alpha_{23},\alpha_{45},\alpha_{56}\right]^{\frac{1}{2}}\left\{ \begin{array}{ccc}
\alpha_{1} & \alpha_{2} & \alpha_{12}\\
\alpha_{3} & \alpha_{123} & \alpha_{23}\end{array}\right\} \left\{ \begin{array}{ccc}
\alpha_{4} & \alpha_{5} & \alpha_{45}\\
\alpha_{6} & \alpha_{456} & \alpha_{56}\end{array}\right\} ,\label{eq:A22}\end{equation}

\begin{equation}
\wp_{22}=\alpha_{1}+\alpha_{2}+\alpha_{3}+\alpha_{4}+\alpha_{5}+\alpha_{6}+\alpha_{123}+\alpha_{456}.\label{eq:A22a}\end{equation}

\[
\epsilon_{23}=(-1)^{\wp_{23}}\left[\alpha_{12},\alpha_{23},\alpha_{45},\alpha_{56},\alpha_{456},\alpha_{1234}\right]^{\frac{1}{2}}\]

\begin{equation}
\times\left\{ \begin{array}{ccc}
\alpha_{1} & \alpha_{2} & \alpha_{12}\\
\alpha_{3} & \alpha_{123} & \alpha_{23}\end{array}\right\} \left\{ \begin{array}{ccc}
\alpha_{4} & \alpha_{5} & \alpha_{45}\\
\alpha_{6} & \alpha_{456} & \alpha_{56}\end{array}\right\} \left\{ \begin{array}{ccc}
\alpha_{4} & \alpha_{56} & \alpha_{456}\\
\alpha & \alpha_{123} & \alpha_{1234}\end{array}\right\} ,\label{eq:A23}\end{equation}

\begin{equation}
\wp_{23}=\alpha_{1}+\alpha_{2}+\alpha_{3}+\alpha_{5}+\alpha_{6}-\alpha_{56}-\alpha+\alpha_{456}.\label{eq:A23a}\end{equation}

\[
\epsilon_{24}=(-1)^{\wp_{24}}\left[\alpha_{12},\alpha_{23},\alpha_{45},\alpha_{56},\alpha_{123},\alpha_{234},\alpha_{456},\alpha_{1234}\right]^{\frac{1}{2}}\]

\begin{equation}
\times\left\{ \begin{array}{ccc}
\alpha_{1} & \alpha_{2} & \alpha_{12}\\
\alpha_{3} & \alpha_{123} & \alpha_{23}\end{array}\right\} \left\{ \begin{array}{ccc}
\alpha_{1} & \alpha_{23} & \alpha_{123}\\
\alpha_{4} & \alpha_{1234} & \alpha_{234}\end{array}\right\} \left\{ \begin{array}{ccc}
\alpha_{4} & \alpha_{5} & \alpha_{45}\\
\alpha_{6} & \alpha_{456} & \alpha_{56}\end{array}\right\} \left\{ \begin{array}{ccc}
\alpha_{4} & \alpha_{56} & \alpha_{456}\\
\alpha & \alpha_{123} & \alpha_{1234}\end{array}\right\} ,\label{eq:A24}\end{equation}

\begin{equation}
\wp_{24}=\alpha_{2}+\alpha_{3}-\alpha_{4}+\alpha_{5}+\alpha_{6}-\alpha_{23}-\alpha_{56}+\alpha_{456}-\alpha-\alpha_{1234}.\label{eq:A24a}\end{equation}

\[
\epsilon_{25}=(-1)^{\wp_{25}}\left[\alpha_{12},\alpha_{23},\alpha_{45},\alpha_{56},\alpha_{123},\alpha_{234},\alpha_{456},\alpha_{23456}\right]^{\frac{1}{2}}\]

\begin{equation}
\times\left\{ \begin{array}{ccc}
\alpha_{1} & \alpha_{2} & \alpha_{12}\\
\alpha_{3} & \alpha_{123} & \alpha_{23}\end{array}\right\} \left\{ \begin{array}{ccc}
\alpha_{1} & \alpha_{23} & \alpha_{123}\\
\alpha_{456} & \alpha & \alpha_{23456}\end{array}\right\} \left\{ \begin{array}{ccc}
\alpha_{4} & \alpha_{5} & \alpha_{45}\\
\alpha_{6} & \alpha_{456} & \alpha_{56}\end{array}\right\} \left\{ \begin{array}{ccc}
\alpha_{4} & \alpha_{56} & \alpha_{456}\\
\alpha_{23456} & \alpha_{23} & \alpha_{234}\end{array}\right\} ,\label{eq:A25}\end{equation}

\begin{equation}
\wp_{25}=\alpha_{2}+\alpha_{3}-\alpha_{123}+\alpha_{5}+\alpha_{6}-\alpha_{56}+\alpha+\alpha_{23456}.\label{eq:A25a}\end{equation}

\[
\epsilon_{26}=(-1)^{\wp_{26}}\left[\alpha_{12},\alpha_{23},\alpha_{45},\alpha_{56},\alpha_{123},\alpha_{23456}\right]^{\frac{1}{2}}\]

\begin{equation}
\times\left\{ \begin{array}{ccc}
\alpha_{1} & \alpha_{2} & \alpha_{12}\\
\alpha_{3} & \alpha_{123} & \alpha_{23}\end{array}\right\} \left\{ \begin{array}{ccc}
\alpha_{1} & \alpha_{23} & \alpha_{123}\\
\alpha_{456} & \alpha & \alpha_{23456}\end{array}\right\} \left\{ \begin{array}{ccc}
\alpha_{4} & \alpha_{5} & \alpha_{45}\\
\alpha_{6} & \alpha_{456} & \alpha_{56}\end{array}\right\} ,\label{eq:A26}\end{equation}

\begin{equation}
\wp_{26}=\alpha_{2}+\alpha_{3}+\alpha_{4}+\alpha_{5}+\alpha_{6}-\alpha_{23}+\alpha_{123}-\alpha.\label{eq:A26a}\end{equation}

\[
\epsilon_{27}=(-1)^{\wp_{27}}\Delta\left(\alpha_{1},\alpha_{2},\alpha_{12}\right)\Delta\left(\alpha_{12},\alpha_{3},\alpha_{123}\right)\]

\begin{equation}
\times\left[\alpha_{45},\alpha_{456},\alpha_{1234},\alpha_{12345}\right]^{\frac{1}{2}}\left\{ \begin{array}{ccc}
\alpha_{4} & \alpha_{5} & \alpha_{45}\\
\alpha_{12345} & \alpha_{123} & \alpha_{1234}\end{array}\right\} \left\{ \begin{array}{ccc}
\alpha_{45} & \alpha_{6} & \alpha_{456}\\
\alpha & \alpha_{123} & \alpha_{12345}\end{array}\right\} ,\label{eq:A27}\end{equation}

\begin{equation}
\wp_{27}=-\alpha+\alpha_{12345}+\alpha_{4}+\alpha_{5}-\alpha_{6}-\alpha_{45}.\label{eq:A27a}\end{equation}

\[
\epsilon_{28}=(-1)^{\wp_{28}}\left[\alpha_{12},\alpha_{23},\alpha_{45},\alpha_{456},\alpha_{1234},\alpha_{12345}\right]^{\frac{1}{2}}\]

\begin{equation}
\times\left\{ \begin{array}{ccc}
\alpha_{1} & \alpha_{2} & \alpha_{12}\\
\alpha_{3} & \alpha_{123} & \alpha_{23}\end{array}\right\} \left\{ \begin{array}{ccc}
\alpha_{4} & \alpha_{5} & \alpha_{45}\\
\alpha_{12345} & \alpha_{123} & \alpha_{1234}\end{array}\right\} \left\{ \begin{array}{ccc}
\alpha_{45} & \alpha_{6} & \alpha_{456}\\
\alpha & \alpha_{123} & \alpha_{12345}\end{array}\right\} ,\label{eq:A28}\end{equation}

\begin{equation}
\wp_{28}=\alpha_{1}+\alpha_{2}+\alpha_{3}+\alpha_{4}+\alpha_{5}+\alpha_{6}+\alpha_{45}-\alpha_{123}+\alpha+\alpha_{12345}.\label{eq:A28a}\end{equation}

\[
\epsilon_{29}=(-1)^{\wp_{29}}\left[\alpha_{12},\alpha_{23},\alpha_{45},\alpha_{123},\alpha_{234},\alpha_{456},\alpha_{1234},\alpha_{12345}\right]^{\frac{1}{2}}\]

\begin{equation}
\times\left\{ \begin{array}{ccc}
\alpha_{1} & \alpha_{2} & \alpha_{12}\\
\alpha_{3} & \alpha_{123} & \alpha_{23}\end{array}\right\} \left\{ \begin{array}{ccc}
\alpha_{1} & \alpha_{23} & \alpha_{123}\\
\alpha_{4} & \alpha_{1234} & \alpha_{234}\end{array}\right\} \left\{ \begin{array}{ccc}
\alpha_{4} & \alpha_{5} & \alpha_{45}\\
\alpha_{12345} & \alpha_{123} & \alpha_{1234}\end{array}\right\} \left\{ \begin{array}{ccc}
\alpha_{45} & \alpha_{6} & \alpha_{456}\\
\alpha & \alpha_{123} & \alpha_{12345}\end{array}\right\} ,\label{eq:A29}\end{equation}

\begin{equation}
\wp_{29}=\alpha_{2}+\alpha_{3}-\alpha_{23}+\alpha_{123}+\alpha_{1234}+\alpha_{12345}+\alpha-\alpha_{5}+\alpha_{6}-\alpha_{45}.\label{eq:A29a}\end{equation}

\[
\epsilon_{30}=(-1)^{\wp_{30}}\left[\alpha_{12},\alpha_{23},\alpha_{45},\alpha_{123},\alpha_{234},\alpha_{456},\alpha_{2345},\alpha_{23456}\right]^{\frac{1}{2}}\]

\begin{equation}
\times\left\{ \begin{array}{ccc}
\alpha_{1} & \alpha_{2} & \alpha_{12}\\
\alpha_{3} & \alpha_{123} & \alpha_{23}\end{array}\right\} \left\{ \begin{array}{ccc}
\alpha_{1} & \alpha_{23} & \alpha_{123}\\
\alpha_{456} & \alpha & \alpha_{23456}\end{array}\right\} \left\{ \begin{array}{ccc}
\alpha_{4} & \alpha_{5} & \alpha_{45}\\
\alpha_{2345} & \alpha_{23} & \alpha_{234}\end{array}\right\} \left\{ \begin{array}{ccc}
\alpha_{45} & \alpha_{6} & \alpha_{456}\\
\alpha_{23456} & \alpha_{23} & \alpha_{2345}\end{array}\right\} ,\label{eq:A30}\end{equation}

\begin{equation}
\wp_{30}=\alpha_{2}+\alpha_{3}+\alpha_{4}+\alpha_{5}+\alpha_{6}-\alpha_{23}-\alpha_{45}-\alpha_{123}-\alpha_{2345}+\alpha+\alpha_{23456}+\alpha_{456}.\label{eq:A30a}\end{equation}

\[
\epsilon_{31}=(-1)^{\wp_{31}}\left[\alpha_{12},\alpha_{23},\alpha_{45},\alpha_{123},\alpha_{234},\alpha_{456},\alpha_{2345},\alpha_{12345}\right]^{\frac{1}{2}}\]

\begin{equation}
\times\left\{ \begin{array}{ccc}
\alpha_{1} & \alpha_{2} & \alpha_{12}\\
\alpha_{3} & \alpha_{123} & \alpha_{23}\end{array}\right\} \left\{ \begin{array}{ccc}
\alpha_{1} & \alpha_{23} & \alpha_{123}\\
\alpha_{45} & \alpha_{12345} & \alpha_{2345}\end{array}\right\} \left\{ \begin{array}{ccc}
\alpha_{4} & \alpha_{5} & \alpha_{45}\\
\alpha_{2345} & \alpha_{23} & \alpha_{234}\end{array}\right\} \left\{ \begin{array}{ccc}
\alpha_{45} & \alpha_{6} & \alpha_{456}\\
\alpha & \alpha_{123} & \alpha_{12345}\end{array}\right\} ,\label{eq:A31}\end{equation}

\begin{equation}
\wp_{31}=\alpha_{2}+\alpha_{3}-\alpha_{4}-\alpha_{5}+\alpha_{6}+\alpha_{2345}+\alpha+\alpha_{12345}.\label{eq:A31a}\end{equation}

\[
\epsilon_{32}=(-1)^{\wp_{32}}\left[\alpha_{12},\alpha_{34},\alpha_{45},\alpha_{123},\alpha_{234},\alpha_{456},\alpha_{1234},\alpha_{12345}\right]^{\frac{1}{2}}\]

\begin{equation}
\times\left\{ \begin{array}{ccc}
\alpha_{1} & \alpha_{2} & \alpha_{12}\\
\alpha_{34} & \alpha_{1234} & \alpha_{234}\end{array}\right\} \left\{ \begin{array}{ccc}
\alpha_{12} & \alpha_{3} & \alpha_{123}\\
\alpha_{4} & \alpha_{1234} & \alpha_{34}\end{array}\right\} \left\{ \begin{array}{ccc}
\alpha_{4} & \alpha_{5} & \alpha_{45}\\
\alpha_{12345} & \alpha_{123} & \alpha_{1234}\end{array}\right\} \left\{ \begin{array}{ccc}
\alpha_{45} & \alpha_{6} & \alpha_{456}\\
\alpha & \alpha_{123} & \alpha_{12345}\end{array}\right\} ,\label{eq:A32}\end{equation}

\begin{equation}
\wp_{32}=\alpha_{1}+\alpha_{2}+\alpha_{34}+\alpha_{6}+\alpha_{45}+\alpha+\alpha_{5}+\alpha_{12345}+\alpha_{3}+\alpha_{12}.\label{eq:A32a}\end{equation}

\[
\epsilon_{33}=(-1)^{\wp_{33}}\left[\alpha_{12},\alpha_{34},\alpha_{45},\alpha_{123},\alpha_{234},\alpha_{456},\alpha_{2345},\alpha_{23456}\right]^{\frac{1}{2}}\]

\begin{equation}
\times\left[\begin{array}{cccccccccc}
\alpha_{1} &  & \alpha_{2} &  & \alpha_{234} &  & \alpha_{2345} &  & \alpha_{23456}\\
 & \alpha_{12} &  & \alpha_{34} &  & \alpha_{5} &  & \alpha_{6} &  & \alpha\\
\alpha_{123} &  & \alpha_{3} &  & \alpha_{4} &  & \alpha_{45} &  & \alpha_{456}\end{array}\right],\label{eq:A33}\end{equation}

\begin{equation}
\wp_{33}=\alpha_{1}+\alpha_{2}-\alpha_{12}+\alpha_{34}-\alpha_{3}-\alpha_{4}.\label{eq:A33a}\end{equation}

\[
\epsilon_{34}=(-1)^{\wp_{34}}\left[\alpha_{12},\alpha_{34},\alpha_{45},\alpha_{123},\alpha_{234},\alpha_{456},\alpha_{2345},\alpha_{12345}\right]^{\frac{1}{2}}\]

\begin{equation}
\times\left\{ \begin{array}{ccc}
\alpha_{45} & \alpha_{6} & \alpha_{456}\\
\alpha & \alpha_{123} & \alpha_{12345}\end{array}\right\} \left[\begin{array}{cccc}
\alpha_{2} & \alpha_{2345} & \alpha_{3} & \alpha_{45}\\
\alpha_{12} & \alpha_{34} & \alpha_{12345} & \alpha_{5}\\
\alpha_{1} & \alpha_{123} & \alpha_{234} & \alpha_{4}\end{array}\right],\label{eq:A34}\end{equation}

\begin{equation}
\wp_{34}=-\alpha_{234}+\alpha_{2345}+2\alpha_{4}+\alpha+\alpha_{6}+\alpha_{12}-\alpha_{34}.\label{eq:A34a}\end{equation}

\[
\epsilon_{35}=(-1)^{\wp_{35}}\left[\alpha_{12},\alpha_{34},\alpha_{45},\alpha_{123},\alpha_{345},\alpha_{456},\alpha_{3456},\alpha_{23456}\right]^{\frac{1}{2}}\]

\begin{equation}
\times\left\{ \begin{array}{ccc}
\alpha_{1} & \alpha_{2} & \alpha_{12}\\
\alpha_{3456} & \alpha & \alpha_{23456}\end{array}\right\} \left\{ \begin{array}{ccc}
\alpha_{12} & \alpha_{3} & \alpha_{123}\\
\alpha_{456} & \alpha & \alpha_{3456}\end{array}\right\} \left\{ \begin{array}{ccc}
\alpha_{4} & \alpha_{5} & \alpha_{45}\\
\alpha_{345} & \alpha_{3} & \alpha_{34}\end{array}\right\} \left\{ \begin{array}{ccc}
\alpha_{45} & \alpha_{6} & \alpha_{456}\\
\alpha_{3456} & \alpha_{3} & \alpha_{345}\end{array}\right\} ,\label{eq:A35}\end{equation}

\begin{equation}
\wp_{35}=\alpha_{345}-\alpha_{3}+\alpha_{4}+\alpha_{5}+\alpha_{45}+\alpha_{6}+\alpha_{456}-\alpha_{1}-\alpha_{2}+\alpha_{12}.\label{eq:A35a}\end{equation}

\[
\epsilon_{36}=(-1)^{\wp_{36}}\left[\alpha_{12},\alpha_{34},\alpha_{45},\alpha_{123},\alpha_{345},\alpha_{456},\alpha_{2345},\alpha_{12345}\right]^{\frac{1}{2}}\]

\begin{equation}
\times\left\{ \begin{array}{ccc}
\alpha_{1} & \alpha_{2} & \alpha_{12}\\
\alpha_{345} & \alpha_{12345} & \alpha_{2345}\end{array}\right\} \left\{ \begin{array}{ccc}
\alpha_{12} & \alpha_{3} & \alpha_{123}\\
\alpha_{45} & \alpha_{12345} & \alpha_{345}\end{array}\right\} \left\{ \begin{array}{ccc}
\alpha_{4} & \alpha_{5} & \alpha_{45}\\
\alpha_{345} & \alpha_{3} & \alpha_{34}\end{array}\right\} \left\{ \begin{array}{ccc}
\alpha_{45} & \alpha_{6} & \alpha_{456}\\
\alpha & \alpha_{123} & \alpha_{12345}\end{array}\right\} ,\label{eq:A36}\end{equation}

\begin{equation}
\wp_{36}=\alpha_{1}+\alpha_{2}-\alpha_{4}-\alpha_{5}+\alpha+\alpha_{12}-\alpha_{123}+\alpha_{6}.\label{eq:A36a}\end{equation}

\[
\epsilon_{37}=(-1)^{\wp_{37}}\left[\alpha_{12},\alpha_{34},\alpha_{45},\alpha_{123},\alpha_{345},\alpha_{456},\alpha_{2345},\alpha_{23456}\right]^{\frac{1}{2}}\]

\begin{equation}
\times\left\{ \begin{array}{ccc}
\alpha_{4} & \alpha_{5} & \alpha_{45}\\
\alpha_{345} & \alpha_{3} & \alpha_{34}\end{array}\right\} \left[\begin{array}{cccc}
\alpha_{2} & \alpha_{23456} & \alpha_{3} & \alpha_{456}\\
\alpha_{12} & \alpha_{345} & \alpha & \alpha_{6}\\
\alpha_{1} & \alpha_{123} & \alpha_{2345} & \alpha_{45}\end{array}\right],\label{eq:A37}\end{equation}

\begin{equation}
\wp_{37}=\alpha_{4}+\alpha_{5}+\alpha_{2345}-\alpha_{123}+\alpha_{3}-\alpha_{23456}+\alpha_{456}+\alpha_{12}.\label{eq:A37a}\end{equation}

\[
\epsilon_{38}=(-1)^{\wp_{38}}\Delta\left(\alpha_{4},\alpha_{5},\alpha_{45}\right)\left[\alpha_{12},\alpha_{123},\alpha_{345},\alpha_{456},\alpha_{3456},\alpha_{23456}\right]^{\frac{1}{2}}\]

\begin{equation}
\times\left\{ \begin{array}{ccc}
\alpha_{1} & \alpha_{2} & \alpha_{12}\\
\alpha_{3456} & \alpha & \alpha_{23456}\end{array}\right\} \left\{ \begin{array}{ccc}
\alpha_{12} & \alpha_{3} & \alpha_{123}\\
\alpha_{456} & \alpha & \alpha_{3456}\end{array}\right\} \left\{ \begin{array}{ccc}
\alpha_{45} & \alpha_{6} & \alpha_{456}\\
\alpha_{3456} & \alpha_{3} & \alpha_{345}\end{array}\right\} ,\label{eq:A38}\end{equation}

\begin{equation}
\wp_{38}=\alpha_{6}+\alpha_{45}-\alpha_{456}+2\alpha-\alpha_{1}-\alpha_{2}-\alpha_{12}.\label{eq:A38a}\end{equation}

\[
\epsilon_{39}=(-1)^{\wp_{39}}\Delta\left(\alpha_{4},\alpha_{5},\alpha_{45}\right)\left[\alpha_{12},\alpha_{123},\alpha_{345},\alpha_{456},\alpha_{2345},\alpha_{12345}\right]^{\frac{1}{2}}\]

\begin{equation}
\times\left\{ \begin{array}{ccc}
\alpha_{1} & \alpha_{2} & \alpha_{12}\\
\alpha_{345} & \alpha_{12345} & \alpha_{2345}\end{array}\right\} \left\{ \begin{array}{ccc}
\alpha_{12} & \alpha_{3} & \alpha_{123}\\
\alpha_{45} & \alpha_{12345} & \alpha_{345}\end{array}\right\} \left\{ \begin{array}{ccc}
\alpha_{45} & \alpha_{6} & \alpha_{456}\\
\alpha & \alpha_{123} & \alpha_{12345}\end{array}\right\} ,\label{eq:A39}\end{equation}

\begin{equation}
\wp_{39}=\alpha_{1}+\alpha_{2}+\alpha_{345}+\alpha-\alpha_{12}+\alpha_{123}-\alpha_{3}+\alpha_{6}.\label{eq:A39a}\end{equation}

\begin{equation}
\epsilon_{40}=(-1)^{\wp_{40}}\Delta\left(\alpha_{4},\alpha_{5},\alpha_{45}\right)\left[\alpha_{12},\alpha_{123},\alpha_{345},\alpha_{456},\alpha_{2345},\alpha_{23456}\right]^{\frac{1}{2}}\left[\begin{array}{cccc}
\alpha_{1} & \alpha_{2345} & \alpha_{123} & \alpha_{45}\\
\alpha_{12} & \alpha & \alpha_{345} & \alpha_{6}\\
\alpha_{2} & \alpha_{3} & \alpha_{23456} & \alpha_{456}\end{array}\right],\label{eq:A40}\end{equation}

\begin{equation}
\wp_{40}=\alpha_{123}+\alpha_{2345}+2\alpha_{2}+\alpha_{23456}-\alpha_{456}-\alpha_{12}-\alpha_{345}.\label{eq:A40a}\end{equation}

\[
\epsilon_{41}=(-1)^{\wp_{41}}\Delta\left(\alpha_{4},\alpha_{5},\alpha_{45}\right)\Delta\left(\alpha_{45},\alpha_{6},\alpha_{456}\right)\]

\begin{equation}
\times\left[\alpha_{12},\alpha_{123},\alpha_{3456},\alpha_{23456}\right]^{\frac{1}{2}}\left\{ \begin{array}{ccc}
\alpha_{1} & \alpha_{2} & \alpha_{12}\\
\alpha_{3456} & \alpha & \alpha_{23456}\end{array}\right\} \left\{ \begin{array}{ccc}
\alpha_{12} & \alpha_{3} & \alpha_{123}\\
\alpha_{456} & \alpha & \alpha_{3456}\end{array}\right\} ,\label{eq:A41}\end{equation}

\begin{equation}
\wp_{41}=\alpha_{12}-\alpha_{1}-\alpha_{2}+\alpha_{3}+\alpha_{456}-\alpha_{3456}.\label{eq:A41a}\end{equation}

\[
\epsilon_{42}=(-1)^{\wp_{42}}\left[\alpha_{12},\alpha_{45},\alpha_{56},\alpha_{123},\alpha_{3456},\alpha_{23456}\right]^{\frac{1}{2}}\]

\begin{equation}
\times\left\{ \begin{array}{ccc}
\alpha_{1} & \alpha_{2} & \alpha_{12}\\
\alpha_{3456} & \alpha & \alpha_{23456}\end{array}\right\} \left\{ \begin{array}{ccc}
\alpha_{12} & \alpha_{3} & \alpha_{123}\\
\alpha_{456} & \alpha & \alpha_{3456}\end{array}\right\} \left\{ \begin{array}{ccc}
\alpha_{4} & \alpha_{5} & \alpha_{45}\\
\alpha_{6} & \alpha_{456} & \alpha_{56}\end{array}\right\} ,\label{eq:A42}\end{equation}

\begin{equation}
\wp_{42}=-\alpha_{4}-\alpha_{5}-\alpha_{6}+\alpha_{12}-\alpha_{1}-\alpha_{2}+\alpha_{3}-\alpha_{3456}.\label{eq:A42a}\end{equation}

}

\noindent{}The quantities $\left\{ \ldots\right\} $ denote $6j$-coefficients
of $\mathrm{SU}\left(2\right)$. The definition of $12j$-coefficient
of the second kind (see Eqs. (\ref{eq:A12}), (\ref{eq:A34}), (\ref{eq:A37}),
(\ref{eq:A40})) one can find in \cite{Jucys,Vanagas}. The definition
of $15j$-coefficient of the second kind (see Eq. (\ref{eq:A33}))
is presented in \cite{Jucys,Alisauskas,Alisauskas2}.

\newpage{}\section{Classes}\label{C}

In this Appendix the data, required for the study of quantities in
Sec. \ref{shells}, are presented. In tables operators $\langle x_{\pi}\rangle$,
pertaining to the specific class $X_{n}(\Delta_{1},\Delta_{2},\ldots,\Delta_{n})$
with $n\in\{2,3,\ldots,6\}$, are listed. The explicit expressions
for $\widehat{\pi}$ (see Eq. (\ref{eq:5.3.2})) and $\widehat{\tilde{\pi}}$
(see Eqs. (\ref{eq:5.3.11})-(\ref{eq:5.3.12})) are given.

{\footnotesize\begin{multicols}{2}

\begin{center}
\begin{tabular}{|c|c|c|}
\hline 
\multicolumn{1}{|c}{} & \multicolumn{1}{c}{$X_{2}\left(0,0\right)$} & \multicolumn{1}{c|}{}\tabularnewline
\hline 
$\left\langle x_{\pi}\right\rangle $ & $\widehat{\pi}$ & $\left\langle x\right\rangle $\tabularnewline
\hline
\hline 
$\left\langle 112112\right\rangle $ & $\left(35\right)$ & \tabularnewline
\cline{1-2} 
$\left\langle 121112\right\rangle $ & $\left(25\right)$ & \multicolumn{1}{c|}{}\tabularnewline
\cline{1-2} 
$\left\langle 211112\right\rangle $ & $\left(15\right)$ & \multicolumn{1}{c|}{$\left\langle 111122\right\rangle $}\tabularnewline
\cline{1-2} 
$\left\langle 121121\right\rangle $ & $-\left(26\right)$ & \multicolumn{1}{c|}{}\tabularnewline
\cline{1-2} 
$\left\langle 112121\right\rangle $ & $\left(36\right)$ & \multicolumn{1}{c|}{}\tabularnewline
\cline{1-2} 
$\left\langle 112211\right\rangle $ & $\left(35\right)\left(46\right)$ & \multicolumn{1}{c|}{}\tabularnewline
\hline
$\left\langle 122122\right\rangle $ & $-\left(24\right)$ & \tabularnewline
\cline{1-2} 
$\left\langle 122212\right\rangle $ & $\left(25\right)$ & \tabularnewline
\cline{1-2} 
$\left\langle 122221\right\rangle $ & $-\left(26\right)$ & $\left\langle 112222\right\rangle $\tabularnewline
\cline{1-2} 
$\left\langle 212122\right\rangle $ & $\left(14\right)$ & \tabularnewline
\cline{1-2} 
$\left\langle 212212\right\rangle $ & $-\left(15\right)$ & \tabularnewline
\cline{1-2} 
$\left\langle 221122\right\rangle $ & $\left(13\right)\left(24\right)$ & \tabularnewline
\hline
\end{tabular}%
\begin{table}[H]
\caption{\label{TabX1}Operators of class $X_{2}\left(0,0\right)$.}

\end{table}

\par\end{center}

\begin{center}
\begin{tabular}{|c|c|c|}
\hline 
\multicolumn{1}{|c}{} & \multicolumn{1}{c}{$X_{2}\left(+1,-1\right)$} & \multicolumn{1}{c|}{}\tabularnewline
\hline 
$\left\langle x_{\pi}\right\rangle $ & $\widehat{\pi}$ & $\left\langle x\right\rangle $\tabularnewline
\hline
\hline 
$\left\langle 111112\right\rangle $ & $1_{6}$ & \tabularnewline
\cline{1-2} 
$\left\langle 111121\right\rangle $ & $\left(56\right)$ & \multicolumn{1}{c|}{$\left\langle 111112\right\rangle $}\tabularnewline
\cline{1-2} 
$\left\langle 111211\right\rangle $ & $-\left(46\right)$ & \multicolumn{1}{c|}{}\tabularnewline
\hline
$\left\langle 211122\right\rangle $ & $\left(14\right)$ & \multicolumn{1}{c|}{}\tabularnewline
\cline{1-2} 
$\left\langle 121122\right\rangle $ & $-\left(24\right)$ & \multicolumn{1}{c|}{}\tabularnewline
\cline{1-2} 
$\left\langle 112122\right\rangle $ & $\left(34\right)$ & \multicolumn{1}{c|}{}\tabularnewline
\cline{1-2} 
$\left\langle 211212\right\rangle $ & $-\left(15\right)$ & \tabularnewline
\cline{1-2} 
$\left\langle 121212\right\rangle $ & $\left(25\right)$ & $\left\langle 111222\right\rangle $\tabularnewline
\cline{1-2} 
$\left\langle 112212\right\rangle $ & $-\left(35\right)$ & \tabularnewline
\cline{1-2} 
$\left\langle 211221\right\rangle $ & $\left(16\right)$ & \tabularnewline
\cline{1-2} 
$\left\langle 121221\right\rangle $ & $-\left(26\right)$ & \tabularnewline
\cline{1-2} 
$\left\langle 112221\right\rangle $ & $\left(36\right)$ & \tabularnewline
\hline
$\left\langle 122222\right\rangle $ & $1_{6}$ & \tabularnewline
\cline{1-2} 
$\left\langle 212222\right\rangle $ & $\left(12\right)$ & $\left\langle 122222\right\rangle $\tabularnewline
\cline{1-2} 
$\left\langle 221222\right\rangle $ & $-\left(13\right)$ & \tabularnewline
\hline
\end{tabular}%
\begin{table}[H]
\caption{\label{TabX2}Operators of class $X_{2}\left(+1,-1\right)$.}

\end{table}

\par\end{center}

\end{multicols}

\begin{center}
\begin{tabular}{|c|c|c|}
\hline 
\multicolumn{1}{|c}{} & \multicolumn{1}{c}{$X_{2}\left(+2,-2\right)$} & \multicolumn{1}{c|}{}\tabularnewline
\hline 
$\left\langle x_{\pi}\right\rangle $ & $\widehat{\pi}$ & $\left\langle x\right\rangle $\tabularnewline
\hline
\hline 
$\left\langle 111122\right\rangle $ & $1_{6}$ & \tabularnewline
\cline{1-2} 
$\left\langle 111212\right\rangle $ & $\left(45\right)$ & \multicolumn{1}{c|}{$\left\langle 111122\right\rangle $}\tabularnewline
\cline{1-2} 
$\left\langle 111221\right\rangle $ & $-\left(46\right)$ & \multicolumn{1}{c|}{}\tabularnewline
\hline
$\left\langle 112222\right\rangle $ & $1_{6}$ & \multicolumn{1}{c|}{}\tabularnewline
\cline{1-2} 
$\left\langle 121222\right\rangle $ & $\left(23\right)$ & \multicolumn{1}{c|}{$\left\langle 112222\right\rangle $}\tabularnewline
\cline{1-2} 
$\left\langle 211222\right\rangle $ & $-\left(13\right)$ & \multicolumn{1}{c|}{}\tabularnewline
\hline
\end{tabular}%
\begin{table}[H]
\caption{\label{TabX3}Operators of class $X_{2}\left(+2,-2\right)$.}

\end{table}

\par\end{center}

}The class $X_{2}\left(+3,-3\right)$ is of order $d=1$. The only
operator which belongs to $X_{2}\left(+3,-3\right)$, is $\left\langle 111222\right\rangle $.

{\footnotesize

\begin{center}
\begin{tabular}{|c|c|c||c|c|c|}
\hline 
\multicolumn{1}{|c}{} & \multicolumn{1}{c}{$X_{3}\left(0,0,0\right)$} & \multicolumn{1}{c||}{} & \multicolumn{1}{c}{} & \multicolumn{1}{c}{$X_{3}\left(0,0,0\right)$} & \multicolumn{1}{c|}{}\tabularnewline
\hline 
$\left\langle x_{\pi}\right\rangle $ & $\widehat{\pi}$ & $\left\langle x\right\rangle $ & $\left\langle x_{\pi}\right\rangle $ & $\widehat{\pi}$ & $\left\langle x\right\rangle $\tabularnewline
\hline
\hline 
$\left\langle 123\left\{ 123\right\} \right\rangle $ & $-\left(24\right)\left(35\right)\widehat{\vartheta}$ & \multicolumn{1}{c||}{} & $\left\langle 213213\right\rangle $ & $-\left(135\right)$ & \tabularnewline
\cline{1-2} \cline{4-5} 
$\left\langle 132123\right\rangle $ & $\left(254\right)$ & \multicolumn{1}{c||}{} & $\left\langle 213312\right\rangle $ & $-\left(135\right)\left(46\right)$ & \multicolumn{1}{c|}{}\tabularnewline
\cline{1-2} \cline{4-5} 
$\left\langle 132132\right\rangle $ & $-\left(264\right)$ & \multicolumn{1}{c||}{} & $\left\langle 231123\right\rangle $ & $\left(13\right)\left(254\right)$ & \multicolumn{1}{c|}{}\tabularnewline
\cline{1-2} \cline{4-5} 
$\left\langle 132213\right\rangle $ & $\left(25\right)$ & \multicolumn{1}{c||}{$\left\langle 112233\right\rangle $} & $\left\langle 231132\right\rangle $ & $-\left(13\right)\left(264\right)$ & \multicolumn{1}{c|}{$\left\langle 112233\right\rangle $}\tabularnewline
\cline{1-2} \cline{4-5} 
$\left\langle 132231\right\rangle $ & $-\left(26\right)$ & \multicolumn{1}{c||}{} & $\left\langle 231213\right\rangle $ & $\left(13\right)\left(25\right)$ & \multicolumn{1}{c|}{}\tabularnewline
\cline{1-2} \cline{4-5} 
$\left\langle 132312\right\rangle $ & $\left(25\right)\left(46\right)$ & \multicolumn{1}{c||}{} & $\left\langle 312123\right\rangle $ & $-\left(154\right)$ & \multicolumn{1}{c|}{}\tabularnewline
\cline{1-2} \cline{4-5} 
$\left\langle 213123\right\rangle $ & $\left(14\right)\left(35\right)$ & \multicolumn{1}{c||}{} & $\left\langle 312213\right\rangle $ & $-\left(15\right)$ & \multicolumn{1}{c|}{}\tabularnewline
\cline{1-2} \cline{4-5} 
$\left\langle 213132\right\rangle $ & $-\left(14\right)\left(36\right)$ & \multicolumn{1}{c||}{} & $\left\langle 321123\right\rangle $ & $-\left(154\right)\left(23\right)$ & \multicolumn{1}{c|}{}\tabularnewline
\hline
\end{tabular}%
\begin{table}[H]
\caption{\label{TabX4}Operators of class $X_{3}\left(0,0,0\right)$.}

\end{table}

\par\end{center}

\begin{center}
\begin{tabular}{|c|c|c||c|c|c|}
\hline 
\multicolumn{1}{|c}{} & \multicolumn{1}{c}{$X_{3}\left(+2,-1,-1\right)$} &  & \multicolumn{1}{c}{} & \multicolumn{1}{c}{$X_{3}\left(+2,-1,-1\right)$} & \tabularnewline
\hline 
$\left\langle x_{\pi}\right\rangle $ & $\widehat{\pi}$ & $\left\langle x\right\rangle $ & $\left\langle x_{\pi}\right\rangle $ & $\widehat{\pi}$ & $\left\langle x\right\rangle $\tabularnewline
\hline
\hline 
$\left\langle 111\left\{ 123\right\} \right\rangle $ & $\widehat{\vartheta}$ & \multicolumn{1}{c||}{$\left\langle 111123\right\rangle $} & $\left\langle 113233\right\rangle $ & $\left(34\right)$ & \tabularnewline
\cline{1-5} 
$\left\langle 112223\right\rangle $ & $1_{6}$ &  & $\left\langle 113323\right\rangle $ & $-\left(35\right)$ & \tabularnewline
\cline{1-2} \cline{4-5} 
$\left\langle 112232\right\rangle $ & $\left(56\right)$ & \multicolumn{1}{c||}{} & $\left\langle 113332\right\rangle $ & $\left(36\right)$ & \multicolumn{1}{c|}{}\tabularnewline
\cline{1-2} \cline{4-5} 
$\left\langle 112322\right\rangle $ & $-\left(46\right)$ & \multicolumn{1}{c||}{} & $\left\langle 131233\right\rangle $ & $\left(243\right)$ & \multicolumn{1}{c|}{}\tabularnewline
\cline{1-2} \cline{4-5} 
$\left\langle 121223\right\rangle $ & $\left(23\right)$ & \multicolumn{1}{c||}{$\left\langle 112223\right\rangle $} & $\left\langle 131323\right\rangle $ & $-\left(253\right)$ & \multicolumn{1}{c|}{$\left\langle 112333\right\rangle $}\tabularnewline
\cline{1-2} \cline{4-5} 
$\left\langle 121232\right\rangle $ & $\left(23\right)\left(56\right)$ & \multicolumn{1}{c||}{} & $\left\langle 131332\right\rangle $ & $\left(263\right)$ & \multicolumn{1}{c|}{}\tabularnewline
\cline{1-2} \cline{4-5} 
$\left\langle 121322\right\rangle $ & $-\left(23\right)\left(46\right)$ & \multicolumn{1}{c||}{} & $\left\langle 311233\right\rangle $ & $-\left(143\right)$ & \multicolumn{1}{c|}{}\tabularnewline
\cline{1-2} \cline{4-5} 
$\left\langle 211223\right\rangle $ & $-\left(13\right)$ & \multicolumn{1}{c||}{} & $\left\langle 311323\right\rangle $ & $\left(153\right)$ & \multicolumn{1}{c|}{}\tabularnewline
\cline{1-2} \cline{4-5} 
\multicolumn{1}{|c|}{$\left\langle 211232\right\rangle $} & $-\left(13\right)\left(56\right)$ & \multicolumn{1}{c||}{} & $\left\langle 311332\right\rangle $ & $-\left(163\right)$ & \multicolumn{1}{c|}{}\tabularnewline
\cline{1-2} \cline{4-5} 
$\left\langle 211322\right\rangle $ & $\left(13\right)\left(46\right)$ & \multicolumn{1}{c||}{} &  &  & \multicolumn{1}{c|}{}\tabularnewline
\hline
\hline 
\multicolumn{1}{|c}{} & \multicolumn{1}{c}{$X_{3}\left(\Delta_{1}^{\prime},\Delta_{2}^{\prime},\Delta_{3}^{\prime}\right)$} &  & \multicolumn{1}{c}{} & \multicolumn{1}{c}{$X_{3}\left(\Delta_{1}^{\prime},\Delta_{2}^{\prime},\Delta_{3}^{\prime}\right)$} & \tabularnewline
\hline
\hline 
\multicolumn{1}{|c}{} & \multicolumn{1}{c}{$X_{3}\left(-1,+2,-1\right)$} &  & \multicolumn{1}{c}{} & \multicolumn{1}{c}{$X_{3}\left(-1,-1,+2\right)$} & \tabularnewline
\hline 
$\left\langle x\right\rangle $ & $\widehat{\tilde{\pi}}$ & $\left\langle y\right\rangle $ & $\left\langle x\right\rangle $ & $\widehat{\tilde{\pi}}$ & $\left\langle y\right\rangle $\tabularnewline
\hline
\hline 
$\left\langle 111123\right\rangle $ & $-\left(15\right)$ & $\left\langle 122223\right\rangle $ & $\left\langle 111123\right\rangle $ & $-\left(16\right)\left(25\right)$ & $\left\langle 123333\right\rangle $\tabularnewline
\hline 
$\left\langle 112223\right\rangle $ & $\left(14\right)\left(25\right)$ & $\left\langle 111223\right\rangle $ & $\left\langle 112223\right\rangle $ & $-\left(16\right)\left(25\right)$ & $\left\langle 122233\right\rangle $\tabularnewline
\hline 
$\left\langle 112333\right\rangle $ & $-\left(13\right)$ & $\left\langle 122333\right\rangle $ & $\left\langle 112333\right\rangle $ & $\left(16\right)\left(25\right)$ & $\left\langle 111233\right\rangle $\tabularnewline
\hline
\end{tabular}%
\begin{table}[H]
\caption{\label{TabX5}Operators of classes $X_{3}\left(\Delta_{1},\Delta_{2},\Delta_{3}\right)$
with $\Delta_{x}=-1,+1,+2$.}

\end{table}

\par\end{center}

\begin{center}
\begin{tabular}{|c|c|c|}
\hline 
\multicolumn{1}{|c}{} & \multicolumn{1}{c}{$X_{3}\left(+3,-2,-1\right)$} & \multicolumn{1}{c|}{}\tabularnewline
\hline 
$\left\langle x_{\pi}\right\rangle $ & $\widehat{\pi}$ & $\left\langle x\right\rangle $\tabularnewline
\hline
\hline 
$\left\langle 111223\right\rangle $ & $1_{6}$ & \tabularnewline
\cline{1-2} 
$\left\langle 111232\right\rangle $ & $\left(56\right)$ & \multicolumn{1}{c|}{$\left\langle 111223\right\rangle $}\tabularnewline
\cline{1-2} 
$\left\langle 111322\right\rangle $ & $-\left(46\right)$ & \multicolumn{1}{c|}{}\tabularnewline
\hline
\hline 
$X_{3}\left(\Delta_{1}^{\prime},\Delta_{2}^{\prime},\Delta_{3}^{\prime}\right)$ & \multicolumn{1}{c|}{$\widehat{\tilde{\pi}}$} & \multicolumn{1}{c|}{$\left\langle y\right\rangle $}\tabularnewline
\hline
\hline 
$X_{3}\left(-2,+3,-1\right)$ & $\left(15\right)\left(24\right)$ & $\left\langle 112223\right\rangle $\tabularnewline
\hline 
$X_{3}\left(-1,-2,+3\right)$ & $\left(16\right)\left(24\right)\left(35\right)$ & $\left\langle 122333\right\rangle $\tabularnewline
\hline 
$X_{3}\left(+3,-1,-2\right)$ & $-\left(46\right)$ & $\left\langle 111233\right\rangle $\tabularnewline
\hline 
$X_{3}\left(-1,+3,-2\right)$ & $\left(146\right)$ & $\left\langle 122233\right\rangle $\tabularnewline
\hline 
$X_{3}\left(-2,-1,+3\right)$ & $\left(14\right)\left(25\right)\left(36\right)$ & $\left\langle 112333\right\rangle $\tabularnewline
\hline
\end{tabular}%
\begin{table}[H]
\caption{\label{TabX8}Operators of classes $X_{3}\left(\Delta_{1},\Delta_{2},\Delta_{3}\right)$
with $\Delta_{x}=-2,-1,+3$.}

\end{table}

\par\end{center}

\begin{center}
\begin{tabular}{|c|c|c||c|c|c|}
\hline 
\multicolumn{1}{|c}{} & \multicolumn{1}{c}{$X_{3}\left(+1,-1,0\right)$} &  & \multicolumn{1}{c}{} & \multicolumn{1}{c}{$X_{3}\left(+1,-1,0\right)$} & \tabularnewline
\hline 
$\left\langle x_{\pi}\right\rangle $ & $\widehat{\pi}$ & $\left\langle x\right\rangle $ & $\left\langle x_{\pi}\right\rangle $ & $\widehat{\pi}$ & $\left\langle x\right\rangle $\tabularnewline
\hline
\hline 
$\left\langle 133233\right\rangle $ & $-\left(24\right)$ &  & $\left\langle 313332\right\rangle $ & $-\left(162\right)$ & \tabularnewline
\cline{1-2} \cline{4-5} 
$\left\langle 133323\right\rangle $ & $\left(25\right)$ & \multicolumn{1}{c||}{} & $\left\langle 331233\right\rangle $ & $\left(13\right)\left(24\right)$ & \multicolumn{1}{c|}{}\tabularnewline
\cline{1-2} \cline{4-5} 
$\left\langle 133332\right\rangle $ & $-\left(26\right)$ & \multicolumn{1}{c||}{$\left\langle 123333\right\rangle $} & $\left\langle 331323\right\rangle $ & $-\left(13\right)\left(25\right)$ & \multicolumn{1}{c|}{$\left\langle 123333\right\rangle $}\tabularnewline
\cline{1-2} \cline{4-5} 
$\left\langle 313233\right\rangle $ & $-\left(142\right)$ & \multicolumn{1}{c||}{} & $\left\langle 331332\right\rangle $ & $\left(13\right)\left(26\right)$ & \tabularnewline
\cline{1-2} \cline{4-5} 
$\left\langle 313323\right\rangle $ & $\left(152\right)$ & \multicolumn{1}{c||}{} &  &  & \tabularnewline
\hline
$\left\langle 113\left\{ 123\right\} \right\rangle $ & $\left(354\right)\widehat{\vartheta}$ &  & $\left\langle \left\{ 123\right\} 223\right\rangle $ & $-\left(35\right)\widehat{\eta}$ & \tabularnewline
\cline{1-2} \cline{4-5} 
$\left\langle 131\left\{ 123\right\} \right\rangle $ & $-\left(254\right)\widehat{\vartheta}$ & \multicolumn{1}{c||}{$\left\langle 111233\right\rangle $} & $\left\langle \left\{ 123\right\} 232\right\rangle $ & $\left(36\right)\widehat{\eta}$ & \multicolumn{1}{c|}{$\left\langle 122233\right\rangle $}\tabularnewline
\cline{1-2} \cline{4-5} 
$\left\langle 311\left\{ 123\right\} \right\rangle $ & $\left(154\right)\widehat{\vartheta}$ & \multicolumn{1}{c||}{} & $\left\langle \left\{ 123\right\} 322\right\rangle $ & $\left(35\right)\left(46\right)\widehat{\eta}$ & \multicolumn{1}{c|}{}\tabularnewline
\hline
\hline 
\multicolumn{1}{|c}{} & \multicolumn{1}{c}{$X_{3}\left(\Delta_{1}^{\prime},\Delta_{2}^{\prime},\Delta_{3}^{\prime}\right)$} &  & \multicolumn{1}{c}{} & \multicolumn{1}{c}{$X_{3}\left(\Delta_{1}^{\prime},\Delta_{2}^{\prime},\Delta_{3}^{\prime}\right)$} & \tabularnewline
\hline
\hline 
\multicolumn{1}{|c}{} & \multicolumn{1}{c}{$X_{3}\left(+1,0,-1\right)$} &  & \multicolumn{1}{c}{} & \multicolumn{1}{c}{$X_{3}\left(0,+1,-1\right)$} & \tabularnewline
\hline 
$\left\langle x\right\rangle $ & $\widehat{\tilde{\pi}}$ & $\left\langle y\right\rangle $ & $\left\langle x\right\rangle $ & $\widehat{\tilde{\pi}}$ & $\left\langle y\right\rangle $\tabularnewline
\hline
\hline 
$\left\langle 123333\right\rangle $ & $-\left(162\right)$ & $\left\langle 122223\right\rangle $ & $\left\langle 123333\right\rangle $ & $\left(15\right)\left(26\right)$ & $\left\langle 111123\right\rangle $\tabularnewline
\hline 
$\left\langle 111233\right\rangle $ & $-\left(46\right)$ & $\left\langle 111223\right\rangle $ & $\left\langle 111233\right\rangle $ & $\left(15\right)\left(246\right)$ & $\left\langle 112223\right\rangle $\tabularnewline
\hline 
$\left\langle 122233\right\rangle $ & $\left(25\right)\left(36\right)$ & $\left\langle 122333\right\rangle $ & $\left\langle 122233\right\rangle $ & $-\left(135\right)\left(26\right)$ & $\left\langle 112333\right\rangle $\tabularnewline
\hline
\end{tabular}%
\begin{table}[H]
\caption{\label{TabX9}Operators of classes $X_{3}\left(\Delta_{1},\Delta_{2},\Delta_{3}\right)$
with $\Delta_{x}=-1,0,+1$.}

\end{table}

\par\end{center}

\begin{center}
\begin{tabular}{|c|c|c||c|c|c|}
\hline 
\multicolumn{1}{|c}{} & \multicolumn{1}{c}{$X_{3}\left(+2,-2,0\right)$} &  & \multicolumn{1}{c}{} & \multicolumn{1}{c}{$X_{3}\left(+2,-2,0\right)$} & \tabularnewline
\hline 
$\left\langle x_{\pi}\right\rangle $ & $\widehat{\pi}$ & $\left\langle x\right\rangle $ & $\left\langle x_{\pi}\right\rangle $ & $\widehat{\pi}$ & $\left\langle x\right\rangle $\tabularnewline
\hline
\hline 
$\left\langle 113223\right\rangle $ & $-\left(35\right)$ &  & $\left\langle 131322\right\rangle $ & $\left(253\right)\left(46\right)$ & \tabularnewline
\cline{1-2} \cline{4-5} 
$\left\langle 113232\right\rangle $ & $\left(36\right)$ &  & $\left\langle 311223\right\rangle $ & $\left(153\right)$ & \tabularnewline
\cline{1-2} \cline{4-5} 
$\left\langle 113322\right\rangle $ & $\left(35\right)\left(46\right)$ & $\left\langle 112233\right\rangle $ & $\left\langle 311232\right\rangle $ & $-\left(163\right)$ & $\left\langle 112233\right\rangle $\tabularnewline
\cline{1-2} \cline{4-5} 
$\left\langle 131223\right\rangle $ & $-\left(253\right)$ &  & $\left\langle 311322\right\rangle $ & $-\left(153\right)\left(46\right)$ & \tabularnewline
\cline{1-2} \cline{4-5} 
$\left\langle 131232\right\rangle $ & $\left(263\right)$ &  &  &  & \tabularnewline
\hline
\hline 
$X_{3}\left(\Delta_{1}^{\prime},\Delta_{2}^{\prime},\Delta_{3}^{\prime}\right)$ & $\widehat{\tilde{\pi}}$ & $\left\langle y\right\rangle $ & $X_{3}\left(\Delta_{1}^{\prime},\Delta_{2}^{\prime},\Delta_{3}^{\prime}\right)$ & $\widehat{\tilde{\pi}}$ & $\left\langle y\right\rangle $\tabularnewline
\hline
\hline 
$X_{3}\left(+2,0,-2\right)$ & $\left(35\right)\left(46\right)$ & $\left\langle 112233\right\rangle $ & $X_{3}\left(0,+2,-2\right)$ & $\left(135\right)\left(246\right)$ & $\left\langle 112233\right\rangle $\tabularnewline
\hline
\end{tabular}%
\begin{table}[H]
\caption{\label{TabX10}Operators of classes $X_{3}\left(\Delta_{1},\Delta_{2},\Delta_{3}\right)$
with $\Delta_{x}=-2,0,+2$.}

\end{table}

\par\end{center}

\begin{center}
\begin{tabular}{|c|c|c||c|c|c|}
\hline 
\multicolumn{1}{|c}{} & \multicolumn{1}{c}{$X_{4}\left(+1,+1,-1,-1\right)$} &  & \multicolumn{1}{c}{} & \multicolumn{1}{c}{$X_{4}\left(+1,+1,-1,-1\right)$} & \tabularnewline
\hline 
$\left\langle x_{\pi}\right\rangle $ & $\widehat{\pi}$ & $\left\langle x\right\rangle $ & $\left\langle x_{\pi}\right\rangle $ & $\widehat{\pi}$ & $\left\langle x\right\rangle $\tabularnewline
\hline
\hline 
$\left\langle 112\left\{ 134\right\} \right\rangle $ & $\left(34\right)\widehat{\vartheta}$ &  & $\left\langle 122\left\{ 234\right\} \right\rangle $ & $\widehat{\vartheta}$ & \tabularnewline
\cline{1-2} \cline{4-5} 
$\left\langle 121\left\{ 134\right\} \right\rangle $ & $-\left(24\right)\widehat{\vartheta}$ & $\left\langle 111234\right\rangle $ & $\left\langle 212\left\{ 234\right\} \right\rangle $ & $\left(12\right)\widehat{\vartheta}$ & $\left\langle 122234\right\rangle $\tabularnewline
\cline{1-2} \cline{4-5} 
$\left\langle 211\left\{ 134\right\} \right\rangle $ & $\left(14\right)\widehat{\vartheta}$ &  & $\left\langle 221\left\{ 234\right\} \right\rangle $ & $-\left(13\right)\widehat{\vartheta}$ & \tabularnewline
\hline
$\left\langle \left\{ 123\right\} 334\right\rangle $ & $\widehat{\eta}$ &  & $\left\langle \left\{ 124\right\} 344\right\rangle $ & $\left(34\right)\widehat{\eta}$ & \tabularnewline
\cline{1-2} \cline{4-5} 
$\left\langle \left\{ 123\right\} 343\right\rangle $ & $\left(56\right)\widehat{\eta}$ & $\left\langle 123334\right\rangle $ & $\left\langle \left\{ 124\right\} 434\right\rangle $ & $-\left(35\right)\widehat{\eta}$ & $\left\langle 123444\right\rangle $\tabularnewline
\cline{1-2} \cline{4-5} 
$\left\langle \left\{ 123\right\} 433\right\rangle $ & $-\left(46\right)\widehat{\eta}$ &  & $\left\langle \left\{ 124\right\} 443\right\rangle $ & $\left(36\right)\widehat{\eta}$ & \tabularnewline
\hline
\hline 
\multicolumn{1}{|c}{} & \multicolumn{1}{c}{$X_{4}\left(\Delta_{1}^{\prime},\Delta_{2}^{\prime},\Delta_{3}^{\prime},\Delta_{4}^{\prime}\right)$} &  & \multicolumn{1}{c}{} & \multicolumn{1}{c}{$X_{4}\left(\Delta_{1}^{\prime},\Delta_{2}^{\prime},\Delta_{3}^{\prime},\Delta_{4}^{\prime}\right)$} & \tabularnewline
\hline
\hline 
\multicolumn{1}{|c}{} & \multicolumn{1}{c}{$X_{4}\left(+1,-1,+1,-1\right)$} &  & \multicolumn{1}{c}{} & \multicolumn{1}{c}{$X_{4}\left(+1,-1,-1,+1\right)$} & \tabularnewline
\hline 
$\left\langle x\right\rangle $ & $\widehat{\tilde{\pi}}$ & $\left\langle y\right\rangle $ & $\left\langle x\right\rangle $ & $\widehat{\tilde{\pi}}$ & $\left\langle y\right\rangle $\tabularnewline
\hline
\hline 
$\left\langle 111234\right\rangle $ & $\left(45\right)$ & $\left\langle 111234\right\rangle $ & $\left\langle 111234\right\rangle $ & $\left(46\right)$ & $\left\langle 111234\right\rangle $\tabularnewline
\hline 
$\left\langle 122234\right\rangle $ & $\left(25\right)$ & $\left\langle 123334\right\rangle $ & $\left\langle 122234\right\rangle $ & $-\left(26\right)\left(35\right)$ & $\left\langle 123444\right\rangle $\tabularnewline
\hline 
$\left\langle 123334\right\rangle $ & $\left(25\right)$ & $\left\langle 122234\right\rangle $ & $\left\langle 123334\right\rangle $ & $\left(26\right)$ & $\left\langle 123334\right\rangle $\tabularnewline
\hline 
$\left\langle 123444\right\rangle $ & $\left(23\right)$ & $\left\langle 123444\right\rangle $ & $\left\langle 123444\right\rangle $ & $-\left(26\right)\left(35\right)$ & $\left\langle 122234\right\rangle $\tabularnewline
\hline
\end{tabular}%
\begin{table}[H]
\caption{\label{TabX12}Operators of classes $X_{4}\left(\Delta_{1},\Delta_{2},\Delta_{3},\Delta_{4}\right)$
with $\Delta_{x}=-1,+1$.}

\end{table}

\par\end{center}

\begin{center}
\begin{tabular}{|c|c|c||c|c|c|}
\hline 
\multicolumn{1}{|c}{} & \multicolumn{1}{c}{$X_{4}\left(+2,-2,+1,-1\right)$} &  & \multicolumn{1}{c}{} & \multicolumn{1}{c}{$X_{4}\left(+2,-2,+1,-1\right)$} & \tabularnewline
\hline 
$\left\langle x_{\pi}\right\rangle $ & $\widehat{\pi}$ & $\left\langle x\right\rangle $ & $\left\langle x_{\pi}\right\rangle $ & $\widehat{\pi}$ & $\left\langle x\right\rangle $\tabularnewline
\hline
\hline 
$\left\langle 113224\right\rangle $ & $-\left(35\right)$ &  & $\left\langle 131422\right\rangle $ & $\left(253\right)\left(46\right)$ & \tabularnewline
\cline{1-2} \cline{4-5} 
$\left\langle 113242\right\rangle $ & $-\left(356\right)$ &  & $\left\langle 311224\right\rangle $ & $-\left(1532\right)$ & \tabularnewline
\cline{1-2} \cline{4-5} 
$\left\langle 113422\right\rangle $ & $\left(35\right)\left(46\right)$ & $\left\langle 112234\right\rangle $ & $\left\langle 311242\right\rangle $ & $\left(1563\right)$ & $\left\langle 112234\right\rangle $\tabularnewline
\cline{1-2} \cline{4-5} 
$\left\langle 131224\right\rangle $ & $-\left(253\right)$ &  & $\left\langle 311422\right\rangle $ & $-\left(153\right)\left(46\right)$ & \tabularnewline
\cline{1-2} \cline{4-5} 
$\left\langle 131242\right\rangle $ & $-\left(2563\right)$ &  &  &  & \tabularnewline
\hline
\hline 
$X_{4}\left(\Delta_{1}^{\prime},\Delta_{2}^{\prime},\Delta_{3}^{\prime},\Delta_{4}^{\prime}\right)$ & $\widehat{\tilde{\pi}}$ & $\left\langle y\right\rangle $ & $X_{4}\left(\Delta_{1}^{\prime},\Delta_{2}^{\prime},\Delta_{3}^{\prime},\Delta_{4}^{\prime}\right)$ & $\widehat{\tilde{\pi}}$ & $\left\langle y\right\rangle $\tabularnewline
\hline
\hline 
$X_{4}\left(+2,-2,-1,+1\right)$ & $\left(56\right)$ & $\left\langle 112234\right\rangle $ & $X_{4}\left(+1,+2,-2,-1\right)$ & $\left(135\right)$ & $\left\langle 122334\right\rangle $\tabularnewline
\hline 
$X_{4}\left(+2,+1,-2,-1\right)$ & $-\left(35\right)$ & $\left\langle 112334\right\rangle $ & $X_{4}\left(+1,+2,-1,-2\right)$ & $-\left(135\right)\left(46\right)$ & $\left\langle 122344\right\rangle $\tabularnewline
\hline 
$X_{4}\left(+2,+1,-1,-2\right)$ & $\left(35\right)\left(46\right)$ & $\left\langle 112344\right\rangle $ & $X_{4}\left(+1,-2,+2,-1\right)$ & $\left(15\right)\left(24\right)$ & $\left\langle 122334\right\rangle $\tabularnewline
\hline 
$X_{4}\left(+2,-1,-2,+1\right)$ & $-\left(356\right)$ & $\left\langle 112334\right\rangle $ & $X_{4}\left(+1,-2,-1,+2\right)$ & $-\left(15\right)\left(264\right)$ & $\left\langle 122344\right\rangle $\tabularnewline
\hline 
$X_{4}\left(+2,-1,+1,-2\right)$ & $-\left(36\right)\left(45\right)$ & $\left\langle 112344\right\rangle $ & $X_{4}\left(+1,-1,+2,-2\right)$ & $\left(135\right)\left(246\right)$ & $\left\langle 123344\right\rangle $\tabularnewline
\hline 
 &  &  & $X_{4}\left(+1,-1,-2,+2\right)$ & $\left(15\right)\left(26\right)$ & $\left\langle 123344\right\rangle $\tabularnewline
\hline
\end{tabular}%
\begin{table}[H]
\caption{\label{TabX13}Operators of classes $X_{4}\left(\Delta_{1},\Delta_{2},\Delta_{3},\Delta_{4}\right)$
with $\Delta_{x}=-2,-1,+1,+2$.}

\end{table}

\par\end{center}

\begin{multicols}{2}

\begin{center}
\begin{tabular}{|c|c|c|}
\hline 
\multicolumn{1}{|c}{} & \multicolumn{1}{c}{$X_{4}\left(+3,-1,-1,-1\right)$} & \tabularnewline
\hline 
$\left\langle x_{\pi}\right\rangle $ & $\widehat{\pi}$ & $\left\langle x\right\rangle $\tabularnewline
\hline
\hline 
$\left\langle 111\left\{ 234\right\} \right\rangle $ & $\widehat{\vartheta}$ & $\left\langle 111234\right\rangle $\tabularnewline
\hline
\hline 
$X_{4}\left(\Delta_{1}^{\prime},\Delta_{2}^{\prime},\Delta_{3}^{\prime},\Delta_{4}^{\prime}\right)$ & $\widehat{\tilde{\pi}}$ & $\left\langle y\right\rangle $\tabularnewline
\hline
\hline 
$X_{4}\left(-1,+3,-1,-1\right)$ & $\left(14\right)$ & $\left\langle 122234\right\rangle $\tabularnewline
\hline 
$X_{4}\left(-1,-1,+3,-1\right)$ & $-\left(15\right)\left(24\right)$ & $\left\langle 123334\right\rangle $\tabularnewline
\hline 
$X_{4}\left(-1,-1,-1,+3\right)$ & $\left(16\right)\left(24\right)\left(35\right)$ & $\left\langle 123444\right\rangle $\tabularnewline
\hline
\end{tabular}%
\begin{table}[H]
\caption{\label{TabX14}Operators of classes $X_{4}\left(\Delta_{1},\Delta_{2},\Delta_{3},\Delta_{4}\right)$
with $\Delta_{x}=-1,+3$.}

\end{table}

\par\end{center}

\begin{center}
\begin{tabular}{|c|c|c|}
\hline 
\multicolumn{1}{|c}{} & \multicolumn{1}{c}{$X_{4}\left(+1,-1,0,0\right)$} & \tabularnewline
\hline 
$\left\langle x_{\pi}\right\rangle $ & $\widehat{\pi}$ & $\left\langle x\right\rangle $\tabularnewline
\hline
\hline 
$\left\langle \left\{ 134\right\} \left\{ 234\right\} \right\rangle $ & $-\left(24\right)\left(35\right)\widehat{\eta}\widehat{\vartheta}$ & $\left\langle 123344\right\rangle $\tabularnewline
\hline
\hline 
$X_{4}\left(\Delta_{1}^{\prime},\Delta_{2}^{\prime},\Delta_{3}^{\prime},\Delta_{4}^{\prime}\right)$ & $\widehat{\tilde{\pi}}$ & $\left\langle y\right\rangle $\tabularnewline
\hline
\hline 
$X_{4}\left(+1,0,-1,0\right)$ & $-\left(24\right)$ & $\left\langle 122344\right\rangle $\tabularnewline
\hline 
$X_{4}\left(+1,0,0,-1\right)$ & $\left(26\right)\left(35\right)$ & $\left\langle 122334\right\rangle $\tabularnewline
\hline 
$X_{4}\left(0,+1,-1,0\right)$ & $\left(13\right)\left(24\right)$ & $\left\langle 112344\right\rangle $\tabularnewline
\hline 
$X_{4}\left(0,+1,0,-1\right)$ & $-\left(135\right)\left(26\right)$ & $\left\langle 112334\right\rangle $\tabularnewline
\hline 
$X_{4}\left(0,0,+1,-1\right)$ & $\left(15\right)\left(26\right)$ & $\left\langle 112234\right\rangle $\tabularnewline
\hline
\end{tabular}%
\begin{table}[H]
\caption{\label{TabX15}Operators of classes $X_{4}\left(\Delta_{1},\Delta_{2},\Delta_{3},\Delta_{4}\right)$
with $\Delta_{x}=-1,0,+1$.}

\end{table}

\par\end{center}

\end{multicols}

\begin{center}
\begin{tabular}{|c|c|c|}
\hline 
\multicolumn{1}{|c}{} & \multicolumn{1}{c}{$X_{4}\left(+2,-1,-1,0\right)$} & \tabularnewline
\hline 
$\left\langle x_{\pi}\right\rangle $ & $\widehat{\pi}$ & $\left\langle x\right\rangle $\tabularnewline
\hline
\hline 
$\left\langle 114\left\{ 234\right\} \right\rangle $ & $\left(354\right)\widehat{\vartheta}$ & \tabularnewline
\cline{1-2} 
$\left\langle 141\left\{ 234\right\} \right\rangle $ & $\left(2543\right)\widehat{\vartheta}$ & $\left\langle 112344\right\rangle $\tabularnewline
\cline{1-2} 
$\left\langle 411\left\{ 234\right\} \right\rangle $ & $-\left(1543\right)\widehat{\vartheta}$ & \tabularnewline
\hline
\hline 
$X_{4}\left(\Delta_{1}^{\prime},\Delta_{2}^{\prime},\Delta_{3}^{\prime},\Delta_{4}^{\prime}\right)$ & $\widehat{\tilde{\pi}}$ & $\left\langle y\right\rangle $\tabularnewline
\hline
\hline 
$X_{4}\left(-1,+2,-1,0\right)$ & $-\left(13\right)$ & $\left\langle 122344\right\rangle $\tabularnewline
\hline 
$X_{4}\left(+2,-1,0,-1\right)$ & $-\left(46\right)$ & $\left\langle 112334\right\rangle $\tabularnewline
\hline 
$X_{4}\left(+2,0,-1,-1\right)$ & $-\left(36\right)\left(45\right)$ & $\left\langle 112234\right\rangle $\tabularnewline
\hline 
$X_{4}\left(0,+2,-1,-1\right)$ & $-\left(145\right)\left(236\right)$ & $\left\langle 112234\right\rangle $\tabularnewline
\hline 
$X_{4}\left(0,-1,+2,-1\right)$ & $-\left(15\right)\left(2436\right)$ & $\left\langle 112334\right\rangle $\tabularnewline
\hline 
$X_{4}\left(-1,+2,0,-1\right)$ & $\left(1364\right)$ & $\left\langle 122334\right\rangle $\tabularnewline
\hline 
$X_{4}\left(-1,-1,+2,0\right)$ & $\left(13\right)\left(24\right)$ & $\left\langle 123344\right\rangle $\tabularnewline
\hline 
$X_{4}\left(-1,-1,0,+2\right)$ & $-\left(164\right)\left(253\right)$ & $\left\langle 123344\right\rangle $\tabularnewline
\hline 
$X_{4}\left(-1,0,+2,-1\right)$ & $\left(14\right)\left(25\right)\left(36\right)$ & $\left\langle 122334\right\rangle $\tabularnewline
\hline 
$X_{4}\left(-1,0,-1,+2\right)$ & $-\left(1634\right)\left(25\right)$ & $\left\langle 122344\right\rangle $\tabularnewline
\hline 
$X_{4}\left(0,-1,-1,+2\right)$ & $\left(15\right)\left(26\right)$ & $\left\langle 112344\right\rangle $\tabularnewline
\hline
\end{tabular}%
\begin{table}[H]
\caption{\label{TabX16}Operators of classes $X_{4}\left(\Delta_{1},\Delta_{2},\Delta_{3},\Delta_{4}\right)$
with $\Delta_{x}=-1,0,+2$.}

\end{table}

\par\end{center}

\begin{center}
\begin{tabular}{|c|c|c|}
\hline 
\multicolumn{1}{|c}{} & \multicolumn{1}{c}{$X_{5}\left(+2,+1,-1,-1,-1\right)$} & \tabularnewline
\hline 
$\left\langle x_{\pi}\right\rangle $ & $\widehat{\pi}$ & $\left\langle x\right\rangle $\tabularnewline
\hline
\hline 
$\left\langle 112\left\{ 345\right\} \right\rangle $ & $\widehat{\vartheta}$ & \tabularnewline
\cline{1-2} 
$\left\langle 121\left\{ 345\right\} \right\rangle $ & $\left(23\right)\widehat{\vartheta}$ & $\left\langle 112345\right\rangle $\tabularnewline
\cline{1-2} 
$\left\langle 211\left\{ 345\right\} \right\rangle $ & $-\left(13\right)\widehat{\vartheta}$ & \tabularnewline
\hline
\hline 
$X_{5}\left(\Delta_{1}^{\prime},\Delta_{2}^{\prime},\Delta_{3}^{\prime},\Delta_{4}^{\prime},\Delta_{5}^{\prime}\right)$ & $\widehat{\tilde{\pi}}$ & $\left\langle y\right\rangle $\tabularnewline
\hline
\hline 
$X_{5}\left(+2,-1,+1,-1,-1\right)$ & $\left(34\right)$ & \tabularnewline
\cline{1-2} 
$X_{5}\left(+2,-1,-1,+1,-1\right)$ & $\left(35\right)$ & $\left\langle 112345\right\rangle $\tabularnewline
\cline{1-2} 
$X_{5}\left(+2,-1,-1,-1,+1\right)$ & $\left(36\right)$ & \tabularnewline
\hline 
$X_{5}\left(+1,+2,-1,-1,-1\right)$ & $-\left(13\right)$ & \tabularnewline
\cline{1-2} 
$X_{5}\left(-1,+2,+1,-1,-1\right)$ & $-\left(134\right)$ & $\left\langle 122345\right\rangle $\tabularnewline
\cline{1-2} 
$X_{5}\left(-1,+2,-1,+1,-1\right)$ & $-\left(135\right)$ & \tabularnewline
\cline{1-2} 
$X_{5}\left(-1,+2,-1,-1,+1\right)$ & $-\left(136\right)$ & \tabularnewline
\hline 
$X_{5}\left(-1,-1,+2,+1,-1\right)$ & $\left(135\right)\left(24\right)$ & \tabularnewline
\cline{1-2} 
$X_{5}\left(-1,+1,+2,-1,-1\right)$ & $-\left(14\right)\left(23\right)$ & $\left\langle 123345\right\rangle $\tabularnewline
\cline{1-2} 
$X_{5}\left(+1,-1,+2,-1,-1\right)$ & $\left(13\right)\left(24\right)$ & \tabularnewline
\cline{1-2} 
$X_{5}\left(-1,-1,+2,-1,+1\right)$ & $\left(136\right)\left(24\right)$ & \tabularnewline
\hline 
$X_{5}\left(-1,+1,-1,+2,-1\right)$ & $\left(14\right)\left(253\right)$ & \tabularnewline
\cline{1-2} 
$X_{5}\left(+1,-1,-1,+2,-1\right)$ & $-\left(153\right)\left(24\right)$ & $\left\langle 123445\right\rangle $\tabularnewline
\cline{1-2} 
$X_{5}\left(-1,-1,+1,+2,-1\right)$ & $-\left(15\right)\left(24\right)$ & \tabularnewline
\cline{1-2} 
$X_{5}\left(-1,-1,-1,+2,+1\right)$ & $\left(1436\right)\left(25\right)$ & \tabularnewline
\hline 
$X_{5}\left(+1,-1,-1,-1,+2\right)$ & $\left(1543\right)\left(26\right)$ & \tabularnewline
\cline{1-2} 
$X_{5}\left(-1,-1,+1,-1,+2\right)$ & $-\left(16\right)\left(254\right)$ & $\left\langle 123455\right\rangle $\tabularnewline
\cline{1-2} 
$X_{5}\left(-1,+1,-1,-1,+2\right)$ & $\left(154\right)\left(263\right)$ & \tabularnewline
\cline{1-2} 
$X_{5}\left(-1,-1,-1,+1,+2\right)$ & $\left(15\right)\left(2634\right)$ & \tabularnewline
\hline
\end{tabular}%
\begin{table}[H]
\caption{\label{TabX18}Operators of classes $X_{5}\left(\Delta_{1},\Delta_{2},\Delta_{3},\Delta_{4},\Delta_{5}\right)$
with $\Delta_{x}=-1,+1,+2$.}

\end{table}

\par\end{center}

\begin{center}
\begin{tabular}{|c|c|c|}
\hline 
\multicolumn{1}{|c}{} & \multicolumn{1}{c}{$X_{5}\left(+1,+1,-1,-1,0\right)$} & \tabularnewline
\hline 
$\left\langle x_{\pi}\right\rangle $ & $\widehat{\pi}$ & $\left\langle x\right\rangle $\tabularnewline
\hline
\hline 
$\left\langle \left\{ 125\right\} \left\{ 345\right\} \right\rangle $ & $\left(354\right)\widehat{\eta}\widehat{\vartheta}$ & $\left\langle 123455\right\rangle $\tabularnewline
\hline
\hline 
$X_{5}\left(\Delta_{1}^{\prime},\Delta_{2}^{\prime},\Delta_{3}^{\prime},\Delta_{4}^{\prime},\Delta_{5}^{\prime}\right)$ & $\widehat{\tilde{\pi}}$ & $\left\langle y\right\rangle $\tabularnewline
\hline
\hline 
$X_{5}\left(+1,-1,+1,-1,0\right)$ & $\left(23\right)$ & $\left\langle 123455\right\rangle $\tabularnewline
\cline{1-2} 
$X_{5}\left(+1,-1,-1,+1,0\right)$ & $\left(24\right)$ & \tabularnewline
\hline 
$X_{5}\left(+1,+1,-1,0,-1\right)$ & $-\left(46\right)$ & \tabularnewline
\cline{1-2} 
$X_{5}\left(+1,-1,+1,0,-1\right)$ & $-\left(23\right)\left(46\right)$ & $\left\langle 123445\right\rangle $\tabularnewline
\cline{1-2} 
$X_{5}\left(+1,-1,-1,0,+1\right)$ & $-\left(264\right)$ & \tabularnewline
\hline 
$X_{5}\left(+1,+1,0,-1,-1\right)$ & $\left(35\right)\left(46\right)$ & \tabularnewline
\cline{1-2} 
$X_{5}\left(+1,-1,0,+1,-1\right)$ & $-\left(254\right)\left(36\right)$ & $\left\langle 123345\right\rangle $\tabularnewline
\cline{1-2} 
$X_{5}\left(+1,-1,0,-1,+1\right)$ & $-\left(263\right)\left(45\right)$ & \tabularnewline
\hline 
$X_{5}\left(+1,0,+1,-1,-1\right)$ & $\left(245\right)\left(36\right)$ & \tabularnewline
\cline{1-2} 
$X_{5}\left(+1,0,-1,+1,-1\right)$ & $\left(25\right)\left(346\right)$ & $\left\langle 122345\right\rangle $\tabularnewline
\cline{1-2} 
$X_{5}\left(+1,0,-1,-1,+1\right)$ & $-\left(26\right)\left(345\right)$ & \tabularnewline
\hline 
$X_{5}\left(0,+1,+1,-1,-1\right)$ & $-\left(146\right)\left(235\right)$ & \tabularnewline
\cline{1-2} 
$X_{5}\left(0,+1,-1,+1,-1\right)$ & $\left(15\right)\left(2346\right)$ & $\left\langle 112345\right\rangle $\tabularnewline
\cline{1-2} 
$X_{5}\left(0,+1,-1,-1,+1\right)$ & $-\left(16\right)\left(2345\right)$ & \tabularnewline
\hline
\end{tabular}%
\begin{table}[H]
\caption{\label{TabX19}Operators of classes $X_{5}\left(\Delta_{1},\Delta_{2},\Delta_{3},\Delta_{4},\Delta_{5}\right)$
with $\Delta_{x}=-1,0,+1$.}

\end{table}

\par\end{center}

\begin{center}
\begin{tabular}{|c|c|c|}
\hline 
\multicolumn{1}{|c}{} & \multicolumn{1}{c}{$X_{6}\left(+1,+1,+1,-1,-1,-1\right)$} & \tabularnewline
\hline 
$\left\langle x_{\pi}\right\rangle $ & $\widehat{\pi}$ & $\left\langle x\right\rangle $\tabularnewline
\hline
\hline 
$\left\langle \left\{ 123\right\} \left\{ 456\right\} \right\rangle $ & $\widehat{\eta}\widehat{\vartheta}$ & $\left\langle 123456\right\rangle $\tabularnewline
\hline
\hline 
$X_{6}\left(\Delta_{1}^{\prime},\Delta_{2}^{\prime},\Delta_{3}^{\prime},\Delta_{4}^{\prime},\Delta_{5}^{\prime},\Delta_{6}^{\prime}\right)$ & $\widehat{\tilde{\pi}}$ & $\left\langle y\right\rangle $\tabularnewline
\hline
\hline 
$X_{6}\left(+1,+1,-1,+1,-1,-1\right)$ & $\left(34\right)$ & \tabularnewline
\cline{1-2} 
$X_{6}\left(+1,+1,-1,-1,+1,-1\right)$ & $\left(35\right)$ & \tabularnewline
\cline{1-2} 
$X_{6}\left(+1,+1,-1,-1,-1,+1\right)$ & $\left(36\right)$ & \tabularnewline
\cline{1-2} 
$X_{6}\left(+1,-1,+1,+1,-1,-1\right)$ & $\left(24\right)$ & \tabularnewline
\cline{1-2} 
$X_{6}\left(+1,-1,+1,-1,+1,-1\right)$ & $\left(254\right)$ & $\left\langle 123456\right\rangle $\tabularnewline
\cline{1-2} 
$X_{6}\left(+1,-1,+1,-1,-1,+1\right)$ & $\left(264\right)$ & \tabularnewline
\cline{1-2} 
$X_{6}\left(+1,-1,-1,+1,+1,-1\right)$ & $\left(24\right)\left(35\right)$ & \tabularnewline
\cline{1-2} 
$X_{6}\left(+1,-1,-1,+1,-1,+1\right)$ & $\left(24\right)\left(36\right)$ & \tabularnewline
\cline{1-2} 
$X_{6}\left(+1,-1,-1,-1,+1,+1\right)$ & $\left(25\right)\left(36\right)$ & \tabularnewline
\hline
\end{tabular}%
\begin{table}[H]
\caption{\label{TabX21}Operators of classes $X_{6}\left(\Delta_{1},\Delta_{2},\Delta_{3},\Delta_{4},\Delta_{5},\Delta_{6}\right)$
with $\Delta_{x}=-1,+1$.}

\end{table}

\par\end{center}

}

\end{document}